\documentclass[12pt, preprint]{aastex}

\begin{document}

\title{The Behavior of Novae Light Curves Before Eruption}
\author{Andrew C. Collazzi, Bradley E. Schaefer, Limin Xiao and Ashley Pagnotta \affil{Physics and Astronomy, Louisiana State University, Baton Rouge, LA 70803}
Peter Kroll and Klaus L\"{o}chel \affil{Sonneberg Observatory, D 96515 Sonneberg, Germany}}
\author{Arne A. Henden \affil{American Association of Variable Star Observers, Cambridge, MA 02138}}

\begin{abstract}
In 1975, E. R. Robinson conducted the hallmark study of the behavior of classical nova light curves before eruption, and this  work has now become part of the standard knowledge of novae.  He made three points; that 5 out of 11 novae showed pre-eruption rises in the years before eruption, that one nova (V446 Her) showed drastic changes in the variability across eruptions, and that all but one of the novae (excepting BT Mon) have the same quiescent magnitudes before and after the outburst.  This work has not been tested since it came out.  We have now tested these results by going back to the original archival photographic plates and measuring large numbers of pre-eruption magnitudes for many novae using comparison stars on a modern magnitude scale.  We find in particular that four out of five claimed pre-eruption rises are due to simple mistakes in the old literature, that V446 Her has the same amplitude of variations across its 1960 eruption, and that BT Mon has essentially unchanged brightness across its 1939 eruption.  Out of 22 nova eruptions, we find two confirmed cases of significant pre-eruption rises (for V533 Her and V1500 Cyg), while T CrB has a deep pre-eruption dip.  These events are a challenge to theorists.  We find no significant cases of changes in variability across 27 nova eruptions beyond what is expected due to the usual fluctuations seen in novae away from eruptions.  For 30 classical novae plus 19 eruptions from 6 recurrent novae, we find that the average change in magnitude from before the eruption to long after the eruption is 0.0 mag.  However, we do find five novae (V723 Cas, V1500 Cyg, V1974 Cyg, V4633 Sgr, and RW UMi) that have significantly large changes, in that the post-eruption quiescent brightness level is over ten times brighter than the pre-eruption level.  These large post-eruption brightenings are another challenge to theorists.
\end{abstract}

\section{Introduction}

A nova is a cataclysmic variable (CV) binary system in which a white dwarf (WD) is accreting mass usually by Roche lobe overflow from its companion star, usually a late-type dwarf star. The eruption itself is a thermonuclear runaway of the hydrogen accumulated on the surface of the WD. The eruption ejects a fast expanding shell, which causes the temporary brightening, after which the star returns to its original state. In quiescence, the light from these systems is dominated by the hot spot and accretion disk around the WD (King 1989). Therefore, by following the optical light from a nova system, we are also tracking how the matter flow of the system changes over time. The matter flow is not expected to change as a result of the eruption, so the brightness of the system is expected to be the same after it returns to quiescence as it was before the eruption. Likewise, there is no expectation for a nova to `anticipate' the eruption and change the accretion rate in the years leading up to eruption.

A large analysis of historical nova data was done by Robinson (1975, henceforth referred to as R75) with 33 novae events.  Of these 33 events, only 11 novae had sufficient data for proper analysis. Of these 11 novae, R75 identified five (V533 Her, CP Lac, BT Mon, GK Per, LV Vul) as showing a significant brightening in the years leading up to the eruption. In addition, one nova (V446 Her) was identified as showing considerably less variability after the eruption as compared to before the eruption.  (R75 also identified the same behavior in RR Tel, which has since been identified as a symbiotic star.)

The R75 collection of nova light curves presents the unsettling result that of 5 out of 11 novae show a pre-eruption rise. This can lead to the conclusion that anticipation is not only common, but is seen to happen almost half of the time (Warner 2008; R75). Nonetheless, the result is unexpected as it indicates that the donor star somehow knows to increase the accretion rate onto the WD in the years leading up to the eruption. If the pre-eruption rise phenomenon is real, some physical mechanism must cause the donor star to increase matter flow to the accretor years in advance. While there are certainly ways for the matter flow to increase in the system, it is unlikely that such a increase would occur as a result of the system anticipating a upcoming eruption. 

R75 also presented the unexpected result of V446 Her showing variability on the order of 4 magnitudes before the eruption, and only a variability of 0.4 magnitudes after the eruption. CVs are known to vary, even on long time scales such as this (Kafka \& Honeycutt 2004; Honeycutt et al. 1998; Schaefer 2009b), however, the sudden change from large variability to small variability points to the nova being directly responsible for the change. The matter flow would have to become significantly less erratic for such a change to occur, and no mechanism is known to account for such a change.

Finally, R75 asks whether a nova's average quiescent magnitude changes across the eruption.  That is, what is the difference between the pre-eruption magnitude ($m_{pre}$) and the post-eruption magnitude ($m_{post}$)?  This is yet a third way of asking about the nature of the effect of the nova eruption on the mass transfer rate. If the system is undisturbed by the eruption itself, the matter flow should not change, and hence the system will show the same brightness before and after the eruption.  Two competing effects are known, with the mass loss from the nova eruption forcing the system slightly apart so that the rate of Roche lobe overflow should decline and the accretion disk should dim somewhat, versus the hot white dwarf (made hot by the nova event) irradiating the companion star such that the deposited energy puffs up the companion's atmosphere and drives a slightly higher accretion rate which makes the accretion disk somewhat brighter.  For this question, R75 found that ``with the possible exception of BT Mon, the preeruption and posteruption magnitudes are the same for all 18 stars for which both magnitudes are known.''   This result gives good information on the relative size and frequency of the effects due to the separation of the stars and the irradiation of the companion.  For larger issues, the value of $m_{pre}-m_{post}$ is central to testing models of nova hibernation (e.g., Shara 1989; Retter \& Naylor 2000) and to the idea that the turn-on of a supersoft source can raise the accretion rate enough to be self-sustaining (van Teesling \& King 1998; Knigge et al. 2000).

R75 presented three conclusions about classical novae (almost half with pre-eruption rises, one changed its variability, and $m_{pre}\approx m_{post}$) that have now become part of the basic knowledge about novae and are presented with discussion in standard reviews (Warner 1989; 2002; 2008).  Surprisingly, no one has revisited, tested, or extended the original conclusion from R75.  Perhaps part of the reason for this is that pre-eruption magnitudes must come from serendipitous observations residing on archival photographic plates, and the astronomical community has largely lost the knowledge of, and ability to access, the archival data.

As all of the novae described in R75's paper erupted before digital storage of data, the only way to get pre-eruption magnitudes is to go to plate archives.  With the whole sky being covered by archival photographic plates starting around the year 1890, so, ideally, all novae after 1891 should have a pre-eruption magnitude available.  This ideal is seldom achievable because the quiescent magnitudes are often below the plate limits.  Quiescent magnitudes of novae range from 11 to fainter than 21, with a median of around 18 mag.  Plate limiting magnitudes ($m_{lim}$) vary widely, with most plates reaching 11 mag. The median for good quality plates is perhaps 14-15 mag, most of the sky is covered by plates down to 16 or 17 mag, and a few plates record parts of the sky to 18 mag and deeper.  In addition, the original Palomar Sky Survey in the 1950s covers the whole sky north of $-30\degr$ declination to roughly 21 mag in the B-band.  With this, most novae are too faint to get any measure of $m_{pre}$, most novae brighter than $\sim15$ mag should have good pre-eruption light curves available, and a fraction of the novae with $15\lesssim m_{pre} \lesssim 18$ should have spotty coverage.  Many novae which erupted after 1960 have at least one $m_{pre}$ measure from the Palomar plates.

R75's light curves were constructed entirely from the old literature.  This has a substantial and serious problem because the older magnitudes are always in error, often by more than one magnitude.  The general problem is that the old comparison star sequences have systematic errors of 0.8-1.3 mag due to incorrect calibration of the North Polar Sequence and the later Harvard-Groningen Selected Areas (Sandage 2001).  Our own studies of Pluto, many supernovae, some novae, and several eclipsing binaries show errors in old comparison star sequences from 0.3-0.9 mag (Schaefer 1994, 1995, 1996, 1998; Schaefer et al. 2008).  The old literature magnitudes cannot be used for comparison with later values.  Nevertheless, the old published light curves are internally consistent (albeit with some unknown constant offset) and thus can be used by themselves to look for pre-eruption rises or changes in brightness across an eruption.  In some cases, it might be possible to convert older magnitude measurements into modern ones by changing the comparison star sequence into a modern one.  The best option is simply to look at the original plates and measure the nova magnitude with respect to comparison stars with modern magnitude measures.

Since the R75 results are a cornerstone of our modern picture of novae, this result should be tested and extended.  The only real way to do this is to check the original plates and use modern comparison star sequences. The premier sources of archival plates are Harvard College Observatory (HCO) in Cambridge, Massachusetts, and Sonneberg Observatory in Sonneberg, Germany.  These two plate collections have roughly three-quarters of the world's deep direct images from pre-CCD times.  Most of the data used for the 5 novae that show pre-eruption rises (and the 1 which showed a change in variability) in R75's sample comes from these two locales. Therefore, we set out to perform an independent test to see whether these novae really do exhibit the described behavior by examining the original photographic plates ourselves.  We place the pre-eruption light curves onto a modern magnitude scale, and we often can find many more magnitudes than were originally published.  Also, we can extend the investigation of pre-eruption light curves to many more novae.

In this paper, we first start by presenting the new light curves (which were largely constructed by personal examination the archival plates) in Section 2. In Section 3, we address the question of the existence and frequency of pre-eruption rises in novae, and extend this to other types of anticipatory behavior. In Section 4, we address the question of whether there is evidence for variability change across the nova eruption. In Section 5, we go into detail regarding the question of comparing pre- and post-eruption magnitudes of a nova. In Section 6, we summarize our results.

\section{New Data}

In this section, we present our new data. First, we present the data that have been obtained via archival plates (Section 2.1); second, we present data obtained from our extensive work on recurrent novae (Section 2.2); lastly, we present magnitudes we got from digital sky survey plates (such as POSS I and POSS II) and measurements of some of these nova from literature (Section 2.3).

\subsection{Data From Archival Plates}

Ideally, we could extract pre-eruption light curves from Harvard and Sonneberg plates for around two dozen novae and five recurrent novae.  In practice, with limited time, we could not examine all these, so instead we did thorough searches for the ten most important classical novae and for all known recurrent novae.  Here, the important novae are those with reported pre-eruption rises (BT Mon, LV Vul, GK Per, CP Lac, and V533 Her), with a change in variability (V446 Her), and where a pre-eruption orbital period might be discovered (QZ Aur and V368 Aql).  We also constructed pre-eruption light curves for DQ Her and HR Del.  Our magnitudes for BT Mon and DQ Her have already been presented in Schaefer \& Patterson (1983), our magnitudes for QZ Aur are being presented and fully discussed in Xiao et al. (2009), while our exhaustive collection of photometry for all the recurrent novae are presented in Schaefer (2009b).

Harvard has roughly 500,000 plates covering 1890 to 1953 (plus some in the 1980s), while Sonneberg has roughly 300,000 plates covering 1925 to present.  Details of the various series of photographic plates are given in Table \ref{LEGEND}.  The columns give the observatory, the series identifier, the telescope diameter (in inches), the plate scale (in arc-seconds per millimeter on the plate), the typical limiting magnitude for the plates in that series, and the years over which the series was taken.  The coverage for the useful plates that we examined for all ten classical novae is presented in Table \ref{SUM}.  The first column gives the nova designation, the second column gives the year of the nova eruption, the third column gives the number of useful plates that we measured, and the last column gives the years for these plates.  The full details and magnitudes for the recurrent novae are given in Schaefer (2009b).

For each plate, we noted its identification number (consisting of series designation followed by a sequential number) and date.  The date is always on the plate or its storage envelope and is expressed either as a calendar day or as the Julian Date (JD).  Often, the Julian Dates are given to the fraction of a day.  Occasionally, the day fraction is not readily available (although it can always be calculated accurately from the start times recorded in log books), but we have not invested the large amount of time required to extract the day fraction, as it has no utility for this study.  In these cases, we express the JD with only 0.1 day precision, with an accuracy of roughly 0.5 days.

Magnitude measurements were taken by visually examining each (back-illuminated) plate through a handheld loupe or a microscope.   Magnitude estimates were made by directly comparing the radius of the nova image against the radii of nearby comparison stars whose brightness is known from modern measures.  Most of our comparison star magnitudes were provided by our own program for calibration of sequences now carried out as part of the {\it American Association of Variable Star Observers} (AAVSO) program (e.g., Henden \& Honeycutt 1997; Henden \& Munari 2006).  We have extensive experience, and have done a number of quantitative studies, which show that our visual method is comparable in accuracy with methods based on two-dimensional scans of the plates and with use of an Iris Diaphragm Photometer (e.g., Schaefer et al. 2008; Schaefer \& Fried 1991; Schaefer 1981; Schaefer \& Patterson 1983).  With these and other studies, we can quantitatively determine the reproducability and the absolute accuracy of the visual estimates.  The typical measurement accuracy for archival plates varies substantially with the quality of the plate, but in the case of a close comparison star sequence (which we always have in this study) and reasonable quality plates, the typical one-sigma accuracy is 0.15 mag.  In general, we cannot make an error estimate appropriate for one plate, and the uncertainties are likely to be similar (as the plates must have good quality to show the faint novae), so we will adopt 0.15 mag as our general error bar.  Because of this, we have not included the error bars in our tables, as the ``$\pm0.15$ mag" can be taken as a given.

All the archival plates we have used have a sensitivity to the Johnson B-band.  Indeed, the Harvard plates provided the original definition of the B-band.  The old magnitudes reported in the literature with the poorly calibrated sequences were labeled as `photographic magnitudes',  but with modern comparison sequences, the differential magnitudes from the old plates are now exactly in the modern Johnson B-magnitude system.

In the subsections below, we report on the background, magnitudes, and results from our analysis of archival plates.  For each nova, we report three quantities for both pre-eruption and post-eruption light curves.  The first is the average magnitude, either for the pre-eruption ($m_{pre}$) or for the post-eruption ($m_{post}$).  The second is the RMS scatter (i.e., the standard deviation) for the pre-eruption and post-eruption light curves ($\sigma_{pre}$ and $\sigma_{post}$).  The third is the total range of variability observed ($R_{pre}$ and $R_{post}$) in magnitudes.  We will keep track of these results in Table \ref{SUM2}.

\subsubsection{V368 Aql (Nova 1936 Aql)}

V368 Aql was discovered on 1936 October 7 (JD 2428449) by N. Tamm via a photographic plate taken at Kvistaberg Observatory (Tamm 1936). It was observed to get as bright as 6.6 photographic mag on 1936 September 24 (JD 2428436)(Klemola 1968). Recently, the orbital period of V368 Aql has been determined to be 0.6905 hours, twice what was previously reported (Shafter et al. 2008). Several attempts have been made to obtain pre-eruption measurements (Hoffmeister 1936; Beyer 1936; Hinderer 1936), however, these resulted in limiting magnitudes only. 

\begin{figure}[h!]
\begin{center}
\includegraphics[scale=0.5]{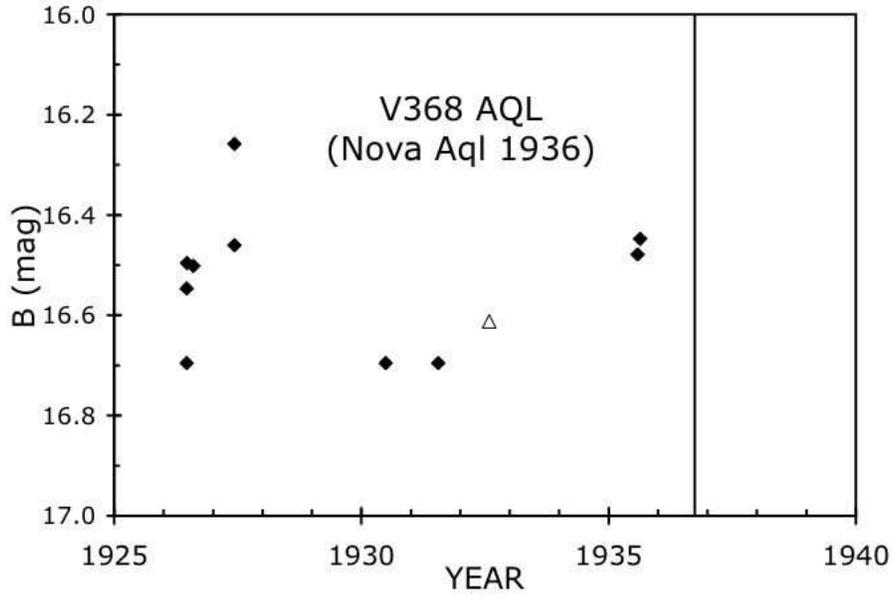}
\end{center}
\caption{V368 Aql light curve. The diamonds represent B magnitude measurements and the triangle represents an upper limit. The vertical line denotes the time of the eruption of V368 Aql. We observe the system to have had an average magnitude of 16.53 before the eruption, an RMS of 0.14, and spanned a range of 0.44 mag.}
\label{V368AQLLC}
\end{figure}

While V368 Aql was not addressed in R75, we have constructed a light curve from archival plates obtained at HCO.  We obtained 11 total measurements, all in Johnson B magnitudes, one of which is a limit (Table \ref{V368AQLMAGS}, Figure \ref{V368AQLLC}). The light curve shows that V368 Aql had an average brightness of $m_{pre} =16.53$. Throughout the nearly 10 years leading up to the eruption, we observe it to have a range of of $R_{pre}=0.44$ mag. The standard deviation of these magnitudes is $\sigma_{pre}=0.14$ mag.  

Szkody (1994) gives V368 Aql to be at B=16.90 and V=16.23 on 1988 August 31 (JD 2447405). The difference in B-band magnitude across the eruption, $\Delta m = m_{pre}-m_{post}$ is -0.37 mag.

Our observations show no evidence of anomalous behavior. The system, while showing a variability that ranges almost half a magnitude, never shows an anticipatory brightening (or dimming) event.  (If we only had the magnitudes after 1930, then an incautious reporter might suggest a pre-eruption rise.  But we do have earlier data, and we see that V368 Aql had other similar minor peaks during quiescence without any nova eruption, so we can readily see that the 1930-1935 brightening is not a pre-eruption rise.  This is a strong lesson that we should not be impressed by small rises in sketchy light curves.) In addition, our $\Delta m$ is not significantly different from zero when compared with the normal range of variations.

\subsubsection{QZ Aur (Nova Aur 1964)}

N. Sanduleak discovered QZ Aur over a decade after its eruption on an objective-prism plate taken at Swasey Observatory on 1964 November 4 (JD 2438704)(Sanduleak 1975).  On checking the archival plates at Sonneberg, the outburst was observed to be as bright as $\sim6.0$ mag (Gessner 1975).  The system has a period of 0.357496 days (Campbell $\&$ Shafter 1995). 

\begin{figure}[h!]
\begin{center}
\includegraphics[scale=0.5]{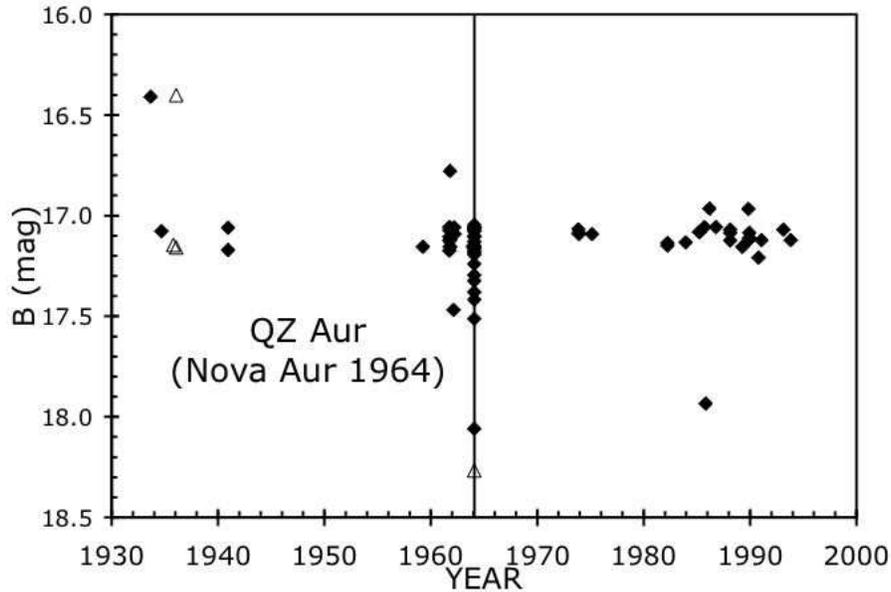}
\end{center}
\caption{QZ Aur light curve. Diamonds denote measured B magnitudes.  The vertical line denotes the date of the eruption.  The two faintest points just before the eruption and the one faintest point after the eruption are due to total eclipses. Ignoring the eclipses, QZ has a similar and fairly small range of variation before and after the eruption (other than two anomolous bright points in 1934 and 1962). Before the eruption, QZ Aur had an average magnitude of B=17.16, an RMS of 0.23, and a range of 1.65. After the eruption, QZ Aur showed an average magnitude of B=17.13, an RMS of 0.18 and a range of 0.97.  The nova shows no pre-eruption rise, which has been tested up to the month before eruption. QZ Aur has almost identical average brightnesses before and after the eruption.}
\label{QZAURLC}
\end{figure}

While not part of R75, we were interested in QZ Aur as a deep eclipsing binary, hoping to use the archival plates to determine an accurate pre-eruption orbital period.  While no positive detections of the quiescent system were made with the Harvard plates, the Sonneberg plates showed QZ Aur near the limiting magnitude on many plates.  The time distribution of Sonneberg plates had many in the month and years just before eruption, which is optimal for looking for any pre-eruption rise.  We obtained 58 magnitude measurements of QZ Aur, four of which were limiting magnitudes.  Here, we only present the light curve (Figure \ref{QZAURLC}), with the individual magnitudes and full analysis presented in Xiao, Schaefer \& Kroll (2009). Of the 54 direct measurements, 40 were before the eruption and 14 were after the eruption.

The pre-eruption light curve shows no anticipatory rise.  The two faintest measures are both at the expected time of eclipse, so we believe that they are indeed eclipses. Without these eclipses, the nova ranges 1.65 mag, and has a average brightness of 17.16 mag. The standard deviation of the measurements before the eruption is measured at 0.23 mag. 

We can get the post-eruption magnitudes (and hence $\Delta m$) by two means.  The first is to use our own post-eruption magnitudes from the Sonneberg plates.  Here, our 14 magnitudes from long after QZ Aur returned to quiescence have an average of 17.13 mag, with an RMS of 0.18 mag and a range of 0.97 mag.  With this, $\Delta m = 0.03$ mag.  The second method is from literature, where Szkody (1994) reports B=17.65 and V=17.18, and Campbell \& Shafter (1995) report $B=17.50\pm0.04$ and $V=16.98\pm0.05$ out of eclipse with variations of roughly a quarter of a magnitude on a few nights, for a literature average of 17.57 mag.  Averaging the two B magnitudes together for the post-eruption value, we get $\Delta m=0.19$ mag.

\subsubsection{V1500 Cyg (Nova Cyg 1975)}

V1500 Cyg was discovered on 1975 August 29 (JD 2442653.98) by K. Osada as a 3.0 mag star (Osada, 1975), as well as hundreds of independent discoverers.  One of us (BES) made visual observations of the nova's magnitude from one day before peak until after peak, pointing to the peak being at visual magnitude 2.0 on 1975 August 30 (JD 2442655.7), with this being the same result as from examining the entire light curve in the database of the AAVSO.  V1500 Cyg has a period of 0.139613 days (Semeniuk et al., 1995). 

V1500 Cyg erupted after R75's report had been submitted, but we include it due to the importance of its pre-eruption behavior.  The pre-eruption star is barely visible on the glass plate of the Palomar Sky Survey blue plate, which places the precursor at B=21.5 in the year 1952 (Duerbeck 1987).  A positive detection of V1500 Cyg was also made in 1970, with the $V\approx20.5$ (Wade 1987).  With B-V=0.79 (Szkody 1994) in post-eruption quiescence, these two magnitudes are consistent with each other, hence implying only small variability from 1952-1970.  From 1952 to 1974, six deep plates have limits, with these showing the pre-eruption magnitude to be consistent with the faint magnitude from the Palomar plates (Kukarkin \& Kholopov 1975; Rosino \& Tempesti 1977; Wade 1987).  V1500 Cyg was then observed to show a distinct rise in 1975 August (Samus 1975; Alksne \& Platais 1975; Kukarkin \& Kholopov 1975). During this rise, V1500 Cyg was seen to get as bright as B=13.5 mag on 1975 August 28.732 (JD 2442653.232) (Kukarkin \& Kholopov 1975).  The light curve with all known pre-eruption magnitudes is presented in Table \ref{V1500CYGMAGS}.  These magnitudes were made using comparison stars from 1975, which should not be too bad.  Any photometric uncertainties will be likely $\sim 0.3$ mag, and as such are greatly smaller than the variations in the light curve that are so important.

We have not personally checked the pre-eruption plates in Russia.  A. Alksnis (2009, private communication) has recently examined the key plates taken with the Baldone Schmidt as used by Alksne \& Platais (1975).  He reports that the three plates show V1500 Cyg as being far above the plate limits, good quality images, near the plate center, and at exactly the right position.  With this, we confidently accept that V1500 Cyg was many magnitudes brighter than its normal pre-eruption level and that it brightened through the month of August.  The plate taken on 1975 August 28 was half a day before the discovery at 3.0 magnitude, so this could well be evidence for the ordinary rise of a fast nova (as opposed to some extended pre-eruption rise).  But V1500 Cyg is an extremely fast nova (second only to U Sco), so we cannot imagine the plates taken on 5-24 August as being part of some ordinary fast rise.  Thus, we are confident that V1500 Cyg shows a spectacular pre-eruption rise.

The pre-eruption behavior shows both a pre-eruption rise as well as a precursor magnitude much fainter than the post-eruption magnitude.  The star appears to have been approximately stable at roughly B=21.5 from at least 1952 to 1974 or so.  Sometime around or after December 1974, V1500 Cyg started its pre-eruption rise.  In August 1975, we can watch it brighten from roughly 17.6 magnitude (23 days before peak)  to 13.5 magnitude (6 days before peak).  This spectacular rise in the light curve of V1500 Cyg is unique to our knowledge out of all novae over all time, and it occurs in the month before the nova eruption.  During the pre-eruption rise, V1500 Cyg rose to 7 mag brighter than its normal quiescent level, and this is also completely unprecedented for any nova outside eruption.  The very close temporal coincidence of this unique event with the nova eruption demonstrates that the two are causally connected.

\subsubsection{HR Del (Nova Del 1967)}

On late 1967 July 8, (JD 2439680) G.E.D. Alcock discovered HR Del as a star of about 5.0 mag (Alcock 1967). It has been estimated that the eruption started a few weeks before the discovery (Solomon 1967). The orbital period is 0.214165 days (K\"{u}rster $\&$ Barwig 1988).

The light curve constructed by R75 for HR Del is based on Solomon (1967), Schweitzer (1968), Fehrenbach et al. (1967), Barnes and Evans (1970), and Stephenson (1967). Solomon (1967) gives pre-discovery estimates from ``Baker-Nunn films exposed for satellite tracking.'' There are seven measurements, all in June of 1967; four before outburst (one of them being a limiting magnitude) and three after. Solomon notes that the data ``establish the approximate beginning date.'' Similarly, Schweitzer (1968) gives eight pre-eruption photographic magnitudes and seven during eruption. Fehrenbach et al. (1967) provide a single photograhpic magnitude of 11.4 in 1935, and note that it appears to be dimmer on the Palomar plates. Barnes and Evans (1970) provide two measurements using Palomar Sky Survey plates, and give the V magnitude and B-V color measured from 1951 July 8 and 1953 October 1. Finally, Stephenson (1967) obtained a photographic measurement of the pre-nova system from the Lick Sky Atlas. R75 also discusses other work done on HR Del.  Wenzel (1967) showed HR Del to have a mean photographic magnitude of 11.9 over 222 plates taken at Sonneberg Observatory between 1928 and 1966. Wenzel also describes a ``small variability'' although no light curve is given. Likewise, Liller (1967) found  ``no evidence'' that HR Del exceeded 10th magnitude between 1890 June and 1952 July.  R75's conclusion for HR Del was that it had normal behavior with no case for a pre-eruption rise. 

Given the date of the eruption, R75 did not have the opportunity to compare the post-magnitude brightness for the nova, which he describes as an `extremely slow nova'. Indeed, the AAVSO light curve shows HR Del returning to a steady level only by 1982.

\begin{figure}[h!]
  \begin{center}
 \includegraphics[scale=0.5]{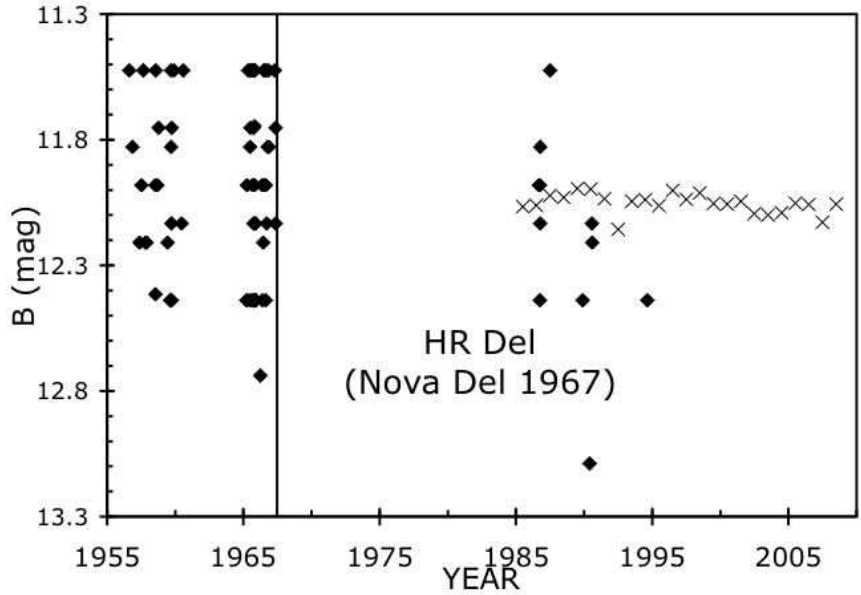}
  \end{center}
  \caption{HR Del light curve. The diamonds are all in B magnitudes, constructed entirely from Sonneberg plates.  The crosses represent the yearly averages from the AAVSO with a color correction to convert to B-band. The vertical line denotes the date of eruption. Before the eruption, HR Del is observed to have an average brightness of 11.97 mag, an RMS of 0.35, and span a range of 1.22 mag. After the eruption, the system spanned a similar RMS and range (0.41 and 1.56 mag, respectively) and has a similar average brightness (12.2 mag). With extensive coverage right up to the explosion in 1964, we see no evidence of a pre-eruption rise.}
 \label{HRDELLC}
\end{figure}

Using the Sonneberg collection, we examined plates from 1956-60, 1965-67, 1986-87, 1989-90, and 1994. The resulting light curve has 69 pre-eruption magnitudes and 11 post-eruption magnitudes (Table \ref{HRDELMAGS}, Figure \ref{HRDELLC}) .  Before the eruption, HR Del has an average magnitude of $m_{pre}= 11.97$ mag with a standard deviation of $\sigma_{pre}=0.35$ mag and a range of $R_{pre}=1.22$ mag. After the eruption, the nova shows an average magnitude of $m_{post}=12.20$ mag with a standard deviation of $\sigma_{post}= 0.41$ mag and a range of $R_{post}=1.56$ mag.  With this, $\Delta m = -0.23$ mag.

Many papers report post-eruption magnitudes.  Bruch \& Engel (1994) summarize many papers with B-V ranging from -0.10 to +0.24 mag, with an average for the B-band values after HR Del has gone to its quiescence level (e.g., Sherrington \& Jameson 1983) of B=12.31.  Kafka \& Honeycutt (2004) report a very impressive campaign of photometry covering hundreds of nights for fifteen years, with the V-band magnitude ranging from 12.06 to 12.56 with an average of near V=12.3.  AAVSO has over 7,000 V-band magnitudes for 1985-2009.1 with an average V=11.95 and a standard deviation of $0.17$ (see Figure \ref{HRDELLC} with a color correction of $B-V=+0.10$). From this, we take the post-eruption average to be B=12.3 with an RMS scatter of 0.2 mag.  With this alternative post-eruption magnitude, we have $\Delta m = -0.33$ mag.

In summary, HR Del shows no pre-eruption rise, and had neither its average brightness nor degree of variability change substantially across the eruption.

\subsubsection{DQ Her (Nova Her 1934)}

DQ Her was discovered by J.P.M. Prentice on 1934, December 12 (JD 2427784) as a star of 3.0 mag (Prentice 1934; Duerbeck 1987). The nova has been seen as bright as 1.3 mag (Beer 1935), and has been extensively studied since eruption. The period of DQ Her has been measured to be 0.193621 days (Horne, Welsh \& Wade 1993).

R75's light curve for DQ Her was constructed using data from Ahnert (1960), Gaposchkin (1956), Kukarkin \& Vorontsov-Velyaminov (1934), and Shapley (1934).  Ahnert (1960) uses Sonneberg plates to provide 27 measurements between 1930 August 27 (JD 2426216.479) and 1934 November 3 (JD 2427745.259). A correction of +0.1 mag must be applied to the the Ahnert (1960) measurements to bring them onto the same scale as Gaposchkin (1956). Gaposchkin (1956) gives 580 measurements of DQ Her, although only eight of these are before the eruption. These measurements are in photographic magnitudes and are taken between 1928 March 22 (JD 2425327.87) and 1934 October 5 (JD 2427715.57). Kukarkin \& Vorontsov-Velyaminov (1934) provide five pre-eruption measurements, four of which are limits, with their sole direct measurement on 1907 October 3 (JD 2417852). Shapley (1934) examined the Harvard plate collection and found DQ Her to be dimmer than 11-12th mag on ``several hundred plates scattered throughout all years since 1890.'' In addition, Shapley (1934) gives several pre-eruption magnitude measurements, although the description of measurements are imprecise (they are likely averages). R75 discusses some other work done on DQ Her, such as DQ Her being visible  on the Franklin-Adams charts (Anonymous, 1935), although this data was not included in his light curve. R75 found no evidence for a pre-eruption anticipation in DQ Her.

\begin{figure}[h!]
  \begin{center}
 \includegraphics[scale=0.5]{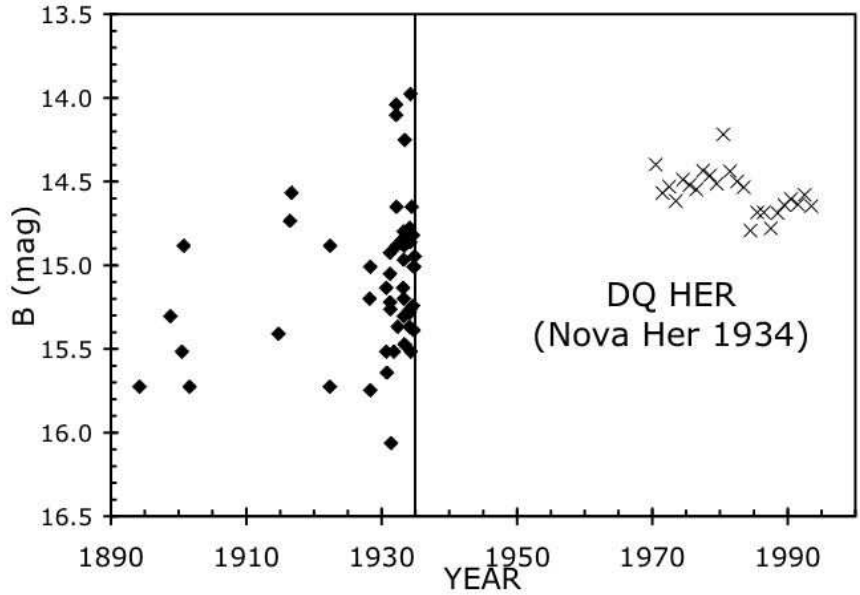}
  \end{center}
  \caption{The DQ Her light curve. All measurements are in B magnitudes. Diamonds mark points of the light curve that come from the literature.  Yearly averages from AAVSO data are marked with crosses. The vertical line marks the date of the eruption. There is no evidence for a pre-eruption rise. Before the eruption, DQ Her was at B=15.09, an RMS of 0.45, and a range of 2.09. After the eruption, DQ Her has B=14.53, with an RMS of 0.27 and a range of 2.00.}
 \label{DQHERLC}
\end{figure}

While we did not examine DQ Her at the plate archives during our recent visits, we are still able to add some information. Schaefer \& Patterson (1983) found 15 DQ Her measurements from the Harvard plate collection not found by Gaposchkin. These measurements were taken using Gaposchkin's magnitude sequence in order to compare them to previously taken data. Since Gaposchkin's study, modern B magnitudes have been found for his comparison stars; we were thus able to set up a calibration curve to convert Gaposchkin's comparison stars to modern B magnitudes. The equation of this curve is $B = 2.11 B_{G} - 15.88$ where $B$ is the modern B magnitude and $B_G$ is the Gaposchkin magnitude. This calibration is only good for measurements dimmer than $\approx 14$ mag.  We combine the measurements of Ahnert, Gaposchkin and Schaefer \& Patterson for a total of 50 pre-eruption magnitudes (Figure \ref{DQHERLC}). We find B=15.09, with an RMS of 0.45 and a range of 2.09. We agree with the result of R75 that DQ Her shows no evidence of a pre-eruption anticipation event.

Finally, we are also able to make use of the AAVSO data archives to obtain post-eruption coverage of DQ Her. We use the 1570 measurements taken between 1970 and 1994 of DQ Her in the Visual (or V) band. We use these year cutoffs to be sufficiently far away from the eruption so it had reached quiescence, and to avoid the eclipse data which is in the database for more recent observations. Bruch \& Engel (1994) provide a color measurement of B-V=0.14, which we apply to all measurements. We find that B=14.53, with an RMS of 0.27, and a range of 2.00. There is no change in the behavior of DQ Her as a result of the eruption. The average brightness is $\Delta m = 0.56$, and the variability is on the same scale. We do not believe the $\Delta m$ is significant given the large variability in the system.

\subsubsection{V446 Her (Nova Her 1960)}

O. Hassel discovered V446 Her on 1960 March 7 (JD 2437001), measuring it to be 5.0 mag (Hassel 1960).  Bertaud (1962) reports it to have been as bright as 3.02 photographic magnitudes on late 1940 March 4 (JD 2436998).  Duerbeck (1987) describes V446 Her as a ``fast nova'', and later an orbital period of 0.207 days was discovered (Thorstensen $\&$ Taylor 2000).  Honeycutt et al. (1998) discovered that V446 Her started to have dwarf nova eruptions around or shortly before 1990.

The pre-eruption light curve for V446 Her as it appears in R75 is comprised of data from Stienon (1963), Cragg (1960), Lowne (1960), Richter (1961), and Apriamashvili (1960). Cragg (1960) used two visual and two photographic measurements from the Palomar Schmidt telescope, and notes that the nova has two close companion stars of similar brightness. Stienon (1963) provides 100 photographic magnitude measurements taken between 1896 October 7 and 1953 October 11. Stienon's measurements are integrated magnitudes of the whole system as the triple was not resolvable. Lowne (1960) provides four photographic magnitude measurements (one limiting) using the Carte-du-Ciel plates, Franklin-Adams Charts, Ross Atlas, and the Palomar plates. As no mention of the triplet nature of the system is in Lowne's results, it can be assumed that the measurements are that of the nova combined with the two nearby companion stars. Richter (1961) gives 40 photographic magnitude measurements (10 limiting) of the system from Sonneberg plates. Richter is unable to resolve the individual stars in the triple. He also provides a magnitude measurement using Palomar plates to compare his measurements to previous works. Finally, Apriamashvili (1960) provides magnitude measurements using the Palomar Sky Survey from 1952 August 12. The post-eruption light curve comes from Stienon (1971), which used Warner-Swasey Observatory photographs to show that the integrated brightness of the system showed small variability ($\sim 0.4$ mag) over nine measurements taken between 1968 September and 1970 September.

With the historic light curve in R75, there is no evidence for a pre-eruption rise, and the average brightness of the nova has not changed significantly. R75 classified V446 Her as an event that showed significant ($\sim 4$ mag) variability before the eruption, and significantly less ($\sim 0.4$ mag) after the eruption.  With this, R75 points to V446 Her as the only star in his sample that shows a substantial change in its variations across the nova eruption.  R75 argues that the pre-eruption flares are not caused by dwarf nova events.  However, when we look at the same data, we only see variations from 15.1 to 16.9, while the magnitudes in Stienon (1963) have an RMS scatter of 0.55 mag.  With this characterization, the pre-eruption variability appears to be comparable to typical CVs (Kafka \& Honeycutt 2004), although the real range of variation will extend to fainter magnitudes when the light from the nearby stars is subtracted out.

\begin{figure}[h!]
  \begin{center}
  \includegraphics[scale=0.5]{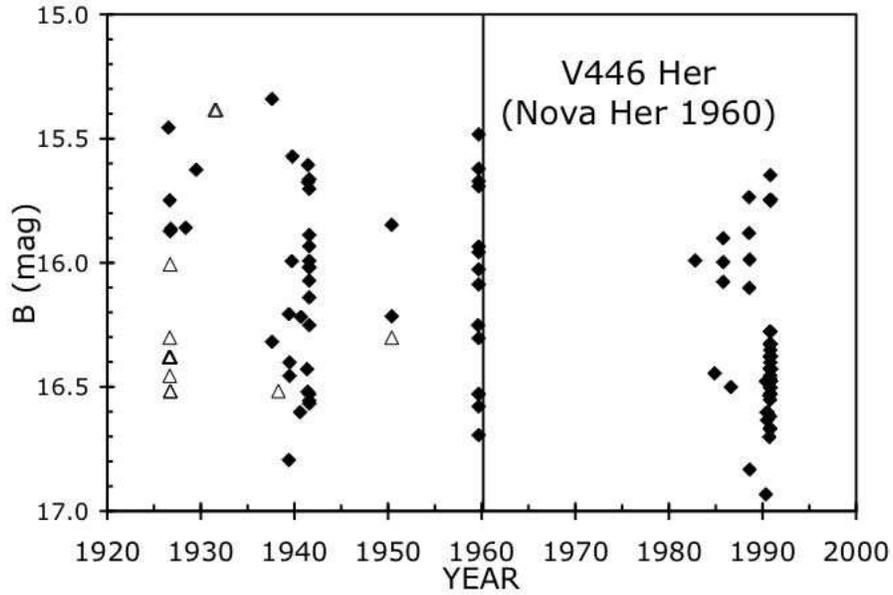}
  \end{center}
  \caption{V446 Her light curve. B-band magnitudes are denoted by diamonds, while triangles denote upper limits. The vertical line is the date of the eruption of the nova. We see no evidence of a pre-eruption rise. Before the eruption, we measure the system to have an average brightness of 16.07, an RMS scatter of 0.37 mag, and a range of 1.33 mag. This is similar to the behavior we observe after the eruption, with an average brightness of 16.31 mag, an RMS scatter of 0.31 mag, and a range of 1.28 mag.  We therefore also observe no change in the average brightness or variability as a result of the eruption.}
  \label{V446HERLC}
\end{figure}

Our pre-eruption light curve of V446 Her is constructed from 24 plates at Harvard (14 of which are limits) plus 39 plates from Sonneberg.  We also have 75 Sonneberg plates from after the nova has returned to quiescence.  We have listed all the measurements in Table \ref{V446HERMAGS} and displayed them in Figure \ref{V446HERLC}.  All these magnitudes are of the combined light of the nova and its two nearby companion stars.  Before the eruption, V446 Her had an average magnitude of 16.07, with a standard deviation of 0.37 mags and a range of 1.33 mag. This is similar to how V446 Her is observed to behave after the eruption, when the average brightness is 16.31 mag, with a standard deviation of 0.31 mags and a range of 1.28 mag. We see V446 Her had essentially identical variations before and after the eruption.  Also, $\Delta m = -0.24$ mag.  We see no sign of an anticipatory event in the V446 light curve.

\subsubsection{V533 Her (Nova Her 1963)}

V533 Her was discovered by L. Peltier in the United States as a 4th magnitude star on 1963 February 6 (JD 2438067)(Peltier 1963); it was independently discovered by E. Dahlgren in Sweden about 8 hours later (van Genderen 1963). The nova has been observed to be as bright as 3.0 photographic mag (G\"{o}tz 1965). It has been described as a moderately fast nova (Deurbeck 1987), and has an orbital period of 0.147 days (Thorstensen $\&$ Taylor 2000)

The historic light curve constructed by R75 for V533 Her comes from G\"{o}tz (1965) and L\"{o}chel (1963), both of which made their observations using Sonneberg plates. L\"{o}chel (1963) reports photographic magnitudes of V533 Her from 120 plates between 1941 and 1962. The reported magnitudes are yearly averages for years with available plates. G\"{o}tz (1962) also provides magnitude measurements in a figure, but no exact numbers are given for the individual measurements. Work by Stephenson \& Herr (1963) showed that V533 Her has a close companion, which means that the measurements from the Sonneberg plates are combined magnitudes. Stephenson \& Herr also provide a photographic magnitude measurement from the Lick Sky Atlas, but R75 did not include as it was of the nova alone, not a combined measurement. A single measurement from a Harvard plate in 1920 was provided by Newsom $\&$ Chester (1963).

\begin{figure}[h!]
  \begin{center}
 \includegraphics[scale=0.5]{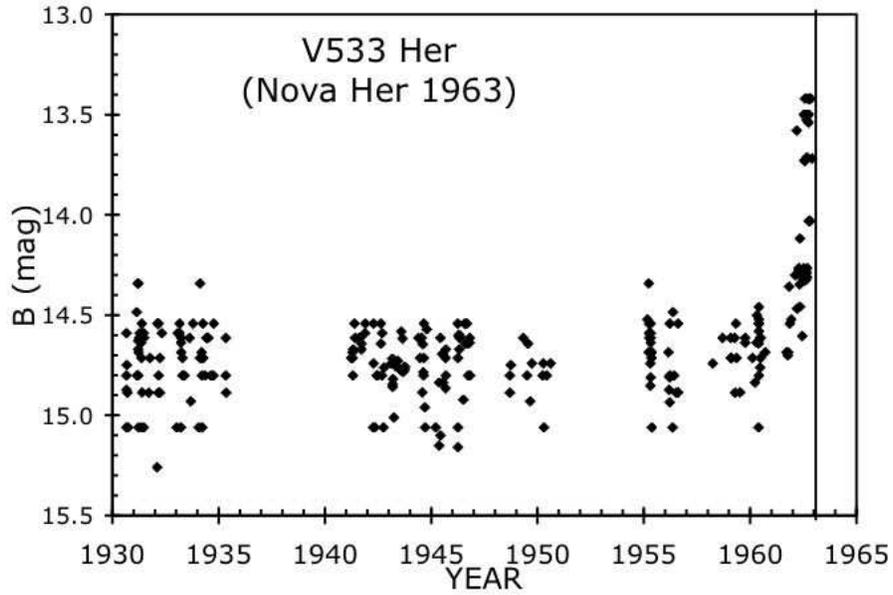}
  \end{center}
  \caption{V533 Her pre-eruption light curve. Diamonds are explicit B magnitude measurements and the vertical line denotes the date of the eruption. The system shows an obvious rise within the $\sim 1.5$ years preceding the eruption. This rise is characterized by the system going from an average magnitude of 14.72 to $\sim 13.4$ mag. Before this rise, the system spanned a range of 0.92 mag. Since this rise is well outside the behavior shown any time before in V533 Her, this rise seems to be causally connected to the eruption.}
\label{V533HERLC}
\end{figure}

\begin{figure}[h!]
  \begin{center}
 \includegraphics[scale=0.5]{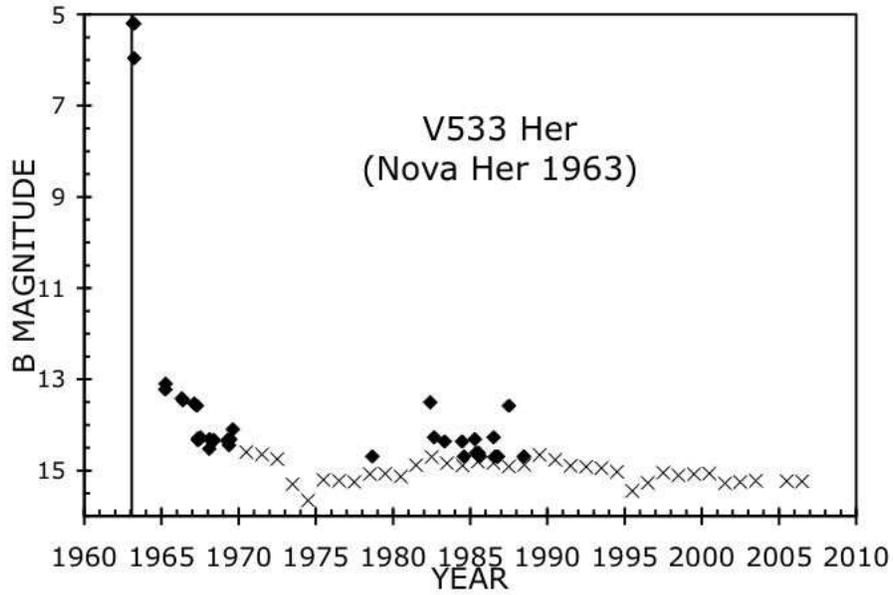}
  \end{center}
  \caption{V533 Her post-eruption light curve. Points before 1970 are in the tail of the eruption.  In addition to our own measurements (marked by filled diamonds) we also display yearly averages from AAVSO data (marked as crosses). This data was taken in V and visual bands, and a color correction term of $B-V = 0.18$ as given in Bruch \& Engel (1994) was applied. The individual data points have an average magnitude of 15.03 mag, and RMS of 0.47, and a range of 1.06 magnitudes.}
\label{V533HERLC2}
\end{figure}

We obtained 309 magnitudes for V533 Her, 268 of which are pre-eruption magnitudes (Table \ref{V533HERMAGS}, Figures \ref{V533HERLC} and \ref{V533HERLC2}). Here, we have separated the before and after light curves for the purposes of highlighting the very distinctive rise in the pre-eruption light curve. From late 1930 through 1961 June, the nova showed a fairly consistent variability, having an average B magnitude of 14.72, with a standard deviation of 0.17 mag. The nova had a range of 0.92 mag during this time. Starting in 1960, and continuing until the eruption in 1963, the nova began rising, until it got as bright as $\sim 13.3$ mag. The rise is significantly outside the normal variability seen beforehand, and seems to be causally connected to the eruption event. In fact, the nova was never seen to go above 14.3 mag during the 1930-1960 period. We therefore confirm the conclusion of R75 that V533 Her is an example of a pre-eruption rise in a nova light curve.

As for our post-eruption data (Figure \ref{V533HERLC2}), we have some measurements up through 1970, before the nova reached quiescence. After 1970, the system leveled off.  We have 16 measurements of V533 Her between 1982 and 1988, during which time it had an average magnitude of 14.25 mag with a standard deviation of 0.47 mag. In addition, the nova ranged over 1.17 mag during this time. Given the variability shown before and after the eruption, we believe that the average magnitude of the nova has not been affected by the eruption. In addition, the variability has not changed significantly enough (especially given the difference in sampling) for us to confidently claim a difference.

We also provide a series of AAVSO yearly photometric averages between 1970 and 2006. This data is composed of over 2,400 measurements, with a varying amount of coverage year to year. As these measurements were taken visually (or sometimes with a V filter), a correction must be added in order to compare these measurements to our B measurements. Bruch \& Engel (1994) give the B-V color term for V533 Her as 0.18, so we have used that term to correct the V band measurements to B band measurements. We present the yearly averages of these measurements in the post-eruption light curve (Figure \ref{V533HERLC2}). These averages show that the system didn't return to quiescence until the mid 1970s. The yearly averages also show some long-term variability. 

In summary, we find that V533 Her is an exquisite case for a nova that shows an anticipation of its eruption. That is, the nova rose $\sim 1.25$ above its previous average brightness during the $\sim 1.5$ years leading up to its eruption. This rise goes well outside  the small variability seen during 65 years away from the eruption.  The uniqueness of the rise and its close temporal connection to the nova eruption argues strongly (but does not prove) a causal connection. There is no evidence for the variability or average brightness being significantly different as a result of the eruption.

\subsubsection{CP Lac (Nova Lac 1936)}

CP Lac was discovered on the night of 1936 June 18 (JD 2428558) by  K. Gomi and ``Dr. Nielsen of Aarhus, Denmark'' (Nielsen 1936). The system was seen to be as bright as V=2.14 (Parenago 1949) . The orbital period has been measured to be 0.145143 days (Peters $\&$ Thorstensen 2006).  Honeycutt et al. (1998) discovered that CP Lac has started to display ``stunted outbursts" in the 1990s, with these presumably having some relation to dwarf nova events.

The R75 light curve for CP Lac was constructed from Hoffleit (1936) and Parenago (1936, 1949).  Hoffleit (1936) provides 38 photographic measurements (1898 August 13 to 1933 October 14) from the Harvard plate collection, five of which are limiting magnitudes.  Hoffleit warns that ``the magnitudes of the comparison stars used at Harvard are provisional and may be in error by over half a magnitude.'' Parenago (1936) observes that the system was invisible on 26 Moscow plates taken between 1899 May 20 (JD 2414795) and 1934 December 7 (JD 2427779), putting a limit of 15th magnitude on the nova for those plates. Parenago (1949) provides extensive coverage of the system during eruption, and follows it back to near-quiescence. He gives eight pre-eruption magnitudes from 1933 October 8 (JD 2427397) to 1935 October 6 (JD 2428082), two of which are limits. These measurements were made using the Harvard international scale, and should therefore be directly comparable to the Hoffleit data. R75 does not include the pre-eruption measurements made by Becker (1936), Wachmann (1936), and B\"{o}hme (1936) due to suspicions that they refer to a nearby star, and not CP Lac itself. In addition, R75 also does not include the \textit{approximate} magnitude measurements made by Schewick (1936).

R75's conclusion was that CP Lac has a weak case for a pre-eruption brightening. In the 3 years leading up to the eruption, CP Lac rose about a quarter of a magnitude, with the light curve being apparently constant before and after this transition.  The significance of this rise is questionable for three reasons.  First, even if we take the light curve at face value, the significance of the rise is small.  The pre-transition range is 0.6 mag with an RMS scatter of 0.14 mag, and an average of 15.37 mag, while the `post-transition' interval has an average of 15.15 mag.  The difference (0.22 mag) is small compared to the range of variations.  Second, and more importantly, the small change is much smaller than the typical year-to-year variations seen in most CVs (Kafka \& Honeycutt 2004).  With the ordinary CVs having yearlong variations of 0.5-1.0 mag, any excursion of 0.22 mag by CP Lac has no credibility as being associated with the eruption.  Third, Hoffleit's light curve makes up the entirety of the `pre-rise' portion of the light curve, while the light curve segment after the putative transition are almost entirely from Parenago.  Hoffleit warned us that her light curve might need a systematic offset by up to half a magnitude.  Thus, we expect that the putative pre-eruption rise is simply due to a slight and known miscalibration of the Harvard data.  This provides a ready explanation for why a sharp transition should occur at the time when the light curve is switching from one observatory to another.  We therefore have many strong reasons to know that the claimed pre-eruption rise is not real.

\begin{figure}[h!]
  \begin{center}
 \includegraphics[scale=0.5]{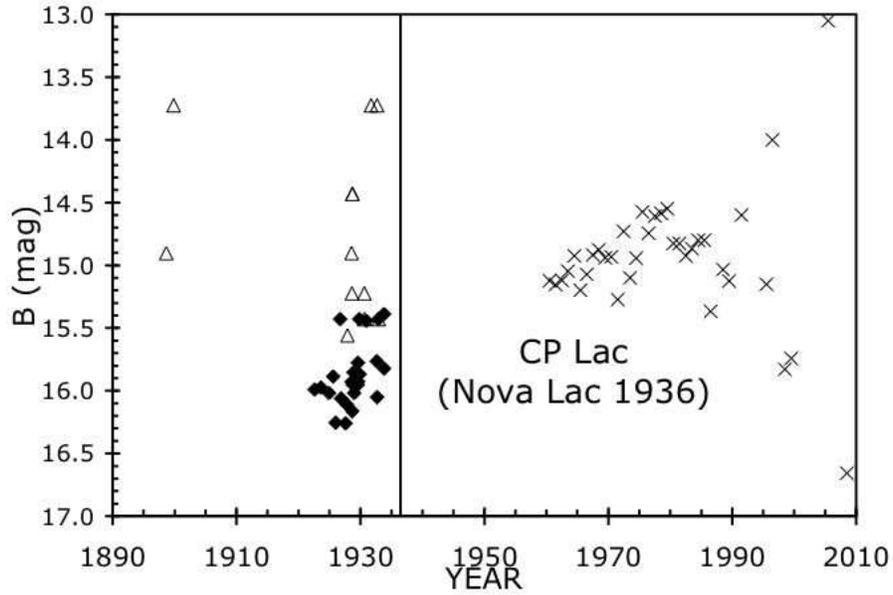}
  \end{center}
  \caption{CP Lac light curve. Triangles denote modern limiting B measurements and diamonds denote B magnitude measurements. AAVSO yearly averages (color corrected with $B-V = 0.2$) are marked as crosses. We find that before the eruption, CP Lac had an average brightness of 15.87, an RMS of 0.26 and a range of 0.87 mag. After the eruption, we find that CP Lac had an average brightness of B=15.5, an RMS of 0.4 and a range of 2.3 mag.}
  \label{CPLACLC}
\end{figure}

We investigated CP Lac on the Harvard plates, obtaining 37 B measurements, 12 of which were limiting magnitudes (Table \ref{CPLACMAGS}, Figure \ref{CPLACLC}). These include all of the later plates examined by Hoffleit, but we have not included some of the earlier plates. We do not have any observations within the time period covered by Parenago at the Moscow plates. As such, we only have two plates within three years of the eruption.  The average pre-eruption magnitude is 15.87 mag, with an RMS scatter of 0.26 mag and a total range of 0.87 mag.

We can again address the question of whether CP Lac had a pre-eruption rise.  At first glance, the first five magnitudes (averaging B=16.02 from JD 2423263-4497) look to be fainter than the last five magnitudes (averaging B=15.69 from JD 2426929-7014).  But again, even with this {\it a posteriori} selection of time intervals, the difference (0.33 mag) is small compared to the range (0.87 mag).  To be quantitative, a K-S test shows the magnitude distributions in the two time intervals to be taken from different parent populations with a confidence at only the 2-sigma level, which is to say that CP Lac does not have a significant change in brightness.  Even if we are to believe a 2-sigma claim, then the ordinary variations of CVs away from any nova event provide a much simpler explanation.  So again, we conclude that CP Lac does not have a pre-eruption rise.

For post-eruption magnitudes measured on a small number of nights, Szkody (1994) gives B=15.76 and B-V=0.29, Hopp (1979) gives B=16.15, and Diaz \& Steiner (1991) gives V=15.5.  Honeycutt et al. (1998) report V-band magnitudes from many hundreds of nights from 1991 to 1997 and find long term trends in which the mag ranges from 15.8 to 16.8, with an average of around V=16.1 (hence B=16.3), even after discounting the ``stunted outbursts.''  The time interval from 1991 to 1995 has CP Lac varying slowly, with an amplitude 0.3 mag, which might look like the variability before the eruption.  The AAVSO V-band light curve from 1960 to 2009 is shown in Figure \ref{CPLACLC} for the yearly averages of 281 measures (with a correction to B-band using B-V=0.2 mag).  The average magnitude has gone from B=15.0 in 1957-1971, to B=14.5 for the next decade, to a decline to B=16.8 in recent years.  Thus it appears that post-eruption CP Lac has a variability of order 0.4 mag superposed on decadal variations from 14.5-16.8.  For constructing the $\Delta m = m_{pre}-m_{post}$ value, we can only guess that the average brightness level from 1957-2009 ($B\approx 15.5$) is the most appropriate to use.  With this, $\Delta m\approx +0.4$ mag. 

\subsubsection{BT Mon (Nova Mon 1939)}

F.L. Whipple (1939) announced his discovery of BT Mon on a spectral plate taken on 1939 December 23 (JD 2429621).  The eruption light curve had a long plateau period, which was likely at the maximum light of $m_{B} = 8.5$ mag (Schaefer $\&$ Patterson 1983). The system shows a period of 0.33381379 days (Smith, Dhillon \& Marsh 1998).  With its deep eclipses, long orbital period, and moderately bright quiescent magnitude, this is the only classical nova yet to have its pre-eruption orbital period measured, with this proving that the binary system separated slightly due to the mass loss of the eruption  (Schaefer \& Patterson 1987).  This result provides the only confident measure of the mass ejected by a nova eruption, with $M_{ejecta}=3\times 10^{-5}$ M$_{\odot}$.

R75 compiled a light curve from Wachmann (1968) and Bertiau (1954). Wachmann (1968) took 276 measurements of BT Mon from Hamburg Observatory. Some of these measurements were of the nova during eruption, and fading back to quiescence; and only ten are pre-erupton measurements, two of them limiting magnitudes. These ten measurements were taken between 1938 December 17 (JD 2429250.49) and 1939 April 11 (JD 2429365.35); the other measurements cover between 1939 December 18 (JD 2429615.56) and 1962 February 8 (JD 2437704.47). In addition, Bertiau (1954) provides 113 photographic daily average measurements between 1934 March 19 (JD 2427516.3), and 1943 May 5 (JD 2430850.2); only two of which are pre-eruption (1934, March 19 and 20). R75 uses a correction of 1.0 mag to bring the Bertiau magnitude scale to be in agreement with the Wachmann scale. In addition, Whipple (1940) and Whipple \& Bok (1940) ``searched 150 Harvard plates of the field of BT Mon taken between 1898 and the eruption without a single positive detection of the prenova'' (R75). As R75 admits, most of these plates had a plate limit of between 11 and 13 (Whipple 1940). However, Whipple (1940) also found BT Mon to be invisible on the ``large-scale plates'' which had limits down to 17.  R75's conclusion is that BT Mon is a weak case for a pre-eruption rise, as the nova was fainter than 17th mag until the $\sim 6$ years leading up to the eruption.  This conclusion is based on the offhand remark of Whipple that he did not find BT Mon on some deep plates.

\begin{figure}[h!]
  \begin{center}
 \includegraphics[scale=0.5]{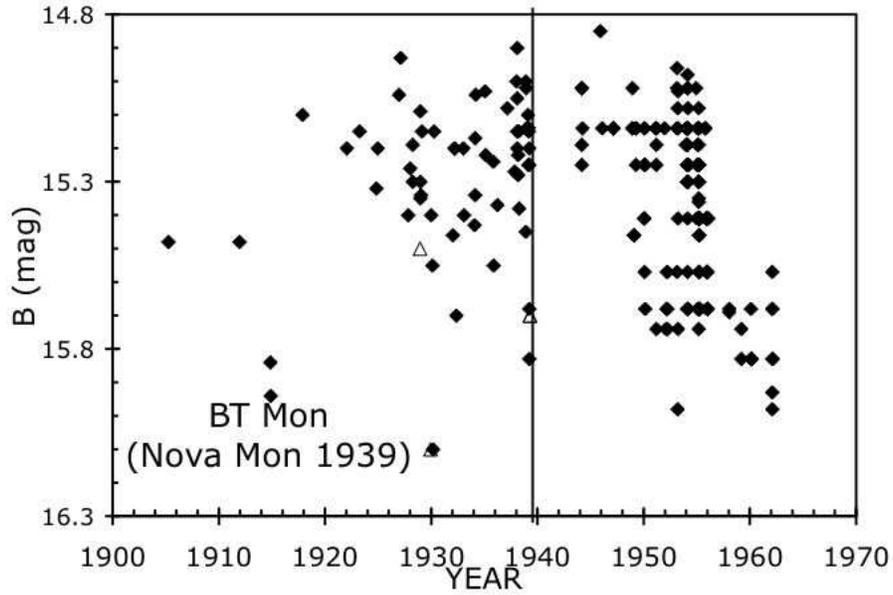}
  \end{center}
  \caption{BT Mon light curve.  These magnitudes here were taken from Schaefer $\&$ Patterson (1983), and Wachmann (1968), and all are on a magnitude scale defined by the comparison stars of Wachmann.  As such, these magnitudes have some unknown offset with respect to the modern B-magnitude scale, but nevertheless are fine for the purposes of this paper.  In particular, we see no pre-eruption rise, while both the variability and average brightness are similar on both sides of the eruption.  Magnitudes are given by the diamonds, limits by triangles, and the date of the eruption is marked by the vertical line.  The fainter magnitudes are all during the phase of the deep eclipse.}
  \label{BTMONLC}
\end{figure}

We did not re-visit BT Mon at the Harvard plate stacks as we have already done this (Schaefer $\&$ Patterson 1983). This previous study yielded 58 new pre-eruption magnitude measurements between 1905 April 3 (JD 2416938.54) and 1939 March 13 (JD 2429335.353).  (This shows that Whipple's search was incomplete.)  These magnitude measurements were made using the Wachmann (1968) comparison stars, so while they are in B, they are not comparable with modern B magnitudes. In addition, these measurements include the nearby comparison star (like Wachmann did), so they are \textit{not} of just the nova itself.  Both the 58 measurements from Schaefer $\&$ Patterson (1983) and the 10 measurements from Wachmann (1968) are displayed in Figure \ref{BTMONLC}.  For the purposes of this paper (looking for pre-eruption rises, changes in variability, and $\Delta m$), it is sufficient to have non-standard magnitude scales, as long as all the magnitudes are consistent.

The measurements show that not only was BT Mon visible on Harvard plates before 1933, but 30 plates with BT Mon present were recorded, the earliest of which was in 1905.  BT Mon is shown to have an average magnitude of 15.38 mag with a standard deviation of 0.29 mag before 1933. In this time period, BT Mon is observed to span a range of 1.2 mag (although this range includes observations of BT Mon in eclipse). From 1933 until the eruption, the nova is relatively unchanged from this previous state, showing an average magnitude of 15.25 mag with a standard deviation of 0.22 mag. The nova continues to show the same level of flickering, ranging 0.9 mag in the 6 years leading to the eruption.  The entire pre-eruption light curve for BT Mon has $m_{pre}=15.28$, $\sigma_{pre}=0.24$, and $R_{pre}=1.20$. Thus, the nova was \textit{not} below 17th magnitude as a reading of Whipple (1940) would suggest, nor did BT Mon rise in brightness in the years before the eruption.

We also include the post-eruption data from Wachmann (1968) for the purposes of comparison to the pre-eruption light curve. These data set is tabulated in that paper and shown in Figure \ref{BTMONLC}. The data are comprised of 176 measurements taken between 1944 February 21 (JD 2431142.39) and 1962 February 8 (JD 2437704.47).  We find that the system shows an average magnitude of 15.37 mag with a standard deviation of 0.25 mag. In addition, BT Mon spans a range of 1.13 mag during this time.  With this, we find $\Delta m= -0.09$ mag.  It seems that the system has not only returned to a state similar to that before the eruption, but did so quickly ( in $\sim 5$ years).  We note that BT Mon also shows decadal secular trends as it faded from roughly 15.1 to 15.8 mag from 1944 to 1962.

In summary, BT Mon is \textit{not} a case for a pre-eruption rise, chiefly because we have found that BT Mon was indeed visible for $\sim 30$ years before was previously observed. In addition, we see BT Mon to show no significant rise in the years where a pre-eruption rise was thought to have been seen previously.  In addition, we find BT Mon to have similar scatter and range both before and after the eruption, so the variability of the system does not appear to have changed as a result of the eruption. Finally, we observe that BT Mon has the same quiescent average magnitude as after the eruption it had before it. BT Mon therefore seems to show no anomalies associated with its eruption.

\subsubsection{GK Per (Nova Per 1901)}
 
T.D. Anderson (1901) discovered GK Per as a star of 2.7 mag on 1901 February 21 (JD 2415437), with a peak at 0.2 mag (Campbell 1903).  As one of the all-time brightest novae, GK Per was intensively studied and is in many way the prototypical nova.  With this close examination, GK Per was system in which four uncommon properties; nearly-periodic large-amplitude brightness oscillations during the transition stage, superluminal expanding light echoes, expanding gas shells, and dwarf nova eruptions, which began in 1967.  The orbital period of GK Per has been measured to be 1.996803 days (Crampton et al. 1986), which is startling as being greatly longer than all other classical novae (i.e., not counting recurrent novae and symbiotic novae).

R75 constructs the pre-eruption light curve of GK Per entirely from Leavitt (1920). Most of the light curve is composed of limits, although 12 are actual detections. R75 notes that  GK Per was seen to  brighten by 0.75 mag in the two years leading up to the eruption. This claim rests primarily on the three plates taken just days before the eruption: AC 1252, AC 1258 and AC 1260. These measurements trace back to low quality plates that were measured by one person (Vaughn) alone. All three measurements are marked as ``barely detected''.

\begin{figure}[h!]
\begin{center}
 \includegraphics[scale=0.5]{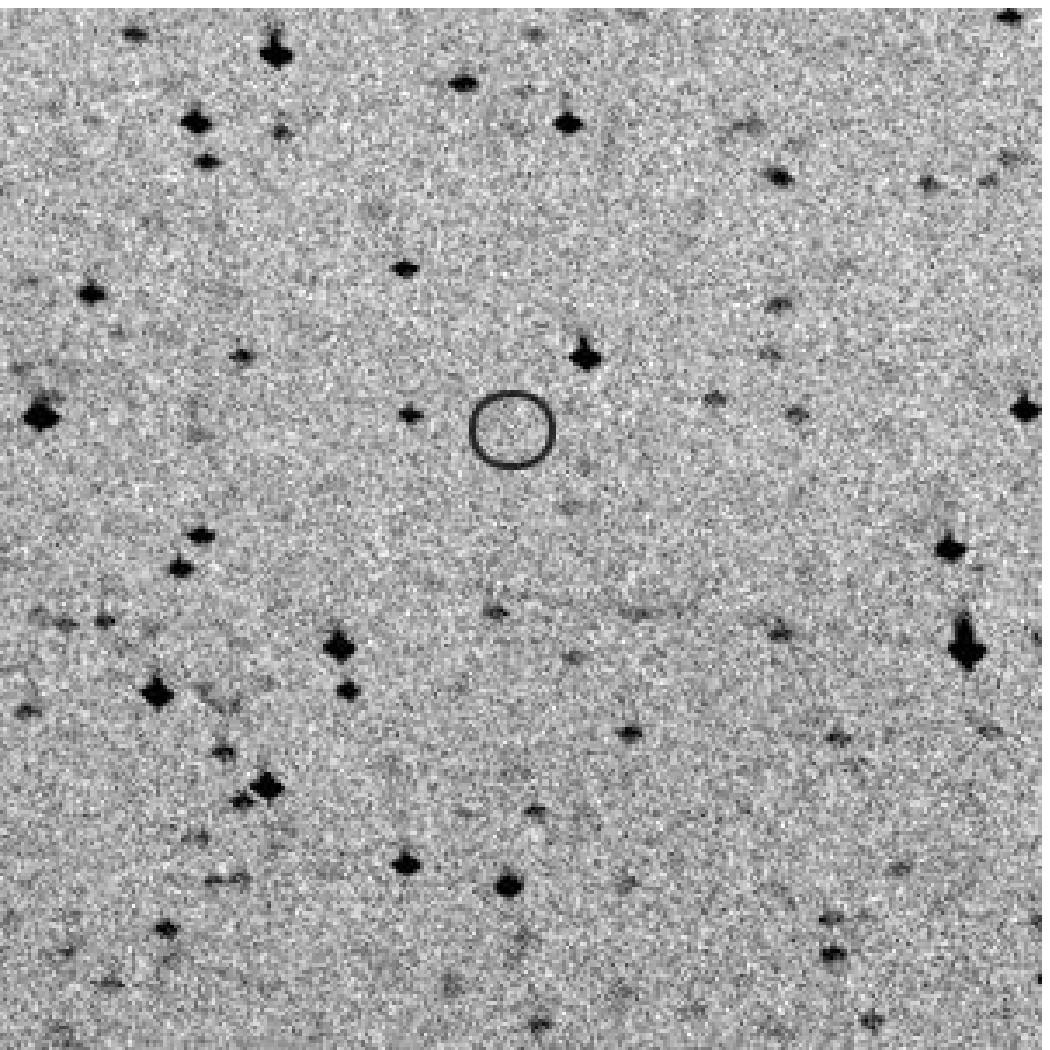}
 \includegraphics[scale=0.5]{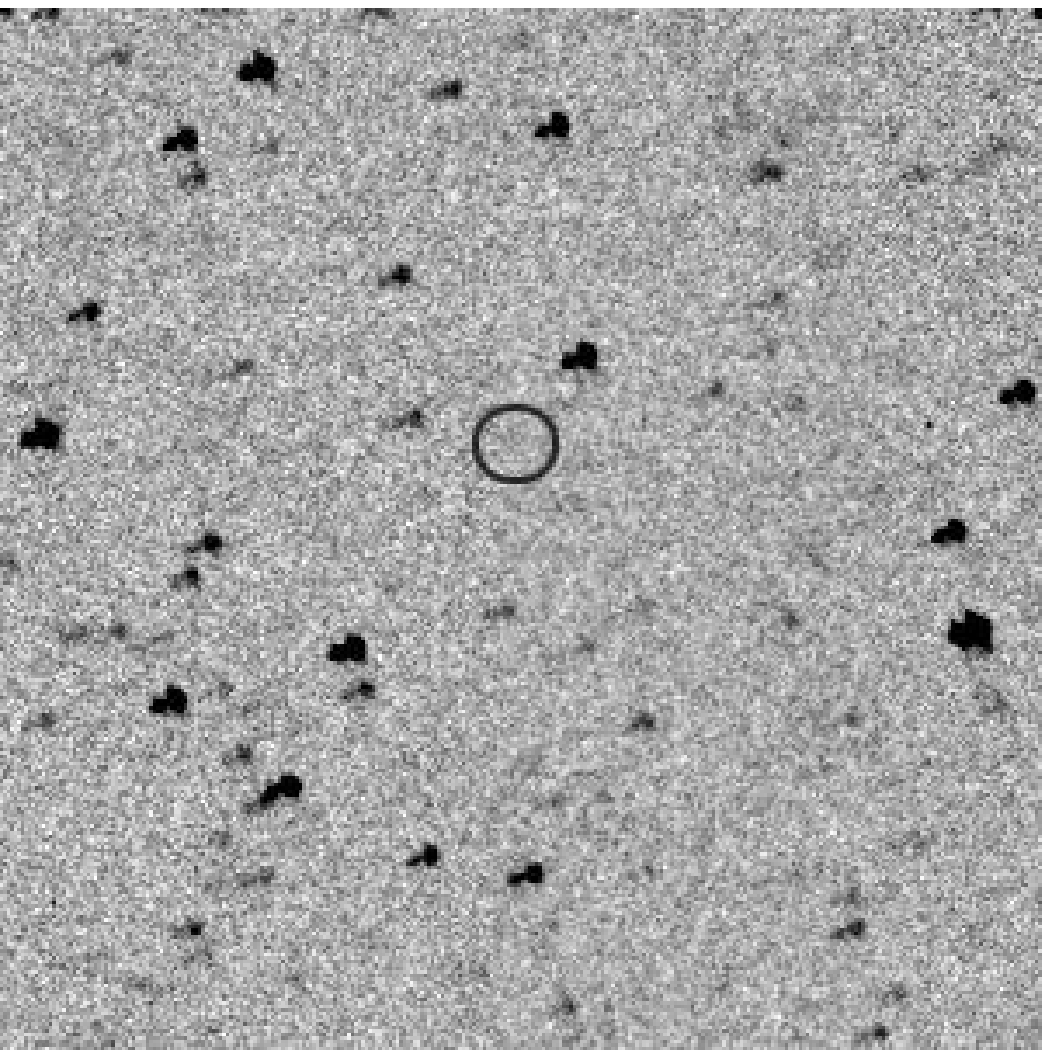}
\includegraphics[scale=0.5]{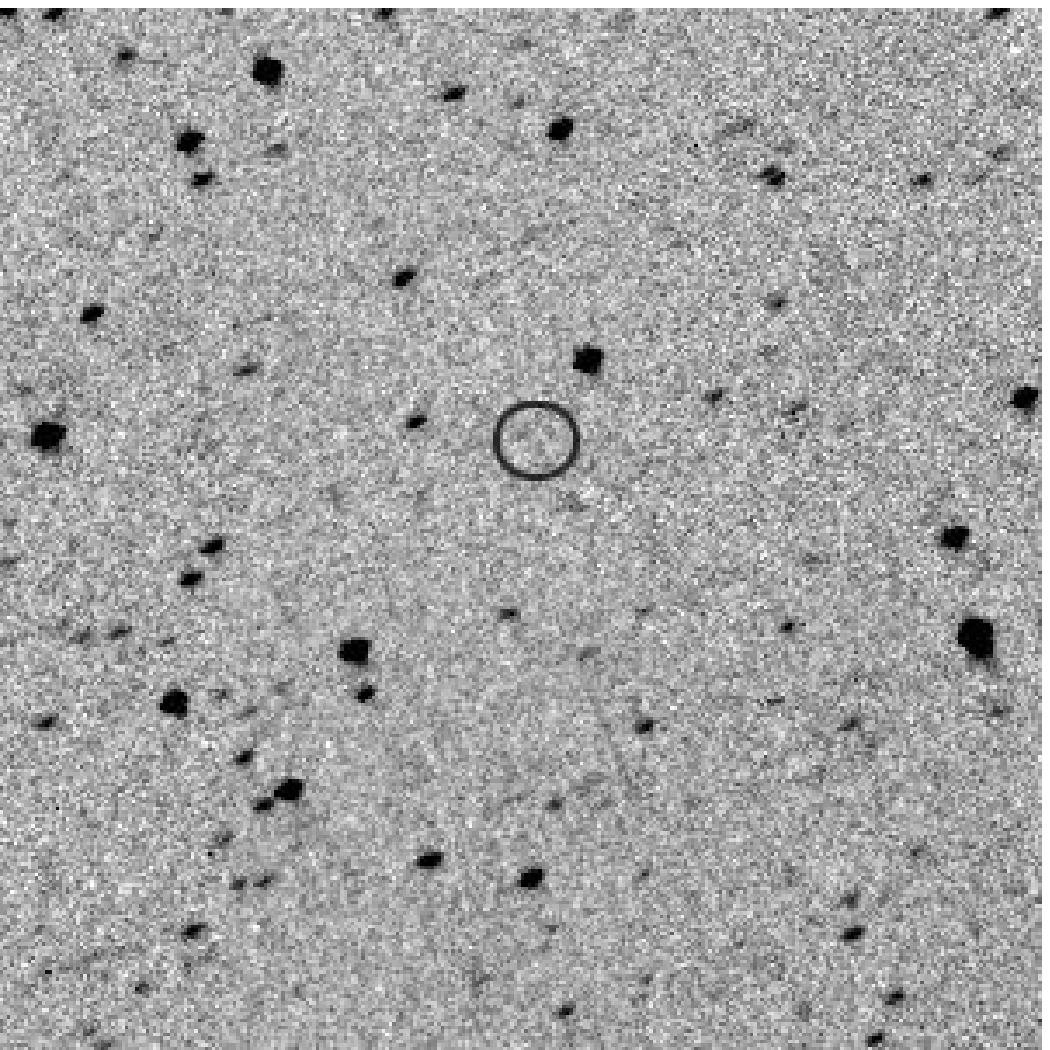}
  \end{center}
  \caption{GK Per plates within a week of its eruption.  These high resolution scans of Harvard plates AC1252, AC1258 and AC1260 (from left to right) show no evidence of GK Per.  Circles are placed in the field to mark the area of the position of GK Per.  North is to the top, east is to the left, and the field of view is just over $1^{\circ}$ wide.  Grains are clearly visible in the background of these plates, so we see that GK Per is not visible at even low significance.  The entire case for a pre-eruption rise is based on these three plates showing a relatively bright pre-eruption nova, and the invisibility of GK Per on these plates demonstrates that no pre-eruption rise actually took place.}
  \label{GKPERPLATES}
\end{figure}

At Harvard, we found and examined the same plates that were the centerpiece of the claim of a slight pre-eruption rise. These plates show no evidence for any positive detection of GK Per even at any low significance level (Figure \ref{GKPERPLATES}). We can see down to the granulation of the emulsion in all three plates, but cannot see the pre-nova.  Archival photographic plates that have been well-devloped and well-stored have zero deterioration over a century, so we are seeing exactly what was seen by Ms. Vaughn.  As such, we can only conclude that Ms. Vaughn made an error in claiming that the nova was visible, and actually that these three key plates only provide limits on GK Per.  The three plates show that the nova was fainter than 12.9-13.2 mag the week before the eruption.  With the old magnitude estimates for these three plates changed to limits, the evidential basis for any claimed pre-eruption rise goes away.

We report on magnitudes for 22 Harvard plates, only nine of which represent positive detections of the pre-nova.  These plates go from 1890 until a week before the eruption.  We measure GK Per to have an average magnitude of 13.87 with a standard deviation of 0.39 mag and a range of 1.11 mag.  While it appears that our last three magnitudes are slightly higher than for previous times, this is \textit{not} outside the system's observed variability.  In particular, these points are roughly as bright as GK Per is observed to be in 1894.

Post-eruption magnitudes are reported in many papers, with Bruch \& Engel (1994) summarizing the results as B=13.81 and B-V=0.79.  The AAVSO light curve from 1918 to present contains over 25,000 V-band magnitudes, which show a fairly constant brightness level (other than the dwarf nova events) with $V\approx13.0$ so $B\approx 13.8$.  The AAVSO data base also includes close to 100 B-band measures from 2006-2008, with an average magnitude of $B=13.94$, a standard deviation of 0.12 mag, and a range of 0.55 mag.  Our $\Delta m$ value is +0.06 mag.

\subsubsection{LV Vul (Nova Vul 1968)}

G.E.D. Alcock (1968) discovered LV Vul on 1968 April 15 (JD 2439962). The nova was seen to get as bright as 4.83 mag (Dorschner et al. 1969).  The AAVSO eruption light curve showed a fast decline and the system was certainly back to quiescence by 1973.

The R75 light curve of LV Vul comes from Meinunger (1968), Herbig (1968), and Kukarkin $\&$ Efremov (1968). Meinunger (1968) used 30 plates to provide six yearly average measurements of the system between the years 1938 and 1967. Herbig (1968) provided a single photographic measurement of the system and a nearby star. Kukarkin $\&$ Efremov (1968) described LV Vul as being between 16th and 17th mag between 1949-1952 with five plates taken on a 40cm astrograph. Additionally, R75  points out that Liller (1968) did not find evidence of LV Vul being above 14th mag between 1898 July and 1952 July using the Harvard plate collection.  In all, the only useful magnitudes in the R75 light curve are the six magnitudes from Meinunger and the one magnitude from Herbig.  These show that LV Vul varied from 16.4 to 16.9 mag between 1935 and 1944.  From 1959 to 1967, the three magnitudes of Meinunger show an increase in brightness from 16.2 to 16.0 to 15.4.  The last point in particular looks like a significant brightening soon before the eruption.  With this, the pre-eruption brightening appears similar to that of V533 Her, where a slow acceleration in the rise occurs over a few years in advance of the explosion.  R75 concludes that LV Vul is a definite case for a pre-eruption rise.

The primary evidence for the pre-eruption rise is the last Meinunger magnitude, which is about one magnitude brighter than the top of the 1935-1944 range.  Unfortunately, there is a typo for this value, as reported in an erratum.  The value should not read ``15.4", but instead should read ``(15.4'', which was standard notation of the time to say that the value is a limit.  That is, LV Vul was \textit{not actually seen} at the elevated brightness level soon before eruption, and the primary evidence for a pre-eruption rise is a simple and certain error.

Using both the Harvard and Sonneberg plate collections, we obtain 206 B-band measurements (61 of which are limits) of LV Vul (Table \ref{LVVULMAGS}, Figure \ref{LVVULLC}). Of these measurements, we have 70 definite measures of LV Vul before the eruption and 75 after. The pre-eruption portion of the light curve is mostly flat, centering around an average magnitude of 16.24, and shows no hint of a pre-eruption rise.  Using individual measurements, we can see the variability of the system far more clearly than we can with the yearly averages in the literature. We find the standard deviation of the pre-eruption light curve to be 0.31 mag, and the measurements span a range of 1.64 mag. After the eruption, we see that the light curve is again flat, centering around 16.10 mag. The standard deviation of the post eruption light curve is 0.21 mag, and it spans a range of 1.25 mag.  From this, we get $\Delta m= +0.14$ mag.

With a larger sample of plates, we were able to observe that LV Vul is a rather ``ordinary'' system, showing no signs of a pre-eruption anticipation. The light curve instead is flat, albeit with the usual significant variability on either side of the eruption. This variability is seen to be of the same order both before and after the eruption. Finally, we observe LV Vul to have similar brightness after as compared to before the eruption. In all, LV Vul exhibits no extraordinary behavior.

\subsection{Recurrent Novae}

The primary reason for the low number and poor quality of pre-eruption light curves is that no one knows which stars will erupt before they go up.  Thus, we are always forced to use serendipitous sky patrol photos with variable depth and scattered timing.  But this limitation does not apply to the subset of novae known as recurrent novae (RNe), which consists of nova systems for which multiple eruptions have been observed.  There are now ten known RNe in the Milky Way, of which four are well known (T Pyx, U Sco, RS Oph, and T CrB), while the others are poorly observed (Schaefer 2009b).  Once a nova is identified as an RNe, then intensive monitoring can provide pre-eruption light curves for the {\it next} eruption.  And with the high profile of RN (as strong candidates to be Type Ia supernova progenitors), intensive efforts have been made to obtain exhaustive information from archival sources (Schaefer 2009b).  So, it is no surprise to realize that we have many good pre-eruption light curves for RN.

A reasonable question is to ask ``How can the pre-eruption behavior of RNe tell us anything about the pre-eruption behavior of novae?''  The answer is simple: RNe are novae, so they provide perfectly good examples.  That is, recurrent and classical novae share an identical system configuration (companion star feeding matter to a white dwarf through Roche Lobe overflow), identical explosion mechanism (thermonuclear runaway of hydrogen-rich material accumulated on the surface of the white dwarf), and identical nova physics (from the nuclear reactions to the gas expulsion to the radiative transfer in the shell).  As such, any phenomenon involving the pre-eruption behavior should apply equally to RNe as to all other novae.  The RNe are distinguished from other novae only because they have a short recurrence time, and this property is a continuum from roughly a decade to perhaps a million years.  To get such a fast recurrence time scale, the RN systems must have a relatively high accretion rate ($\sim 10^{-7}$ M$_{\odot}$ yr$^{-1}$) and a white dwarf near the Chandrasekhar mass.  

Schaefer (2009b) provides an exhaustive analysis of virtually all photometric data taken on recurrent novae.  Indeed, that database now provides essentially all photometry at quiescence for seven of the ten galactic RNe.  Here, we will extract the data from Schaefer (2009b) that is relevant for the pre-eruption behavior.  The quantitative measures are summarized in Table \ref{SUM2}.

\subsubsection{CI Aql (Nova Aql 1917, 1941, 2000)}

CI Aql had eruptions in 1917, 1941, and 2000, each reaching a peak of 9.0 mag.  Mennickent \& Honeycutt (1995) found that CI Aql has deep eclipses and a long orbital period of 0.62 days.  After its eruption in 2000, we have been thoroughly measuring the light curve of CI Aql to measure the change in orbital period across its recent eruption.  The light curve of CI Aql in quiescence is displayed in Figure \ref{CIAQLLC}.

For CI Aql, we have ignored the two magnitudes that are during eclipses.  Also, we have converted the B-band magnitudes to V-band by applying the long term average $B-V=1.03$ mag (Schaefer 2009b).  The magnitudes in Table \ref{SUM2} are in the V-band because most of the original observations were in this band.  Before the 1917 eruption, we have only one measure, with B=17.22 which implies V=16.19.  Between 1917 and 1941, we have eight measures, with an average of B=17.12 (which implies V=16.09) with an RMS scatter of 0.11 mag and a range of 0.34 mag.  Between 1941 and 2000, we have data from Szkody (1994) and from various deep Schmidt plates, with an average of V=16.21, a standard deviation of 0.12 mag, and a total range of 0.33 mag.  Between the years 1991 and 1996, we have 240 magnitudes from RoboScope (that might have a calibration error of up to half a magnitude, Schaefer (2009b)) with an average of V=16.15, an RMS scatter of 0.07 mag, and a range of 0.34 mag.  Early in 2002, after the end of the 2000 eruption, we have a very large amount of photometry, but most of this was taken during eclipses with the goal of measuring eclipse times.  We have constructed 39 nightly averages from outside eclipses, with an average of V=16.12, a standard deviation of 0.09 mag, and a range of 0.38 mag.

We do not have any magnitudes within four years before any eruption, so we have no useful constraint on pre-eruption rises.

\subsubsection{T CrB (Nova CrB 1866, 1946)}

T CrB was discovered as a star of 2.0 mag by J. Birmingham on 1866 May 12 (Lynn 1866). The orbital period has been measured to be 228.57 days (Fekel et al. 2000). Schaefer (2009b) notes that the orbital light curve shows ellipsoidal variations in the red giant companion star. Using folded light curves from Zamanov et al. (2004) and Leibowitz et al. (1997), the inclination angle can be constrained to $<68^{\circ}$. T CrB has two known eruptions, and the 1866 eruption has no prior observations for us to construct a pre-eruption light curve from. However, we have very good coverage of the nova before and after the 1946 eruption in both the B and V bands (Figure \ref{TCRBLC}).

Before the 1946 eruption, we see that T CrB in quiescence was at B=10.39, with an RMS scatter of 0.17 and a range of 0.51. This is not including the unique anticipation event we see in T CrB leading into the 1946 eruption; T CrB  does not show a pre-eruption rise, but a pre-eruption dip! The event began one year before the 1946 eruption, and the system gets as dim as B=12.23, the dimmest T CrB has been seen (Schaefer 2009b).  Just 29 days before eruption, T CrB was \textit{two magnitudes} dimmer than normal. The dip is not a new discovery, and was identified as odd behavior even even while it was happening (Peltier 1945). What is new is that the close temporal coincidence (29 days) for an event that is unique out of a long observational history (143 years) makes an obvious connection between this pre-eruption rise and the eruption of 1946.

After the nova returned to quiescence (after the eruption), the nova underwent a sudden brightening. This brightening lasted $\sim 100$ days, and started after T CrB had been back at quiescence for 50 days. The nova is blue in color during this secondary maximum, as opposed to the red color it has when in quiescence. Schaefer (2009b) details that no other nova event has ever shown this kind of behavior. The cause of this rebrightening is unknown.

AAVSO has extensive coverage of T CrB after the 1946 eruption in the visual band (Figure \ref{TCRBLC2}), and some recent coverage in the B band. Taking the 86 B measurements from AAVSO between 2007 April 19 (JD 2454210.45) and 2009 June 1 (JD 2454984.40), we find T CrB to have an average B=11.48, with RMS scatter of 0.25 and a range of 1.16.  With this, we have $\Delta m=-1.09$ . This is in reasonable agreement with the measurements of Bruch \& Engel (1994), who put T CrB at B=11.25 and Schaefer (2009b) who gives B to be 11.60, 11.44 and 11.64.  However, we see from Figure \ref{TCRBLC2} that we could have chosen other post-eruption epochs that would lead to $\Delta m$ ranging over half a magnitude in each direction, with the large negative $\Delta m$ coming as a consequence of T CrB slowly fading long after the eruption is over.

When examining the history of T CrB through the AAVSO database, this kind of $\Delta m$ is very normal (Figure \ref{TCRBLC2}). We can see large-scale variability on a long term scale in the V band. This indicates a number of issues, the biggest of which being that it is not clear \textit{when}  a post-eruption quiescent magnitude should be measured. Depending on where in the light curve of T CrB the measurement is taken, the difference between the pre- and post- eruption magnitude may be small or large. It is clear in the case of T CrB that one magnitude is not a significant change, but for novae with far less coverage, this kind of trend may not be visible.  This heavily-sampled light curve exquisitely demonstrates the perpetual and substantial variations of novae in quiescence on all time scales.

\subsubsection{RS Oph (Nova Oph 1898, 1907, 1933, 1945, 1958, 1967, 1985, 2006)}

W.P. Flemming discovered RS Oph on Harvard spectral plates in 1901 (Pickering 1901). It has been observed to be as bright as 4.3 mag (Campbell 1933). The orbital period of RS Oph has been measured at 455.72 days (Fekel et al. 2000). RS Oph has been extensively studied; we have gathered nearly 47,000 observations of RS Oph in quiescence from the AAVSO database. These measurements are in V, and are binned into 0.01 year time intervals (Figure \ref{RSOPHLC}). This data covers RS Oph from 1914-2004, and has magnificent coverage throughout. We do not have any pre-eruption data for the first two known outbursts of RS Oph.

Between 1914 June and 1933 June, we have 126 measurements of RS Oph, with V=11.11, an RMS scatter of 0.18, and a range of 1.40. Between 1935 January and 1945 October, we have 635 measurements, with V=11.28, an RMS scatter of 0.41 and a range of 2.7. Between 1947 January and 1958 July, we show an average of V=11.28 over 766 measurements, with an RMS scatter of 0.36 and a range of 2.38. Between 1959 July and 1967 October, we have 583 measurements, with V=11.00, an RMS scatter of 0.58, and a range of 2.7. We have 1172 measurements between 1969 February and 1984 November, with V=11.39, and RMS scatter of 0.49, and a range of 2.60. We have 1614 measurements of RS Oph between the years 1986 April and 2006 February, with an average brightness of V=11.34 over the years , with an RMS scatter of 0.30 and a range of 2.00. Finally, between 2007 July and 2009 June, we have 154 measurements, with V=11.22, an RMS scatter of 0.29, and a range of 1.41.

We see no evidence of any pre-eruption anticipation in any of the eruptions for which we have pre-eruption coverage. RS Oph does, however, highlight the very important point that nova systems vary on all time scales. We see variations on timescales ranging from as short as daily to as long as centennially. In the case of RS Oph, no one type of phenomenon (rises or dips) appears immediately before any eruption without similar variations occurring often at other times without any eruption immediately following, i.e., we see many rises and dips throughout the light curve which do not immediately preceede eruptions.  The 2006 eruption has a wiggle eight months before the eruption (Starrfield 2008), but an examination of the AAVSO light curve (see also Figure \ref{RSOPHBREAKDOWN}) shows a long gap between a perfectly ordinary fluctuation and the nova eruption.  Another possibly interesting event is the rise shortly before the 1985 outburst, but again the larger picture shows that this is a perfectly ordinary variation, while a close up of the light curve shows that the rise actually reversed itself so that the quiescent nova faded for three weeks by over half a magnitude.  Without the whole picture of RS Oph's behavior, we could have been fooled into claiming pre-eruption anticipation events. The importance of establishing some sort of history before declaring an anticipation event is clear when seeing the large changes in RS Oph over time.

\subsubsection{T Pyx (Nova Pyx 1890, 1902, 1920,  1944, 1967)}

The 1890 and 1902 eruptions of T Pyx were discovered on the Harvard plates in 1913 by H. Leavitt (Pickering 1913). Schaefer et al. (1992) discovered the orbital period to be 0.076 days, and this has had extensive confirmation from photometry (Patterson et al. 1998) and spectroscopy (Uthas 2009).  The long term light curve shows that T Pyx is undergoing a secular dimming, from 13.8 mag before the 1890 eruption to 15.5 in 2004 (Schaefer 2005) to 15.7 in 2009.  This decline can only be from a drop in the accretion rate, with the 2009 rate being just 3\% of the 1890 rate, and this turn-off of accretion provides a ready explanation of why T Pyx did not undergo an eruption in the 1980s (Schaefer 2005).  T Pyx has a nova shell, for which recent {\it Hubble Space Telescope} images show to be expanding with a velocity of close to 600 km s$^{-1}$, with a mass of $\sim$10$^{-4.5}$ M$_{\odot}$, and expanding from an eruption in 1866$\pm$6 (Schaefer et al. 2009).  With the 1866 eruption having such a large mass and low ejection velocity, it cannot be a recurrent nova event, but is instead an ordinary nova eruption where the material had accreted onto the white dwarf over $\sim$750,000 at the usual gravitational wave radiation angular momentum loss rate of $\sim$4 $\times$ 10$^{-11}$ M$_{\odot}$ yr$^{-1}$.  With this, the pre-nova in 1865 would have $m_{pre}=18.5$ mag, while we observe $m_{post}=13.8$ mag in early 1890, for $\Delta m=4.7$ mag across the 1866 ordinary nova eruption (Schaefer et al. 2009).  While this large value of $\Delta m$ is confident, it is not directly observed, and it will not be considered further in this paper.

In addition to the long-term decline, T Pyx has been the cause of interest because of its high accretion rate of $> 10^{-8}$ M$_{\odot}$ yr$^{-1}$ (Patterson et al. 1998; Selvelli et al. 2008). This accretion rate is more than three orders of magnitude higher than what would be expected from its short orbital period. Knigge et al. (2000) provide an explanation for the high accretion rate, proposing that T Pyx is a wind-driven super-soft source, where the luminous soft X-rays from sustained nuclear burning on the surface of the white dwarf are heating up the atmosphere of the nearby companion star and driving the high accretion rate.  However, Sevelli et al. (2008) looked at T Pyx with {\it XMM-Newton} and found that it is not longer a super-soft X-ray source.  A simple reconciliation of these two claims is that T Pyx started out as a very luminous super-soft source in 1866, which powers the high mass transfer rate (as described by Knigge et al. (2000)), but that the secular decline in brightness (as described by Schaefer (2005)) indicates that the current accretion rate is now so low that the super-soft source has effectively turned off (as described by Schaefer et al. (2009)), so that soft X-ray flux cannot currently be seen (as described by Selvelli et al. 2008).

The quiescent light curve of T Pyx goes as far back as $\sim$23 days before the 1890 eruption (Figure \ref{TPYXLC}). The light curve is composed of measurements from Harvard plates, photometry from the literature, and our own photometry, for a total of 223 B measurements (Schaefer 2009b). Before the 1890 eruption, we have five magnitudes for T Pyx with an average of B=13.8. Between 1890 May and 1901 March, we have 12 measurements, with B=14.38, an RMS scatter of 0.18 and a range of 0.60. Between 1903 May and 1910 May, we have 7 measurements, with B=14.74, an RMS scatter of 0.13 and a range of 0.40. Between 1924 February and 1944 March, we have 68 measurements, showing B=14.88, with an RMS of 0.25 and a range of 1.30. Between 1946 May and 1954 January, we have 28 measurements, with B=14.70, an RMS scatter of 0.20, and a range of 0.90. Finally, between 1969 April and 2009 April, we have 107 measurements, with B=15.51, an RMS=0.12, and a range of 0.72. We do not have sufficient coverage before the 1890, 1920 or 1967 eruptions to observe any anticipatory events.  The 1902 and 1944 eruptions did not have any anticipation events. The secular decline in the light curve of T Pyx that occurs over the last century, crossing over multiple eruptions, and cannot be connected to individual eruptions.

\subsubsection{V3890 Sgr (Nova Sgr 1962, 1990)}

H. Dinnerstein discovered V3890 Sgr on Maria-Mitchel (MMO) plates taken by D. Hoffleit on 1962 June 2 (JD 2437818) and seen as bright as 8.4 photographic mag (Dinnerstein \& Hoffleit 1973).  The companion star has been identified to be a red giant based on its infrared brightness and colors (Harrison, Johnson \& Spyromilio 1993).  The orbital period has recently been discovered to be 519.7 days, based on photometric modulations displaying a shallow eclipse plus ellipsoidal variations on Harvard plates, MMO plates, AAVSO magnitudes, ROTSE images, and SMARTS images from 1899 to 2009 (Schaefer 2009a).

Our light curve for V3890 Sgr comes from Harvard College Observatory, MMO, AAVSO observations and CTIO observations taken from SMARTS telescopes (Figure \ref{V3890SGRLC}). Observations are either in the V-band, or an appropriate correction is applied to correct the measurement into the V-band (Schaefer 2009b). One problem that is prevalent in the light curve is that the first three data sets (HCO, MMO, and AAVSO) have clear detection thresholds with many of the observations showing V3890 Sgr to be below the threshold. This makes seeking secular changes somewhat difficult, as we are only seeing the bright portions of the light curve.  However, both the pre-eruption and post-eruption light curve for the 1962 eruption come from only one source of data, MMO, so threshold effects are uniform. Thus, any pre-eruption rise would be visible, and variability can be compared across the eruption.

We have 31 measurements of V3890 Sgr before the 1962 eruption, going in as close as 250 days before the peak. During this time, we find that V3890 Sgr averaged a brightness of B=16.23 with an RMS of 0.31 and a range of 1.70. There is no evidence for a pre-eruption anticipation event. After the eruption, we have 150 measurements, with B=16.11, an RMS scatter of 0.32, and a range of 1.50.

\subsubsection{U Sco (Nova Sco 1863, 1906, 1917, 1936, 1945, 1969, 1979, 1987, 1999)}

N.R. Pogson discovered U Sco as a bright star of  9.1 mag on 1863 May 20 (Pogson, 1908). The nova had been previously unseen by Pogson on prior nights when he was observing the field. After its initial discovery, U Sco was unseen despite many attempts to observe it until its rediscovery on the Harvard plates' where it has been seen as bright as 8.8 mag (during the 1906 eruption; Thomas (1940)). Schaefer (1990) discovered the deep eclipses with an orbital period of 1.23 days. 

Our U Sco light curve comes from Harvard Plates and our own photometry (Schaefer 2009b). These measurements are in B magnitudes, and are presented in Figure \ref{USCOLC}. Despite all the eruptions of U Sco, only the 1979 outburst has observations shortly before the event. The 1987 and 1999 outbursts have good coverage after the event, but none immediately before.

Between 1969 and 1979, we have 2 measurements of U Sco; it is observed to be at B=18.41.  The measurement taken just 89 days before the 1979 eruption is within the normal range of the system, therefore we conclude that the 1979 eruption does not have an anticipatory reaction.  Between 1979 and 1987, we have four measurements, with B=18.27, an RMS scatter of 0.11 and a range of 0.25. Between 1987 and 1999, we have 26 measurements, with B=18.52, an RMS scatter of 0.19, and a range of 0.94. Finally, between 1999 and 2009, we measure U Sco 30 times, with B=18.45, an RMS scatter of 0.31 and a range of 1.35. The range of U Sco's brightness apparently has a substantial increase from 1979 and 1987 to 1999 and 2009, but this is caused by the large difference in the number of observations between the two time intervals.

\subsection{DSS Images}

Many novae are too faint at quiescence to be seen in the usual archival plate collections.  For novae after the year 1955 or so, the obvious solution is to use the Palomar Observatory Sky Survey (POSS).  This goes down to $B\sim21$ and can cover many quiescent novae.  For these cases, the big disadvantage is that we only get {\it one} pre-eruption magnitude, and this is not enough to look for pre-eruption rises (or dips) nor is it adequate to determine the variability before the outburst.  Nevertheless, even one pre-eruption magnitude will allow for a measure of $\Delta m$.  Admittedly, the $m_{pre}$ value will not have been averaged over flickering and ordinary fluctuations, but we can still pick out cases of large $\Delta m$ values despite this uncertainty.  With many novae, we can produce an average $\Delta m$ that has averaged over the flickering and fluctuations.

The basic data appears in the various sky surveys made with the big Schmidt telescopes, typified by the first POSS survey (POSS I) in the middle 1950s and the second POSS survey (POSS II) in the late 1980s through the late 1990s.  Fortunately, these deep images have already been scanned and are digitally available from many sources.  In principle, we can examine these images and compare the stellar radii between the novae and nearby comparison stars to produce a magnitude with typically a quarter of a magnitude accuracy.  In practice, we often have no sequence of comparison stars that goes deep enough.  Fortunately, a large program has calibrated both the POSS I and POSS II plates and reports the B-band magnitudes (USNO B-1.0 B1 and B2 respectively) for all stars (Monet 2003).  These magnitudes have a quoted uncertainty of 0.3 mag, which is just fine for our nova program as there is no use in having accuracy much better than the amplitude of normal flickering and fluctuations.  And, when comparing B1 and B2 magnitudes, any uncertainty in the calibration cancels out for $\Delta m$.

We selected novae that appeared on the POSS I and POSS II plates that a) erupted after the first POSS survey, b) were bright enough for reliable measurements off the plates, c) were far enough North for the surveys, and d) in fields not so crowded as to obscure the source.  In these cases, the pre-eruption magnitude measurements came from the USNO B-1.0 catalog, generally the POSS I magnitudes (B1).  In five cases, the pre-eruption magnitude came from the POSS II (B2).  For three novae (V1500 Cyg, V1974 Cyg, and RW UMi), the pre-nova was completely invisible on the POSS I plates, with the limits coming from the faintest nearby stars.  This is an important result because all three novae have flat light curves now long after the eruption is over that are substantially brighter than the Palomar limits, demonstrating that these three nova have significantly large $\Delta m$ values.  These large magnitude differences have been seen in these novae before (Ka{\l}u\.{z}ny \& Chlebowski 1989, Wade 1987; Bianchini et al. 2003; Tamburini et al. 2007). The post-eruption magnitude was generally from the POSS II (B2), but in some cases we used magnitudes from the literature.  With this, we can get $m_{pre}$ and $m_{post}$ for 20 novae.  A summary of the new pre-eruption magnitudes is presented in Table \ref{DSSSUM}.

\section{Pre-Eruption Rises and Dips}

R75 concluded that of 11 novae with sufficient pre-eruption coverage, five novae showed a distinct and significant pre-eruption brightening. Again, this result is surprising as is suggests that nearly half of all novae preemptively brighten beyond their normal variability before an eruption. In addition, it suggests that the donor star anticipates the eruption, and increases matter flow onto the star. There is no clear explanation for why this would occur. 

For four of the five claimed cases of pre-eruption rises, we find that all the claims can be rejected.  For CP Lac, the claimed step rise is neither significant nor reproduced in later data, and the time match of the step rise to the switching of the data source (one of which we are warned has an unknown offset) is highly suspicious.  For BT Mon, the original claimed pre-eruption rise was based on a cursory search of the Harvard plates that has since been shown to be wrong.  For GK Per, the error was entirely in the report on three plates just before eruption, where the nova is certainly invisible (see Figure \ref{GKPERPLATES}).  For LV Vul, the entire case for a pre-eruption rise is based on a typo in one paper (with a later published erratum).  Thus we see that most of the evidence for the phenomenon of `pre-eruption rises' comes down to simple errors in the literature.

Nevertheless, our independent investigation does show that in fact \textit{one} of the R75 novae, V533 Her, showed a significant pre-eruption rise. This nova rose smoothly by nearly 1.3 magnitudes within the $\sim$1.5 years leading up to the eruption. This is far outside the variability range of 0.5 mag that is observed over 65 years away from the eruption.  We take the coincidence in time and the acceleration in the brightening that leads up to the eruption as strong evidence for a causal connection between the rise and the eruption.

V1500 Cyg showed a rise similar to that of V533 Her. The system was observed at close to B=21.5 mag in 1952 and 1970, and was never seen to brighten substantially until the month leading into the eruption. In the month of 1975 August, the system rose nearly 7 magnitudes, getting as bright as B=13.5 (Kukarkin $\&$ Kholopov 1975) just 1 and 5 days before eruption. This makes it an even more impressive rise than that of V533 Her, as V1500 Cyg got substantially brighter in a much shorter span of time.

We also greatly expanded the sample pool by including more novae as well as recurrent novae. The added novae are V368 Aql, QZ Aur, V1500 Cyg, HR Del, DQ Her, and V446 Her; the added recurrent novae are CI Aql, T CrB, RS Oph, T Pyx, V3890 Sgr, and U Sco.  In all, we have 22 nova events for which we have good data to test whether a pre-eruption rise occurred, and only two (V533 Her and V1500 Cyg) showed such a rise.  The statistics are that 18.2\% (two out of eleven) of non-recurrent novae have rises, 11.8\% (two out of 17) of the nova systems have rises, and 9.1\% (two out of 22) of the individual eruptions have rises.  Clearly, pre-eruption rises are an uncommon or even rare behavior.  With {\it two} confident examples of significant pre-eruption rises, we have a hard time making a case that the cause is due to some special or unique property.  This is further emphasized by the utter ordinariness of the nova V533 Her.

T CrB shows a pre-eruption \textit{dip}. This dip was confidently measured in both the B and V bands (and remarked upon as being unique at the time) during the year leading up to the 1946 eruption. The dip behaves in a complex manner in both the B and V bands, with the system being two magnitudes fainter than normal at a time of 29 days before the eruption.  The uniqueness of this event out of the 143 year photometric history of T CrB and the close temporal coincidence provide a strong case that the dip and eruption are causally connected.    The statistics are that 0\% (0 out of ten) of non-recurrent novae have dips, 6.2\% (one out of 16) of the nova systems have dips, and 4.8\% (one out of 22) of the individual eruptions have dips.  Pre-eruption dips are an uncommon or even rare behavior.  Given the association of T CrB with other unusual properties (the post-eruption maximum, the 228 day orbital period, the high mass white dwarf, and the high accretion rate), it is tempting to relegate the pre-eruption dip phenomenon to rare and special circumstances not generally applicable to novae in general.

The cases of V533 Her, V1500 Cyg, and T CrB are unique events in their long observed history.  Both anticipatory events showed a close proximity to the eruption.  It is highly unlikely that these are just random events that just happened to occur within the month before eruption.  This probability argument makes a strong case that the anticipatory behavior is causally connected to the eruption.

For the two pre-eruption rises, we have to ask whether the increase in accretion (with its consequent rise in brightness) was the cause of the eruption happening at that time.  That is, matter is falling onto the white dwarf at a high rate during any rise, so this will be the most likely time for the trigger conditions to be achieved?  (To make a poor analogy, terrestrial dams on rivers are most likely to break during a heavy rainstorm because that is when most of the water comes.)  In this case, the random fluctuations of the accretion mean that most of the matter falls onto the white dwarf during a 'high state', so a randomly occurring rise would be the most likely time for the trigger condition to be passed.  In this view, there is a causal connection between the rise and the eruption, but this would largely be trivial and random.  This idea has substantial problems with the total uniqueness of these anticipatory events within the very long history for V533 Her, where the accelerating rise over a bit more than a year is not just a random or usual fluctuation.  Similarly, for V1500 Cyg, the brightening of any nova by eight magnitudes (from 21.5 to 13.5) is completely outside the realm of any non-eruption fluctuations seen on any other nova.  Another substantial problem is that the extra accreted mass associated with the eruption is miniscule compared to the total mass required to trigger the eruption, so it is incredibly unlikely that a unique rise would happen to be the `final straw' to trigger the nova.  To be quantitative, for V1500 Cyg, an increase in accretion by a factor of 30-1000 over a one month period (see Table 5) over its base rate of $3\times10^{-11}$ M$_{\odot}$ yr$^{-1}$ (Patterson 1984) produces an extra accretion of $\sim10^{-9}$ M$_{\odot}$, which is extremely small compared to the $\sim10^{-5}$ M$_{\odot}$ required to trigger the nova.  For V533 Her, even a doubling of the accretion rate for a year (cf. Fig. 6) above its regular rate of $2\times10^{-9}$ M$_{\odot}$ yr$^{-1}$ (Patterson 1984) only gives an extra $2\times10^{-11}$ M$_{\odot}$  out of a required trigger mass a million times larger.  It is very unlikely (at the $10^{-4}$ to $10^{-6}$ level) that a unique rise would happen to provide the extra mass that pushes the nova over the limit.  So, with these two strong arguments, we reject the idea that the eruption is randomly caused by the higher accretion rate happening to dump more material onto the white dwarf.

There is another effect that might have the increase in accretion cause the nova event.  If there is a random rise in the accretion rate, then the energy from the extra material will somewhat raise the temperature at the base of the accumulated material layer (where the trigger occurs) and this might push the trigger.  The energy from the extra accreted matter ($M_{extra}$) will be $0.5GM_{WD}M_{extra}/R_{WD}$, where the mass and radius of the white dwarf are $M_{WD}$ (taken to be 1 M$_{\odot}$) and $R_{WD}$ (taken to be 5000 km), while $G$ is the usual gravitational constant.  With $M_{extra}$ from the previous paragraph, the extra energy will be $2\times10^{41}$ and $5\times10^{39}$ ergs for V1500 Cyg and V533 Her.  This extra energy will have to diffuse down to the bottom of the hydrogen layer.  (We do not know the time scale for this diffusion, and this might be greatly longer than the observed rises, in which case the idea has further problems.)  In the unphysically optimistic case that this extra energy is spread out uniformly through the hydrogen envelope (of mass $\sim10^{-5}$ M$_{\odot}$), the temperature increase will be 160,000 and 3000 K.  The trigger temperature is $2\times10^7$ K (Shen \& Bildsten 2008).  So any rise caused by the extra accretion associated with a pre-eruption rise is miniscule and will certainly have no effect.  So this possible causal connection can be completely ruled out.

This sets a challenge for theorists. What is the physical mechanism that caused the rises and the dip?  The V533 Her and V1500 Cyg rises likely could only have been from an increase in accretion (since the white dwarf and companion star are both much fainter than the accretion light).  But then, how would the secondary star (which controls the accretion flow) `know' that the base of the accreted material on the surface of the white dwarf is getting near its trigger point?  The T CrB dip might have been caused by a cessation of the accretion flow, but again, how would the red giant companion star `know' to stop pouring matter through the Roche lobe just as the bottom layer on the white dwarf is getting near critical.  Alternatively, the dip might arise from the companion sending out obscuring clouds of material which hide the inner light sources.  For this last possibility, we are still left with the question of how such a unique and unprecedented cloud would be associated with the nova trigger, and the color evolution of the dip.  So, we have no good answer to the challenges posed by the anticipatory behavior of V533 Her, V1500 Cyg, and T CrB.

With only two known rises and one known dip, we clearly need more examples.  But with our work here, we have exhausted the supply of novae for which reasonably well-sampled pre-eruption light curves can be constructed. So the only hope for an observational advance is to await some future nova that happens to have been covered by a CCD sky survey.

\section{Changes in Amplitude of Variations}

In addition to finding pre-eruption rises, R75 also observed that V446 Her showed variability on the order of 4 magnitudes before the eruption, and only a variability of 0.4 magnitudes after the eruption.  He labeled this one-nova phenomenon as ``Class V.''  Although novae vary on all time scales, for the nova to change behavior so closely to the eruption suggests that there is some causal connection between the eruption and the variability of the mass transfer. Only one nova was found to have this kind of behavior, meaning that $\sim 9\%$ of R75's sample (and hence novae) should have this kind of behavior.

A substantial problem with testing this phenomenon is that all CVs (and novae in particular) vary on all time scales and there are often time intervals of relative quiet followed by intervals of relatively high variability at times far from any eruption.  All CVs flicker on time scales of minutes to hours, with amplitudes up to several tenths of a magnitude, and with this flickering displaying a power law power density spectrum (Bruch 1992; Yonehara et al. 1997).  On long time scales, all novae display long term trends and short term fluctuations that are unexplained.  A series of wonderful long-term light curves has been obtained with the RoboScope telescope in Indiana; these light curves are shown in Honeycutt et al. (1998) and Kafka \& Honeycutt (2004). All of the old novae (long after the eruption) show trends on all time scales from a few days to months to a year to a decade with amplitudes from 0.4-2.0 mags.  This behavior is ubiquitous.  A feature of these long, well-sampled light curves is that the novae have occasional long time intervals where $\sigma_{post}$ is small intermixed with intervals where $\sigma_{post}$ is comparatively large.  For example, Q Cyg has $\sigma_{post}\approx 0.1$ in 1992 and 1993, but $\sigma_{post}\approx0.25$ in 1991 and 1994-7 (Honeycutt et al. 1998).  These changes in variation amplitude are independent of any eruption.  The problem arises should one of these changes happen to occur around the time of an eruption.  Taken in isolation, such a normal case could be viewed as evidence for a new class of phenomenon associated with the eruption.  Not taken in isolation, we realize that such a case is common and has nothing to do with any eruption physics.

The realization that the amplitude of variations changes on all time scales brings a further realization that it will be effectively impossible to find any convincing case of ``Class V" novae.  That is, we expect to see changes in range and RMS scatter across all eruptions (or any other randomly specified time).  With this, we cannot use a temporal coincidence to argue for a causal connection.  Thus, even before we start, we must realize that there is no realistic chance of finding a case where we can even suggest that a difference in RMS scatter is caused by the eruption.  

Nevertheless, we have enough data to compare pre- and post-eruption variability on ten novae, and 18 eruptions on six recurrent novae (Table \ref{SUM2}).  We can still look to see if there are any cases where the variations change greatly in amplitude across an eruption.

First, we should look at the prototypical ``Class V" system, V446 Her.  A glance at Figure \ref{V446HERLC} tells the story.  V446 Her does not have any substantial change in amplitude of fluctuations across its nova outburst.  This is confirmed with the quantitative analysis given in Table \ref{SUM2}.  We have the RMS scatter changing from 0.37 mag before to 0.31 mag after, and the range changing from 1.33 mag before to 1.28 mag after.  So the prototype and only ``Class V" nova is certainly not in ``Class V.''

Next, we should consider all other novae.  From Table \ref{SUM2}, we see that in no other case is there any substantial change in the amplitude of variability.  Instead, we see just the normal and expected changes.  So in all, there are zero cases for ``Class V.''  As such, ``Class V'' must be regarded as nonexistent.

\section{Brightness of a Nova Before and After the Eruption}

Finally, R75 covered the question of whether the systems were of the same average brightness after the eruption as compared to before it. The idea was to test whether novae would eventually return to their pre-eruption magnitude, or whether the accretion rate is somehow changed by the eruption itself.  R75 was able to get $\Delta m$ values for 18 systems. Of these, only one, BT Mon, showed a possible case of a system being brighter than it was before the eruption.

A problem with this question is that novae are flickering and fluctuating on all time scales at times far from eruptions, so it is difficult to measure (or even to define) an average $m_{pre}$ and $m_{post}$ value.  This problem is worse when we have only one (or just a few) pre-eruption magnitudes, as is the case for all the novae in Table \ref{DSSSUM}.  The only practical solution is to acknowledge that we expect $\Delta m$ to have a substantial intrinsic scatter and to realize that the question must be handled statitically.  The expected scatter apparently should have an RMS scatter of order 0.5 mag based on many nova light curves away from eruptions.  The task is to measure the average (or median) and the RMS scatter of the distribution of $\Delta m$ values and to seek significant outliers.  If the median $\Delta m$ is consistent with zero, then we conclude that (at least most) of the mass transfer rates in novae are unaffected by the nova explosion.  If we find significant outliers, then we can seek the physics of a mechanism to explain the brightness change.

Another problem with this task is the issue of determining when a nova eruption has finally ended.  That is, normal nova light curves have tails that slowly asymptote to the quiescent level, so it is difficult to decide when the nova is post-eruption.  If $m_{post}$ is chosen too early in time, then $\Delta m$ will be systematically large in the positive direction.  In practice, this problem is both easy to handle and negligibly small.  Most of our novae have well-observed light curves going to quiescence, and these can be used to readily determine a date when the light curve has gone flat.  We find that plotting the light curve with the logarithm of time on the horizontal axis makes the eruption tail appear as a straight line, and so the intersection with the flat post-eruption segment is well determined.  Our $m_{post}$ values all come from long enough after the eruption that we are confident the eruption has ended.  If novae have some hypothetical very slow decline at the end of the usual tail in the light curve, then that is exactly the sort of effect that we seek.

We have $\Delta m$ values for 30 CNe (3 of which are important limits), plus 19 eruptions of six RNe (Table \ref{ROBSUM}).  We have an average of +0.26 mag, a median of +0.14 mag, and a standard deviation of 0.69 mag.  With this, we see that V723 Cas, V1500 Cyg, V1974 Cyg, V4633 Sgr, and RW UMi are all $>3$-sigma outliers.  With these outliers rejected, we have an average of +0.16 mag, a median of +0.13 mag, and a standard deviation of 0.42 mag.  The formal uncertainty on this average is $\pm0.08$ mag, so the average $\Delta m$ is consistent with being zero.  If we include the RN eruptions, the average $\Delta m$ is 0.03 mag, the median is 0.03 mag, and the standard deviation is 0.43 mag.  Again, the average $\Delta m$ is close to zero.  As such, we conclude that most novae have no change in average accretion rates across outbursts. 

But at least some novae show highly significant brightening after eruption.  Five novae (V723 Cas, V1500 Cyg, V1974 Cyg, V4633 Sgr, and RW UMi) all show significantly large $\Delta m$ values (+3.01, +2.71, $>$+3.75, $>$+3.0, and $>$2.67 mag respectively).  (We are also suspicious of V1330 Cyg and QU Vul with $\Delta m$ values of +1.17 and +1.18 mag, respectively, with these now being 2.4-sigma outliers.)  These are very significant rises, and there is no chance of artifacts (say, due to small number of pre-eruption plates, or due to the incredibly long tails in the eruption light curves).  We are left with 5 out of 30 classical novae that have a $\Delta m > 2.5 mag$, which is to say that their quiescent brightness increased by over a factor of ten from before eruption to long after the eruption had completely faded away. 

Apparently we must have some mechanism that (at least occasionally) will make the post-eruption system much brighter than the pre-eruption system.  This must translate into a large increase in the accretion rate caused by the nova.  Explaining why a sixth of classical novae brighten greatly after eruption is another challenge to theorists.

\section{Conclusions}

We have made a nearly-exhaustive study of the pre-eruption behavior of novae light curves, with this being a test, modernization, and extension of the seminal paper of Robinson (1975).  (1) Two novae are found to have a pre-eruption rise, with V533 Her rising by 1.25 mag in the 1.5 years before its eruption.  V1500 Cyg also shows a significant brightening, rising 8 magnitudes during the one month before its eruption. (2) T CrB showed a significant pre-eruption dip before it's 1946 eruption. (3) V445 Her showed no significant change in it's variability following it's eruption, and indeed no nova examined showed a significant change in variability connected to the eruption (4) Most novae have essentially no change in quiescent brightness across their eruption, which leads us to conclude that the eruption does not affect the matter flow. (5) However, we find that 5 out of 30 classical novae (V723 Cas, V1500 Cyg, V1974 Cyg, V4633 Sgr, and RW UMi) have brightened by more than a factor of ten ($\Delta m > 2.5$ mag).

~

We would like to thank Alison Doane and the staff of the Harvard College Observatory for their work in maintaining the plate archive. In addition, we would like to thank the observers of the AAVSO worldwide for all their work. We also would like to acknowledge the help of Downes et al. (2006) and Duerbeck (1987) in identifying possible targets and locating older references. This work is supported by NSF grant \#AST-070 879.

{\it Facilities:} \facility{AAVSO}

{}

\clearpage

\begin{deluxetable}{llllll}
\tabletypesize{\scriptsize}
\tablewidth{0pc}
\tablecaption{Plate Identification - Observatory Legend}
\tablehead{\colhead{Observatory} & \colhead{Series} & \colhead{D (mm)} & \colhead{Scale ("/mm)} & \colhead{$m_{lim}$} & \colhead{Years}}
\startdata
Harvard\tablenotemark{a} 	& A		&	610	&	60	&	18		&	1893-1950	\\
						& MA	&	305	&	97	&	17-18	&	1905-1983\\
						& MC	&	406	&	98	&	17-18	&	1909-1992\\
						& MF		&	254	&	167	&	17		&	1915-1955 \\
						& I 		&	203	&	163	&	17		&	1889-1946\\
\hline
Sonneberg 	& AA\tablenotemark{b}	&	170	&	170						& 16.5	&	1923-1971\\
			& B					&	400	&	110						& 17.5	&	1957-1959\\
			& F					&	140	&	300						& 16		&	1928-1969\\
			& GA 				&	400	&	130						& 17.5	&	1938-1945\\
			& GB 				&	400	&	100						& 17.5	&	1960-1993\\
			& GC 				&	400	&	130						& 17.5	&	1961-1998\\
			& SC 				&	500/700	&	120					& 18		&	1952-1993\\
			& TE					&	55-86 (Varies)	&	690-980 (Varies)	& 14.5	&	1953-present\\
\enddata
\tablenotetext{a}{http://tdc-www.harvard.edu/plates/plates.html}
\tablenotetext{b}{These plates are referred to as A plates at Sonneberg Observatory, but we shall refer to them as AA plates in order to avoid confusion with Harvard's A plates.}
\label{LEGEND}
\end{deluxetable}

\clearpage

\begin{deluxetable}{cccc}
\tabletypesize{\scriptsize}
\tablewidth{0pc}
\tablecaption{Archival Plates Examined}
\tablehead{\colhead{Nova} & \colhead{Year} &  \colhead{$\#$ Plates\tablenotemark{a}} & \colhead{Years Covered}}
\startdata
V368 Aql & 1936 & 11 & 1926-7,1930-32,1935, \\
QZ Aur & 1964 & 58 & 1933-4,1940,1959, \\
& & & 1961-4,1973,1975, \\
& & & 1982-3,1985-6,1988 \\
HR Del & 1967 & 80 & 1956-60, 1965-7,1986-90,1994\\
DQ Her\tablenotemark{b} & 1934 & 50  & 1894-1934  \\
V446 Her & 1960 & 138 & 1926,1931,1937-41,\\
& & & 1950,1959,1982,1984,\\
& & & 1985-6,1988,1990\\
V533 Her & 1963 & 309 & 1930-5,1941-6,1948-50,\\
& & & 1955-6,1958-63,1965-69,1982-88 \\
CP Lac & 1936 & 37 & 1898-9,1922-33 \\
BT Mon\tablenotemark{b} & 1939 & 68 & 1905,1911,1914,1917,\\
& & & 1923-4,1926-30,1932-39 \\
GK Per & 1901 & 22 & 1890-4,1896-1901 \\
LV Vul & 1968 & 206 & 1896,1899,1906,1910,\\
& & &1921,1923,1925-30,1935,\\
& & & 1936,1938-42,1944-5,1948,\\
& & & 1950,1959-67,1974-7,\\
& & & 1979,1982-5,1993\\
\enddata
\tablenotetext{a}{Total number of plates with useful data (including limiting magnitudes)}
\tablenotetext{b}{From previous work done at HCO, see Schaefer \& Patterson (1983)}
\label{SUM}
\end{deluxetable}

\clearpage

\begin{deluxetable}{ccccccc}
\tabletypesize{\scriptsize}
\tablewidth{0pc}
\tablecaption{Light Curve Summmary}
\tablehead{\colhead{Nova} & \colhead{$m_{pre}$} & \colhead{$\sigma_{pre}$} & \colhead{$R_{pre}$} & \colhead{$m_{post}$} & \colhead{$\sigma_{post}$} & \colhead{$R_{post}$}}
\startdata
V368 Aql	&	16.53	&	0.14	&	0.44	&	16.90	&	\ldots	&	\ldots	\\
QZ Aur	&	17.16	&	0.23	&	1.65	&	17.13	&	0.18	&	0.97\\
HR Del	&	11.97	&	0.35	&	1.22	&	12.20	&	0.41	&	1.56	\\
DQ Her	&	15.09	&	0.45	&	2.09	&	14.60	&	0.27 &	2.00	\\
V446 Her	&	16.07	&	0.37	&	1.33	&	16.31	&	0.31	&	1.28	\\
V533 Her	&	14.72	&	0.17	&	0.92	&	14.25	&	0.47	&	1.17	\\
CP Lac	&	15.87	&	0.26	&	0.87	&	15.5	&	0.4	&	2.3	\\
BT Mon	&	15.28	&	0.24	&	1.20	&	15.37	&	0.25	&	1.13	\\
GK Per	&	13.87	&	0.39	&	1.11	&	13.94	&  0.12	&	0.55	\\
LV Vul	&	16.24	&	0.31	&	1.64	&	16.10	&	0.21	&	1.25	\\
CI Aql-1917	&	16.19	&	\ldots	&	\ldots	&	16.09	&	0.11	&	0.34	\\
CI Aql-1941	&	16.09	&	0.11	&	0.34	&	16.21	&	0.07	&	0.34	\\
CI Aql-2000	&	16.21	&	0.07	&	0.34	&	16.12	&	0.09	&	0.38	\\
T CrB-1946	&	10.39	&	0.17	&	0.51	&	11.48	&	0.25	&  1.16\\
RS Oph-1933	&	11.11	&	0.18	&	1.40	&	11.28	&	0.41	&	2.70\\
RS Oph-1945	&	11.28	&	0.41	&	2.70	&	11.28	&	0.36	&	2.38\\
RS Oph-1958	&	11.28	&	0.36	&	2.38	&	11.00	&	0.58	&	2.70\\
RS Oph-1967	&	11.00	&	0.58	&	2.70	&	11.39	&	0.49	&	2.60\\
RS Oph-1985	&	11.39	&	0.49	&	2.60	&	11.34	&	0.30	&	2.00\\
RS Oph-2006	&	11.34	&	0.30	&	2.00	&	11.22	&	0.29	&	1.41\\
T Pyx-1890	&	13.80	&	\ldots	& \ldots	&	14.38	&	0.18	&	0.60\\
T Pyx-1902	&	14.38	&	0.18	&	0.60	&	14.74	&	0.13	&	0.40	\\
T Pyx-1920	&	14.74	&	0.13	&	0.40	&	14.88	&	0.25	&	1.30	\\
T Pyx-1944	&	14.88	&	0.25	&	1.30	&	14.70	&	0.20	&	0.90	\\
T Pyx-1967	&	14.70	&	0.20	&	0.90	&	15.51	&	0.12	&	0.72 \\
V3890 Sgr-1962	&	16.23	&	0.31&	1.70	&	16.11	&	0.32	&	1.50\\
U Sco-1979	&	18.41	&	\ldots	&	\ldots	&	18.27	&	0.11	&	0.25\\
U Sco-1987	&	18.27	&	0.11	&	0.25	&	18.52	&	0.19	&	0.94\\
U Sco-1999	&	18.52	&	0.19	&	0.94	&	18.45	&	0.31	&	1.35\\
\enddata
\label{SUM2}
\end{deluxetable}

\clearpage

\begin{deluxetable}{ccc}
\tabletypesize{\scriptsize}
\tablewidth{0pc}
\tablecaption{V368 Aql Magnitudes}
\tablehead{\colhead{Plate} & \colhead{JD ($\pm$ 0.4)} & \colhead{B (mag)}}
\startdata
MF10383	&	2424683.5	&		16.55 \\
MF 10393	&	2424684.5	&		16.70 \\
MF10401	&	2424686.5	&		16.50 \\
MF10610	&	2424732.5	&		16.50 \\
MC 21957	&	2425035.5	&		16.26 \\
MC22559	&	2425037.5	&		16.46 \\
MF 14479	&	2426154.5	&		16.70 \\
MF 15718	&	2426540.5	&		16.70 \\
MF 17160	&	2426917.5	&	$>$	16.61 \\
A17858	&	2428013.5	&		16.48 \\
A17908	&	2428033.5	&		16.45 \\
\enddata
\label{V368AQLMAGS}
\end{deluxetable}

\clearpage

\begin{deluxetable}{ccccc}
\tabletypesize{\scriptsize}
\tablewidth{0pc}
\tablecaption{V1500 Cyg Magnitudes}
\tablehead{\colhead{Plate} & \colhead{Date} & \colhead{JD} & \colhead{mag} & \colhead{Source}}
\startdata
Palomar Schmidt	&	1952 July 19		&	2434212.91	&	B=21.5	&	Duerbeck (1987)	\\
Palomar Schmidt	&	1952 July 19		&	2434212.87	&	R$>$20	&	Duerbeck (1987)	\\
Asiago				&	1967 October 30	&	2439794	&	B$>$19.5	&	Rosino $\&$ Tempesti (1977) \\
Asiago			&	1967 November 1	&	2439796	&	V$>$19.5	&	Rosino $\&$ Tempesti (1977) \\
Palomar Schmidt	&	1969 August 17	&	2440816	&	B$>$18.0	&	Wade (1987)\\
Palomar Schmidt	&	1970 July 31		&	2440799	&	V$\approx$20.5	&	Wade (1987) \\
Baldone Schmidt	&	1972		&	$\sim$2441560	&	B$>$19	&	Kukarkin $\&$ Kholopov (1975) \\
Baldone Schmidt	&	1974 December		&	$\sim$2442397	&	B$>$17.9	&	Kukarkin $\&$ Kholopov (1975) \\
Baldone Schmidt	&	1975 August 5		&	2442630.41	&	V=15.95	&	Alksne $\&$ Platais (1975) \\
Baldone Schmidt	&	1975 August 7		&	2442632.51	&	B=17.6	&	Alksne $\&$ Platais (1975) \\
40cm Astrograph	&	1975 August 12	&	2442637.48	&	B=17.0	&	Samus (1975) \\
Baldone Schmidt	&	1975 August 24		&	2442649.44	&	R=13.5	&	Alksne $\&$ Platais (1975) \\
Moscow Schmidt	&	1975 August 28		&	244653.232	&	B=13.5	&	Kukarkin $\&$ Kholopov (1975)\\
\enddata
\label{V1500CYGMAGS}
\end{deluxetable}

\clearpage

\begin{deluxetable}{ccc}
\tabletypesize{\scriptsize}
\tablewidth{0pc}
\tablecaption{HR Del Magnitudes.}
\tablehead{\colhead{Series} & \colhead{JD} & \colhead{B (mag)}}
\startdata
TE3	&	2435698.48	&	11.52	\\
TE3	&	2435781.32	&	11.83	\\
TE3	&	2435967.57	&	12.21	\\
TE3	&	2436025.46	&	11.98	\\
TE3	&	2436073.45	&	11.52	\\
TE3	&	2436130.36	&	12.21	\\
TE3	&	2436160.29	&	12.21	\\
TE3	&	2436395.50	&	12.41	\\
TE3	&	2436397.40	&	11.98	\\
TE3	&	2436410.46	&	11.52	\\
TE3	&	2436453.50	&	11.98	\\
TE3	&	2436482.35	&	11.75	\\
TE3	&	2436723.50	&	12.21	\\
TE3	&	2436793.54	&	12.44	\\
TE3	&	2436816.47	&	11.52	\\
TE3	&	2436820.45	&	11.83	\\
TE3	&	2436822.47	&	12.13	\\
TE3	&	2436836.40	&	12.44	\\
TE3	&	2436837.40	&	11.75	\\
TE3	&	2436893.27	&	11.52	\\
TE3	&	2437103.51	&	12.13	\\
TE3	&	2437145.49	&	11.52	\\
TE3	&	2438832.66	&	12.44	\\
TE3	&	2438849.64	&	12.44	\\
TE3	&	2438852.63	&	11.98	\\
TE3	&	2438882.57	&	11.52	\\
TE3	&	2438932.49	&	11.52	\\
TE3	&	2438935.48	&	11.83	\\
TE3	&	2438937.51	&	12.44	\\
TE3	&	2438941.41	&	11.75	\\
TE3	&	2438977.48	&	11.52	\\
TE3	&	2439007.45	&	11.98	\\
TE3	&	2439021.33	&	11.98	\\
TE3	&	2439024.40	&	11.52	\\
TE3	&	2439026.36	&	12.44	\\
TE3	&	2439027.39	&	12.13	\\
TE3	&	2439028.36	&	11.75	\\
TE3	&	2439029.35	&	12.44	\\
TE3	&	2439034.30	&	11.98	\\
TE3	&	2439051.30	&	12.44	\\
TE3	&	2439054.27	&	12.44	\\
TE3	&	2439054.27	&	12.44	\\
TE3	&	2439054.33	&	11.98	\\
TE3	&	2439056.34	&	11.52	\\
TE3	&	2439057.34	&	11.52	\\
TE3	&	2439058.37	&	11.75	\\
TE3	&	2439059.32	&	11.52	\\
TE3	&	2439060.33	&	12.13	\\
TE3	&	2439063.37	&	12.44	\\
TE3	&	2439081.26	&	12.13	\\
TE3	&	2439088.25	&	12.44	\\
TE3	&	2439205.64	&	12.74	\\
TE3	&	2439261.53	&	11.98	\\
TE3	&	2439270.49	&	12.44	\\
TE3	&	2439288.50	&	12.21	\\
TE3	&	2439299.51	&	11.52	\\
TE3	&	2439317.45	&	11.98	\\
TE3	&	2439331.48	&	11.98	\\
TE3	&	2439349.43	&	11.52	\\
TE3	&	2439351.45	&	11.52	\\
TE3	&	2439354.48	&	12.44	\\
TE3	&	2439355.43	&	11.98	\\
TE3	&	2439378.35	&	12.13	\\
TE3	&	2439406.32	&	11.83	\\
TE3	&	2439436.29	&	11.52	\\
TE3	&	2439443.30	&	11.83	\\
TE3	&	2439596.60	&	11.52	\\
TE3	&	2439618.54	&	11.75	\\
TE3	&	2439621.54	&	12.13	\\
TE3	&	2446679.42	&	11.98	\\
TE3	&	2446702.33	&	12.44	\\
TE3	&	2446706.34	&	11.98	\\
TE3	&	2446708.44	&	12.13	\\
TE3	&	2446714.37	&	11.83	\\
TE3	&	2446976.49	&	11.52	\\
TE3	&	2447848.25	&	12.44	\\
TE3	&	2448032.52	&	13.09	\\
TE3	&	2448095.47	&	12.13	\\
TE3	&	2448105.52	&	12.21	\\
TE3	&	2449580.48	&	12.44	\\
\enddata
\label{HRDELMAGS}
\end{deluxetable}

\clearpage

\begin{deluxetable}{ccc}
\tabletypesize{\scriptsize}
\tablewidth{0pc}
\tablecaption{V446 Her Magnitudes}
\tablehead{\colhead{Plate} & \colhead{JD ($\pm$ 0.4)} & \colhead{B (mag)}}
\startdata
MF10553	&	2424714.5	&		15.46	\\
MF10595	&	2424731.5	&	$>$16.38	\\
MF10617	&	2424733.5	&	$>$16.38	\\
MF10674	&	2424755.5	&		15.75	\\
MF10742	&	2424763.5	&	$>$16.46	\\
MF10743	&	2424763.5	&	$>$16.30	\\
MF10794	&	2424767.5	&	$>$16.38	\\
MF10795	&	2424767.5	&	$>$16.38	\\
MF10796	&	2424767.5	&	$>$16.38	\\
MF10797	&	2424767.5	&	$>$16.52	\\
MF10798	&	2424767.5	&	$>$16.01	\\
MF10833	&	2424772.5	&		15.87	\\
MF10853	&	2424786.5	&	$>$16.52	\\
MF10928	&	2424800.5	&		15.86	\\
MC23414	&	2425380.5	&		15.86	\\
MC24324	&	2425795.5	&		15.62	\\
MF15636	&	2426509.5	&	$>$15.38	\\
MF15852	&	2426566.5	&	$>$15.38	\\
MF23570	&	2428746.5	&		16.32	\\
MF23598	&	2428752.5	&		15.34	\\
A20084	&	2428997.5	&	$>$16.52	\\
GA310	&	2429400.5	&		16.79	\\
GA324	&	2429407.5	&		16.21	\\
GA331	&	2429424.5	&		16.46	\\
GA339	&	2429431.5	&		16.40	\\
GA408	&	2429516.3	&		15.99	\\
GA416	&	2429541.3	&		15.57	\\
GA640	&	2429839.4	&		16.60	\\
GA700	&	2429877.4	&		16.22	\\
GA845	&	2430114.5	&		16.43	\\
GA849	&	2430133.5	&		16.52	\\
GA852	&	2430145.5	&		15.67	\\
GA851	&	2430145.5	&		15.61	\\
GA869	&	2430197.5	&		16.56	\\
GA873	&	2430199.4	&		15.99	\\
GA874	&	2430199.5	&		16.53	\\
GA877	&	2430199.5	&		16.14	\\
GA876	&	2430199.5	&		16.57	\\
GA878	&	2430200.5	&		16.07	\\
GA879	&	2430200.5	&		16.02	\\
GA880	&	2430200.5	&		16.02	\\
GA881	&	2430200.5	&		16.25	\\
GA882	&	2430201.4	&		15.93	\\
GA883	&	2430201.4	&		15.66	\\
GA884	&	2430201.5	&		15.70	\\
GA885	&	2430201.5	&		15.89	\\
MF38924	&	2433422.5	&	$>$16.30	\\
MF38926	&	2433422.5	&		15.85	\\
MF38927	&	2433422.5	&		16.21	\\
B8736	&	2436786.5	&		16.25	\\
B8754	&	2436804.5	&		16.58	\\
B8755	&	2436804.5	&		16.53	\\
B8765	&	2436806.5	&		15.96	\\
B8766	&	2436806.5	&		16.03	\\
B8767	&	2436806.5	&		15.67	\\
B8768	&	2436806.5	&		15.93	\\
B8769	&	2436806.5	&		15.48	\\
B8782	&	2436808.5	&		15.69	\\
B8783	&	2436808.5	&		15.62	\\
B8787	&	2436809.5	&		16.30	\\
B8788	&	2436809.5	&		16.09	\\
B8789	&	2436809.5	&		16.69	\\
B8790	&	2436809.5	&		16.53	\\
SC5896	&	2445254.5	&		15.99	\\
MC40446	&	2446022.5	&		16.44	\\
SC5895	&	2446350.5	&		16.08	\\
SC5904	&	2446351.5	&		15.90	\\
SC5905	&	2446351.5	&		16.00	\\
SC6259	&	2446643.5	&		16.50	\\
GC8491	&	2447352.5	&		15.74	\\
GC8492	&	2447352.5	&		15.88	\\
GC8497	&	2447364.5	&		16.10	\\
GC8550	&	2447384.5	&		15.99	\\
GC8566	&	2447386.5	&		16.83	\\
SC7843	&	2448012.5	&		16.48	\\
SC7844	&	2448012.5	&		16.93	\\
SC7897	&	2448064.5	&		16.48	\\
SC7898	&	2448064.5	&		16.48	\\
SC7904	&	2448065.5	&		16.60	\\
SC7905	&	2448065.5	&		16.63	\\
SC7909	&	2448066.5	&		16.60	\\
SC7910	&	2448066.5	&		16.60	\\
SC7997	&	2448150.5	&		16.53	\\
SC7998	&	2448150.5	&		16.70	\\
SC8012	&	2448173.5	&		16.47	\\
SC8013	&	2448173.5	&		16.67	\\
SC8038	&	2448177.5	&		16.62	\\
SC8039	&	2448177.5	&		16.47	\\
SC8042	&	2448178.5	&		16.47	\\
SC8043	&	2448178.5	&		16.55	\\
SC8058	&	2448179.5	&		16.67	\\
SC8059	&	2448179.5	&		16.67	\\
SC8060	&	2448179.5	&		16.38	\\
SC8061	&	2448179.5	&		16.38	\\
SC8062	&	2448179.5	&		16.28	\\
SC8063	&	2448179.5	&		16.33	\\
SC8064	&	2448179.5	&		16.33	\\
SC8083	&	2448183.5	&		16.38	\\
SC8084	&	2448183.5	&		16.43	\\
SC8085	&	2448183.5	&		16.48	\\
SC8086	&	2448183.5	&		16.48	\\
SC8087	&	2448183.5	&		16.43	\\
SC8088	&	2448183.5	&		16.48	\\
SC8089	&	2448183.5	&		16.43	\\
SC8090	&	2448183.5	&		16.43	\\
SC8091	&	2448183.5	&		16.43	\\
SC8092	&	2448183.5	&		16.38	\\
SC8101	&	2448185.5	&		15.75	\\
SC8102	&	2448185.5	&		15.75	\\
SC8103	&	2448185.5	&		15.75	\\
SC8104	&	2448185.5	&		15.75	\\
SC8105	&	2448185.5	&		15.75	\\
SC8106	&	2448185.5	&		15.75	\\
SC8107	&	2448185.5	&		15.75	\\
SC8108	&	2448185.5	&		15.75	\\
SC8109	&	2448185.5	&		15.75	\\
SC8110	&	2448185.5	&		15.75	\\
SC8126	&	2448186.5	&		16.45	\\
SC8127	&	2448186.5	&		16.35	\\
SC8128	&	2448186.5	&		16.43	\\
SC8129	&	2448186.5	&		16.43	\\
SC8130	&	2448186.5	&		16.53	\\
SC8131	&	2448186.5	&		16.53	\\
SC8132	&	2448186.5	&		16.50	\\
SC8133	&	2448186.5	&		16.50	\\
SC8141	&	2448187.5	&		16.40	\\
SC8142	&	2448187.5	&		16.33	\\
SC8143	&	2448187.5	&		16.43	\\
SC8144	&	2448187.5	&		16.43	\\
SC8145	&	2448187.5	&		16.43	\\
SC8146	&	2448187.5	&		16.33	\\
SC8147	&	2448187.5	&		16.38	\\
SC8148	&	2448187.5	&		15.65	\\
SC8157	&	2448188.5	&		16.38	\\
SC8158	&	2448188.5	&		16.28	\\
SC8159	&	2448188.5	&		16.28	\\
SC8160	&	2448188.5	&		16.38	\\
SC8161	&	2448188.5	&		16.38	\\
\enddata
\label{V446HERMAGS}
\end{deluxetable}

\begin{deluxetable}{ccc}
\tabletypesize{\scriptsize}
\tablewidth{0pc}
\tablecaption{V533 Her Magnitudes}
\tablehead{\colhead{Plate} & \colhead{JD ($\pm$ 0.4)} & \colhead{B (mag)}}
\startdata
F654		&	2426214.5	&	14.87	\\
F657		&	2426215.5	&	14.59	\\
F650		&	2426216.5	&	14.80	\\
F662		&	2426217.5	&	15.06	\\
F669		&	2426220.5	&	14.75	\\
F674		&	2426232.5	&	14.89	\\
F678		&	2426243.5	&	15.06	\\
F756		&	2426395.5	&	14.49	\\
F759		&	2426396.5	&	14.80	\\
F767		&	2426413.5	&	14.34	\\
F778		&	2426417.5	&	14.34	\\
F782		&	2426418.5	&	14.63	\\
F786		&	2426419.5	&	14.80	\\
F790		&	2426420.5	&	14.67	\\
F792		&	2426421.5	&	15.06	\\
F795		&	2426423.5	&	14.69	\\
F800		&	2426438.5	&	14.61	\\
F802		&	2426439.5	&	14.61	\\
F805		&	2426443.5	&	14.59	\\
F812		&	2426469.5	&	14.64	\\
F813		&	2426473.5	&	15.06	\\
F814		&	2426474.5	&	14.61	\\
F816		&	2426475.5	&	14.71	\\
F819		&	2426477.5	&	14.54	\\
F822		&	2426483.5	&	14.89	\\
F834		&	2426511.5	&	15.06	\\
F836		&	2426514.5	&	14.59	\\
F841		&	2426516.5	&	14.61	\\
F850		&	2426598.5	&	14.89	\\
F859		&	2426619.5	&	14.71	\\
F948		&	2426743.5	&	15.26	\\
F959		&	2426749.5	&	14.54	\\
F969		&	2426766.5	&	14.89	\\
F977		&	2426769.5	&	14.54	\\
F981		&	2426770.5	&	14.80	\\
F1013	&	2426791.5	&	14.89	\\
F995		&	2426793.5	&	14.71	\\
F1000	&	2426826.5	&	14.59	\\
F1121	&	2427080.5	&	15.06	\\
F1116	&	2427099.5	&	14.59	\\
F1138	&	2427130.5	&	14.59	\\
F1147	&	2427132.5	&	14.54	\\
F1151	&	2427152.5	&	15.06	\\
F1154	&	2427153.5	&	14.64	\\
F1158	&	2427154.5	&	14.61	\\
F1164	&	2427157.5	&	14.69	\\
F1180	&	2427180.5	&	14.71	\\
F1181	&	2427180.5	&	14.80	\\
F1194	&	2427214.5	&	14.80	\\
F1231	&	2427301.5	&	14.61	\\
F1245	&	2427325.5	&	14.93	\\
F1285	&	2427359.5	&	14.54	\\
F1386	&	2427461.5	&	15.06	\\
F1402	&	2427481.5	&	14.71	\\
F1409	&	2427481.5	&	14.34	\\
F1416	&	2427504.5	&	14.69	\\
F1418	&	2427510.5	&	14.80	\\
F1481	&	2427512.5	&	14.71	\\
F1425	&	2427515.5	&	15.06	\\
F1432	&	2427531.5	&	14.54	\\
F1435	&	2427532.5	&	14.71	\\
F1462	&	2427567.5	&	14.80	\\
F1444	&	2427595.5	&	14.61	\\
F1482	&	2427595.5	&	14.61	\\
F1498	&	2427624.5	&	14.61	\\
F1534	&	2427667.5	&	14.80	\\
F1590	&	2427710.5	&	14.80	\\
F1615	&	2427713.5	&	14.54	\\
F1666	&	2427923.5	&	14.61	\\
F1672	&	2427925.5	&	14.80	\\
F1675	&	2427926.5	&	14.89	\\
F2827	&	2430076.5	&	14.71	\\
F2832	&	2430100.5	&	14.71	\\
F2834	&	2430101.5	&	14.69	\\
F2838	&	2430103.5	&	14.80	\\
F2839	&	2430113.5	&	14.67	\\
F2841	&	2430130.5	&	14.54	\\
F2843	&	2430141.5	&	14.61	\\
F2867	&	2430208.5	&	14.61	\\
F2875	&	2430233.5	&	14.61	\\
F2881	&	2430253.5	&	14.67	\\
F2888	&	2430253.5	&	14.64	\\
F2920	&	2430312.5	&	14.59	\\
F2924	&	2430318.5	&	14.54	\\
F6207	&	2430354.42	&	14.44	\\
F6212	&	2430384.46	&	14.31	\\
F6227	&	2430447.43	&	14.09	\\
F3012	&	2430456.5	&	15.06	\\
F3014	&	2430460.5	&	14.74	\\
F3032	&	2430463.5	&	14.54	\\
F3028	&	2430485.5	&	15.06	\\
F3036	&	2430516.5	&	14.80	\\
F3041	&	2430530.5	&	14.80	\\
F3050	&	2430587.5	&	14.64	\\
F3061	&	2430589.5	&	14.54	\\
F3066	&	2430604.5	&	14.80	\\
F3073	&	2430612.5	&	14.59	\\
F3085	&	2430632.5	&	15.06	\\
F3095	&	2430644.5	&	14.76	\\
F3143	&	2430763.5	&	14.74	\\
F3154	&	2430783.5	&	14.72	\\
F3158	&	2430791.5	&	14.84	\\
F3173	&	2430791.5	&	14.86	\\
F3179	&	2430793.5	&	14.82	\\
F3185	&	2430812.5	&	15.01	\\
F3202	&	2430847.5	&	14.76	\\
F3209	&	2430875.5	&	14.73	\\
F3218	&	2430931.5	&	14.77	\\
F3222	&	2430935.5	&	14.58	\\
F3226	&	2430937.5	&	14.78	\\
F3236	&	2430959.5	&	14.62	\\
F3239	&	2430971.5	&	14.78	\\
F3252	&	2430997.5	&	14.77	\\
F3260	&	2431000.5	&	14.76	\\
F3322	&	2431233.5	&	14.61	\\
F3340	&	2431261.5	&	14.71	\\
F3345	&	2431289.5	&	14.61	\\
F3343	&	2431296.5	&	14.89	\\
F3350	&	2431312.5	&	14.64	\\
F3354	&	2431316.5	&	14.79	\\
F3359	&	2431321.5	&	14.80	\\
F3367	&	2431325.5	&	14.71	\\
F3371	&	2431327.5	&	14.54	\\
F3375	&	2431342.5	&	14.96	\\
F3384	&	2431347.5	&	15.06	\\
F3391	&	2431373.5	&	14.57	\\
F3427	&	2431529.5	&	15.06	\\
F3432	&	2431586.5	&	15.15	\\
F3434	&	2431587.5	&	14.84	\\
F3437	&	2431608.5	&	15.10	\\
F3439	&	2431612.5	&	14.61	\\
F3443	&	2431645.5	&	14.69	\\
F3446	&	2431650.5	&	14.69	\\
F3449	&	2431654.5	&	14.84	\\
F3455	&	2431700.5	&	14.86	\\
F3457	&	2431701.5	&	14.71	\\
F3464	&	2431704.5	&	14.67	\\
F3468	&	2431706.5	&	14.80	\\
F3518	&	2431906.5	&	14.54	\\
F3521	&	2431910.5	&	14.71	\\
F3524	&	2431911.5	&	15.06	\\
F3527	&	2431912.5	&	15.16	\\
F3531	&	2431916.5	&	14.60	\\
F3536	&	2431933.5	&	14.67	\\
F3545	&	2431939.5	&	14.61	\\
F3547	&	2431943.5	&	14.60	\\
F3553	&	2432005.5	&	14.92	\\
F3555	&	2432026.5	&	14.54	\\
F3557	&	2432036.5	&	14.64	\\
F3559	&	2432063.5	&	14.54	\\
F3565	&	2432086.5	&	14.64	\\
F3569	&	2432089.5	&	14.80	\\
F3583	&	2432109.5	&	14.64	\\
F3586	&	2432112.5	&	14.61	\\
F3593	&	2432118.5	&	14.80	\\
F3668	&	2432801.5	&	14.80	\\
F3670	&	2432802.5	&	14.89	\\
F3678	&	2432822.5	&	14.75	\\
F3741	&	2433030.5	&	14.61	\\
F3748	&	2433097.5	&	14.80	\\
F3756	&	2433115.5	&	14.64	\\
F3763	&	2433150.5	&	14.93	\\
F3779	&	2433181.5	&	14.74	\\
F3835	&	2433366.5	&	14.80	\\
F3840	&	2433378.5	&	14.74	\\
F3851	&	2433387.5	&	15.06	\\
F3854	&	2433439.5	&	14.80	\\
F3877	&	2433505.5	&	14.74	\\
F4361	&	2435167.5	&	14.52	\\
F4364	&	2435184.5	&	14.69	\\
F4366	&	2435186.5	&	14.34	\\
F4367	&	2435190.5	&	14.54	\\
F4392	&	2435191.5	&	14.59	\\
F4369	&	2435195.5	&	14.61	\\
F4371	&	2435215.5	&	14.74	\\
F4372	&	2435216.5	&	14.85	\\
F4347	&	2435217.5	&	14.61	\\
F4346	&	2435218.5	&	14.64	\\
F4348	&	2435219.5	&	14.54	\\
F4380	&	2435221.5	&	14.81	\\
F4381	&	2435222.5	&	14.59	\\
F4386	&	2435238.5	&	14.71	\\
F4385	&	2435239.5	&	15.06	\\
F4387	&	2435243.5	&	14.69	\\
F4488	&	2435525.5	&	14.69	\\
F4492	&	2435540.5	&	14.87	\\
F4498	&	2435547.5	&	14.54	\\
F4501	&	2435548.5	&	14.93	\\
F4502	&	2435549.5	&	14.81	\\
F4503	&	2435549.5	&	14.80	\\
F4521	&	2435600.5	&	15.06	\\
F4523	&	2435601.5	&	14.49	\\
F4528	&	2435626.5	&	14.80	\\
F4532	&	2435660.5	&	14.89	\\
F4539	&	2435690.5	&	14.54	\\
F4542	&	2435694.5	&	14.89	\\
F4618	&	2436287.5	&	14.74	\\
F4726	&	2436457.5	&	14.61	\\
F4782	&	2436602.5	&	14.71	\\
F4787	&	2436603.5	&	14.61	\\
F4829	&	2436613.5	&	14.71	\\
F4866	&	2436662.5	&	14.61	\\
F4870	&	2436671.5	&	14.89	\\
F4879	&	2436688.5	&	14.54	\\
F4882	&	2436696.5	&	14.71	\\
F4889	&	2436758.5	&	14.89	\\
F4952	&	2436843.5	&	14.64	\\
F4972	&	2436844.5	&	14.61	\\
F4997	&	2436971.5	&	14.71	\\
F5003	&	2437015.5	&	14.84	\\
F5007	&	2437043.5	&	14.64	\\
F5005	&	2437050.5	&	14.50	\\
F5012	&	2437071.5	&	14.63	\\
F5014	&	2437073.5	&	14.54	\\
F5015	&	2437075.5	&	15.06	\\
F5016	&	2437077.5	&	14.64	\\
F5009	&	2437079.5	&	14.52	\\
F5020	&	2437079.5	&	14.46	\\
F5021	&	2437080.5	&	14.80	\\
F5024	&	2437081.5	&	14.58	\\
F5025	&	2437085.5	&	14.54	\\
F5028	&	2437102.5	&	14.76	\\
F5031	&	2437106.5	&	14.61	\\
F5033	&	2437108.5	&	14.71	\\
F5034	&	2437109.5	&	14.61	\\
F5063	&	2437193.5	&	14.69	\\
F5140	&	2437559.5	&	14.69	\\
F5149	&	2437577.33	&	14.70	\\
TE3 2544	&	2437586.32	&	14.69	\\
TE3 2575	&	2437603.32	&	14.36	\\
TE3 2589	&	2437615.27	&	14.54	\\
TE3 2610	&	2437642.24	&	14.52	\\
F5203	&	2437705.67	&	14.30	\\
TE3 2759	&	2437731.65	&	13.58	\\
F5213	&	2437736.5	&	14.47	\\
F5219	&	2437761.61	&	11.27	\\
F5222	&	2437766.58	&	14.28	\\
TE3 2802	&	2437766.61	&	14.27	\\
F5223	&	2437779.43	&	14.35	\\
F5227	&	2437786.5	&	14.46	\\
F5225	&	2437789.45	&	14.12	\\
TE3 2853	&	2437820.5	&	13.50	\\
F5237	&	2437821.5	&	14.31	\\
F5238	&	2437823.5	&	14.60	\\
TE3 2874	&	2437824.45	&	14.27	\\
F5242	&	2437828.5	&	14.31	\\
F5244	&	2437842.49	&	14.31	\\
F5247	&	2437869.45	&	13.73	\\
TE3 2927	&	2437871.47	&	13.72	\\
TE3 2895	&	2437871.5	&	13.42	\\
TE3 2948	&	2437877.49	&	14.33	\\
F5255	&	2437877.5	&	14.30	\\
TE3 2976	&	2437885.45	&	13.50	\\
F5267	&	2437903.50	&	14.29	\\
TE3 3026	&	2437904.39	&	13.53	\\
TE3 3041	&	2437906.40	&	13.71	\\
TE3 3051	&	2437907.38	&	13.50	\\
F5272	&	2437910.50	&	14.31	\\
TE3 3085	&	2437911.37	&	14.27	\\
F5280	&	2437932.50	&	13.42	\\
TE3 3104	&	2437933.34	&	13.42	\\
F5283	&	2437934.34	&	13.54	\\
TE4 3148	&	2437938.38	&	14.03	\\
F5298	&	2437938.5	&	13.50	\\
TE4 3160	&	2437940.29	&	13.42	\\
TE4 3174	&	2437942.36	&	13.42	\\
TE4 3219	&	2437955.29	&	13.42	\\
TE4 3220	&	2437956.27	&	14.03	\\
TE4 3230	&	2437959.32	&	14.03	\\
TE4 3271	&	2437972.28	&	13.42	\\
TE6 2428	&	2437993.34	&	13.72	\\
TE4 3410	&	2438085.65	&	5.19	\\
TE4 3432	&	2438089.64	&	5.19	\\
TE7 672	&	2438106.53	&	5.96	\\
F5391	&	2438111.51	&	5.20	\\
F5728	&	2438849.60	&	13.22	\\
F5433	&	2438852.54	&	13.09	\\
F5856	&	2439242.51	&	13.42	\\
F5863	&	2439259.5	&	13.46	\\
F5965	&	2439536.67	&	13.53	\\
F5983	&	2439596.46	&	13.58	\\
F5999	&	2439618.45	&	14.30	\\
F6001	&	2439619.41	&	14.33	\\
F6004	&	2439619.5	&	14.31	\\
F6008	&	2439637.42	&	14.33	\\
F6015	&	2439673.46	&	14.27	\\
F6074	&	2439887.66	&	14.52	\\
F6077	&	2439894.65	&	14.31	\\
F6111	&	2439967.48	&	14.36	\\
F6122	&	2439994.42	&	14.33	\\
F6203	&	2440321.44	&	14.36	\\
F6205	&	2440326.44	&	14.33	\\
TE4	&	2443747.5	&	14.69	\\
TE4 9497	&	2445053.57	&	13.50	\\
TE4 9575	&	2445203.38	&	14.27	\\
TE4 9736	&	2445486.46	&	14.36	\\
TE4 9963	&	2445871.45	&	14.36	\\
TE4	&	2445913.42	&	14.69	\\
TE4	&	2446173.52	&	14.31	\\
TE4	&	2446200.49	&	14.61	\\
TE4	&	2446264.43	&	14.61	\\
TE4	&	2446292.5	&	14.69	\\
TE4	&	2446613.42	&	14.27	\\
TE4	&	2446650.39	&	14.69	\\
TE4	&	2446704.29	&	14.69	\\
TE4	&	2446708.28	&	14.69	\\
TE4	&	2446709.27	&	14.69	\\
TE4 10755	&	2446975.45	&	13.58	\\
TE4	&	2447331.47	&	14.69	\\
\enddata
\label{V533HERMAGS}
\end{deluxetable}

\begin{deluxetable}{ccc}
\tabletypesize{\scriptsize}
\tablewidth{0pc}
\tablecaption{CP Lac Magnitudes}
\tablehead{\colhead{Plate} & \colhead{JD ($\pm$ 0.4)} & \colhead{B (mag)}}
\startdata
I23963	&	2414515.8	&	$>$14.90	\\
I21047	&	2414960.8	&	$>$13.73	\\
MC18885	&	2423263.5	&		15.99	\\
MC19960	&	2423647.7	&		15.97	\\
MC21116	&	2424104.5	&		16.02	\\
MC21530	&	2424351.8	&		15.89	\\
MC21728	&	2424497.5	&		16.25	\\
MC22087	&	2424772.6	&		15.43	\\
MC22178	&	2424816.7	&		16.06	\\
MC22620	&	2425096.7	&		16.26	\\
MC22841	&	2425181.5	&		16.11	\\
MC22918	&	2425201.5	&	$>$15.56	\\
MC24382	&	2425283.8	&		15.92	\\
MA2155	&	2425436.5	&	$>$14.90	\\
MC23527	&	2425452.8	&		15.93	\\
MA2204	&	2425472.5	&	$>$15.22	\\
MC23587	&	2425474.7	&		16.16	\\
MA2227	&	2425479.5	&	$>$14.43	\\
MA2245	&	2425494.5	&	$>$14.43	\\
MC23642	&	2425511.6	&		15.95	\\
MC23786	&	2425557.5	&		15.85	\\
MC23893	&	2425596.5	&		16.02	\\
I47236	&	2425808.8	&		15.95	\\
I47273	&	2425828.8	&		15.78	\\
MC24536	&	2425893.6	&		15.43	\\
MC24585	&	2425913.6	&		15.87	\\
I48519	&	2426187.8	&	$>$15.22	\\
I48549	&	2426190.8	&	$>$15.43	\\
MC25225	&	2426301.5	&		15.44	\\
MA2597	&	2426585.8	&	$>$13.73	\\
I51187	&	2426929.8	&		15.76	\\
I51262	&	2426944.7	&	$>$13.73	\\
I51304	&	2426949.7	&		16.05	\\
I51679	&	2427040.5	&	$>$15.43	\\
I52381	&	2427354.6	&		15.82	\\
I52394	&	2427360.5	&		15.39	\\
I51540	&	2727014.5	&		15.43	\\
\enddata
\label{CPLACMAGS}
\end{deluxetable}

\clearpage

\begin{deluxetable}{ccc}
\tabletypesize{\scriptsize}
\tablewidth{0pc}
\tablecaption{GK Per Magnitudes}
\tablehead{\colhead{Plate} & \colhead{JD ($\pm$ 0.4)} & \colhead{B (mag)}}
\startdata
I101		&	2411316.5	&	$>$	13.13	\\
I2071	&	2411666.5	&	$>$	11.24	\\
I2401	&	2411703.5	&	$>$	9.39		\\
I4928	&	2412067.5	&	$>$	14.37	\\
I5031	&	2412442.5	&	$>$	12.91	\\
I7681	&	2412444.5	&	$>$	13.13	\\
I7958	&	2412488.5	&	$>$	14.47	\\
I10251	&	2412820.5	&	$>$	13.13	\\
I11614	&	2413112.5	&		13.52	\\
I16163	&	2413841.5	&		13.81	\\
I17098	&	2413925.5	&		14.26	\\
I16965	&	2413925.5	&	$>$	14.37	\\
I19305	&	2414214.5	&		14.12	\\
I21649	&	2414595.5	&		14.44	\\
I22241	&	2414680.5	&	$>$	13.13	\\
I24548	&	2415055.5	&		14.10	\\
I24699	&	2415085.5	&		13.39	\\
I26053	&	2415335.5	&		13.84	\\
I26584	&	2415422.5	&		13.33	\\
AC1252	&	2415433.5	&	$>$	13.2		\\
AC1258	&	2415434.6	&	$>$	12.9		\\
AC1260	&	2415435.5	&	$>$	13.0		\\
\enddata
\label{GKPERMAGS}
\end{deluxetable}

\begin{deluxetable}{ccc}
\tabletypesize{\scriptsize}
\tablewidth{0pc}
\tablecaption{LV Vul Magnitudes}
\tablehead{\colhead{Plate} & \colhead{JD ($\pm$ 0.4)} & \colhead{B (mag)}}
\startdata
AA1990	&	2413784.5	&	$>$	15.84	\\
AA3886	&	2414884.5	&	$>$	15.84	\\
AA7866	&	2417409.5	&	$>$	15.84	\\
MC586	&	2418937.5	&		16.52	\\
MC17801	&	2422925.6	&	$>$	16.52	\\
MC17821	&	2422927.6	&		16.60	\\
MC19918	&	2423621.7	&		16.52	\\
MF9552	&	2424352.5	&		16.52	\\
MF9665	&	2424373.5	&	$>$	15.84	\\
MF9749	&	2424401.5	&	$>$	15.84	\\
MF9761	&	2424402.5	&		16.75	\\
AA13521	&	2424403.5	&		16.44	\\
MF9781	&	2424404.5	&		16.52	\\
MF9786	&	2424405.5	&		15.96	\\
MC22016	&	2424713.8	&		16.47	\\
MF10860	&	2424787.5	&		16.75	\\
MC22850	&	2425183.0	&		16.52	\\
MC23687	&	2425524.5	&	$>$	15.84	\\
A1152	&	2425821.5	&	$>$	16.44	\\
MA2422	&	2425831.5	&		16.44	\\
MA2421	&	2425831.5	&	$>$	15.84	\\
A1143	&	2425837.5	&	$>$	16.22	\\
MF13596	&	2425889.5	&		15.84	\\
MF11990	&	2425889.5	&	$>$	16.68	\\
MC24987	&	2426151.8	&	$>$	15.84	\\
A2542	&	2428094.3	&		16.42	\\
A2550	&	2428108.3	&		16.58	\\
A2627	&	2428249.6	&	$>$	16.22	\\
F2687	&	2428727.5	&	$>$	16.22	\\
A2512	&	2428904.5	&		16.22	\\
F2297	&	2429022.5	&		16.22	\\
AA20185	&	2429044.5	&		16.27	\\
GA61	&	2429050.5	&	$>$	16.22	\\
GA63	&	2429053.5	&	$>$	16.22	\\
GA68	&	2429075.5	&	$>$	15.84	\\
GA74	&	2429087.5	&		16.22	\\
GA75	&	2429099.5	&	$>$	15.84	\\
GA83	&	2429105.6	&	$>$	16.22	\\
GA98	&	2429111.4	&	$>$	16.22	\\
GA113	&	2429130.4	&		16.52	\\
AA20416	&	2429134.3	&		16.75	\\
GA120	&	2429140.4	&		16.65	\\
F2362	&	2429157.5	&		15.94	\\
GA127	&	2429166.4	&		16.22	\\
F2386	&	2429166.5	&		16.52	\\
GA159	&	2429168.5	&	$>$	16.75	\\
GA180	&	2429193.3	&		16.27	\\
GA208	&	2429230.3	&		16.42	\\
GA226	&	2429250.2	&		16.02	\\
GA381	&	2429494.5	&	$>$	16.52	\\
F2704	&	2429783.5	&	$>$	15.39	\\
F2709	&	2429788.5	&		15.98	\\
F2718a	&	2429817.5	&		15.11	\\
F2718b	&	2429839.5	&	$>$	16.22	\\
F2719a	&	2429842.5	&		15.63	\\
F2734	&	2429876.5	&	$>$	16.22	\\
F2743	&	2429879.5	&	$>$	15.39	\\
F2754	&	2429906.5	&	$>$	14.95	\\
F2840	&	2430113.5	&	$>$	16.22	\\
F2850	&	2430163.5	&	$>$	15.84	\\
F2853	&	2430168.5	&	$>$	14.95	\\
F2860	&	2430198.5	&	$>$	16.44	\\
F2869	&	2430224.5	&	$>$	14.95	\\
F2880	&	2430254.5	&	$>$	14.95	\\
F2896	&	2430261.5	&	$>$	16.22	\\
F3029	&	2430425.5	&		16.22	\\
F3032	&	2430499.5	&		16.42	\\
MC32351	&	2430560.7	&		16.52	\\
F3335	&	2431238.5	&		15.94	\\
F3337	&	2431253.5	&		16.42	\\
F3346	&	2431292.5	&		15.84	\\
MC34169	&	2431704.6	&		16.60	\\
F3361	&	2432793.5	&		16.34	\\
F3860	&	2433451.5	&		15.84	\\
A4677	&	2433481.4	&	$>$	15.39	\\
F4871	&	2436672.5	&	$>$	16.22	\\
F4892	&	2436725.5	&		16.44	\\
F4897	&	2436751.5	&		16.03	\\
F4980	&	2436896.5	&	$>$	15.39	\\
F5039	&	2437140.5	&		15.84	\\
F5047	&	2437170.5	&		16.32	\\
F5050	&	2437172.5	&		16.44	\\
F5053	&	2437173.5	&		15.84	\\
A4803	&	2437577.5	&		16.42	\\
F5152	&	2437578.3	&		16.37	\\
F5168	&	2437587.3	&		16.44	\\
A6912	&	2437820.5	&		15.69	\\
F5241	&	2437838.5	&	$>$	16.22	\\
F5259	&	2437885.5	&		16.44	\\
F5269	&	2437907.4	&		16.33	\\
F5295	&	2437935.4	&	$>$	16.33	\\
A7005	&	2437962.3	&		15.84	\\
GC884	&	2438322.4	&		15.76	\\
A7366	&	2438621.4	&	$>$	16.22	\\
F5658	&	2438650.4	&	$>$	16.22	\\
GC1261	&	2438653.4	&		16.05	\\
GC1265	&	2438667.4	&		15.78	\\
A7379	&	2438671.3	&	$>$	16.22	\\
F5675	&	2438672.3	&	$>$	16.22	\\
A7464	&	2438974.4	&	$>$	16.22	\\
F5759	&	2438976.5	&		16.43	\\
A7468	&	2438992.4	&		16.22	\\
A7506	&	2439055.3	&		16.42	\\
F5867	&	2439261.5	&	$>$	15.84	\\
A7588	&	2439299.5	&		16.44	\\
F5879	&	2439317.4	&	$>$	15.39	\\
A7590	&	2439347.4	&		16.29	\\
F5882	&	2439348.5	&		16.44	\\
F5882	&	2439349.4	&	$>$	16.22	\\
F5893	&	2439376.4	&	$>$	16.22	\\
F5894	&	2439378.3	&	$>$	16.22	\\
F5907	&	2439388.3	&	$>$	16.22	\\
F5920	&	2439407.3	&		16.44	\\
A7716	&	2439619.5	&		16.22	\\
F6012	&	2439670.4	&	$>$	16.22	\\
F6020	&	2439684.5	&	$>$	15.39	\\
A7735	&	2439685.5	&		16.22	\\
F6030	&	2439712.4	&	$>$	15.39	\\
A7746	&	2439731.4	&		15.90	\\
F6035	&	2439739.4	&		16.34	\\
F6044	&	2439765.3	&		16.44	\\
GC2933	&	2442277.4	&		16.11	\\
GC2934	&	2442283.4	&		16.42	\\
GC2935	&	2442283.4	&		16.23	\\
GC3111	&	2442717.3	&		16.33	\\
GC3258	&	2443016.4	&		16.12	\\
GC3359	&	2443477.5	&		16.22	\\
GC3714	&	2444132.4	&		16.26	\\
GC3713	&	2444132.4	&		16.03	\\
GC4754	&	2445231.4	&		16.42	\\
GC4755	&	2445231.4	&		15.80	\\
GC4987	&	2445527.5	&		16.14	\\
GC5003	&	2445546.4	&		15.84	\\
GC5004	&	2445546.4	&		15.96	\\
GC5027	&	2445562.5	&		16.03	\\
GC5028	&	2445562.5	&		15.84	\\
GC5085	&	2445612.3	&		16.42	\\
GC5086	&	2445612.4	&		15.39	\\
A1115	&	2445806.5	&	$>$	16.44	\\
GC5612	&	2445913.4	&		16.14	\\
GC5617	&	2445913.5	&		16.22	\\
GC5627	&	2445916.4	&		16.12	\\
GC5641	&	2445919.4	&		16.44	\\
GC5641b	&	2445919.5	&		16.03	\\
GC5644	&	2445930.4	&		16.44	\\
GC5645	&	2445930.4	&		15.94	\\
GC5647	&	2445933.3	&		15.94	\\
GC5648	&	2445933.4	&		16.22	\\
GC5655	&	2445935.4	&		16.22	\\
GC5656	&	2445935.4	&		15.84	\\
GC5658	&	2445936.4	&		16.33	\\
GC5659	&	2445936.4	&		15.96	\\
GC5660	&	2445936.4	&		16.29	\\
GC5672	&	2445940.4	&		16.20	\\
GC5673	&	2445940.5	&		16.64	\\
GC5674	&	2445940.5	&		16.22	\\
GC5679	&	2445942.4	&		16.22	\\
GC5689	&	2445944.4	&		15.61	\\
GC5690	&	2445944.5	&		16.03	\\
GC5692	&	2445945.4	&		16.03	\\
GC5693	&	2445949.5	&		16.04	\\
GC5695	&	2445957.3	&		16.42	\\
GC5708	&	2445973.4	&		15.75	\\
GC5719	&	2446000.3	&		16.22	\\
GC5720	&	2446000.4	&		16.14	\\
GC5732	&	2446001.3	&		15.76	\\
GC5731	&	2446001.4	&		16.03	\\
GC5742	&	2446003.3	&		16.03	\\
GC5766	&	2446004.3	&		16.03	\\
GC5767	&	2446004.3	&		15.94	\\
GC5824	&	2446017.2	&		16.22	\\
GC5827	&	2446018.2	&		16.42	\\
GC5833	&	2446018.3	&		15.84	\\
GC5834	&	2446019.2	&		15.84	\\
GC5840	&	2446019.3	&	$>$	15.84	\\
GC5863	&	2446036.2	&	$>$	16.22	\\
GC5897	&	2446059.2	&		16.22	\\
GC5985	&	2446093.7	&	$>$	15.39	\\
GC6069	&	2446113.7	&		16.08	\\
GC6105	&	2446116.7	&		16.03	\\
GC6106	&	2446116.7	&		16.03	\\
GC6130	&	2446121.6	&	$>$	15.39	\\
GC6222	&	2446200.5	&		16.03	\\
GC6255	&	2446235.5	&		16.03	\\
GC6266	&	2446260.5	&		16.03	\\
GC6280	&	2446264.4	&		15.84	\\
GC6302	&	2446270.4	&		16.22	\\
GC6325	&	2446271.5	&		16.42	\\
GC6352	&	2446287.5	&	$>$	16.22	\\
GC6373	&	2446290.4	&		15.94	\\
GC6390	&	2446291.4	&		16.12	\\
GC6382	&	2446291.4	&	$>$	16.22	\\
GC6405	&	2446293.5	&		16.22	\\
GC6415	&	2446296.4	&		15.84	\\
GC6454	&	2446299.4	&		16.03	\\
GC6464	&	2446301.4	&		16.03	\\
GC6483	&	2446327.4	&		15.96	\\
GC6489	&	2446358.3	&		16.09	\\
GC10184	&	2448450.5	&		16.27	\\
GC10137	&	2448481.4	&		16.09	\\
GC10256	&	2448559.3	&		16.12	\\
GC10839	&	2449125.5	&		16.24	\\
GC10883	&	2449213.5	&	$>$	16.22	\\
GC10896	&	2449215.5	&		16.22	\\
A7606	&	2449379.4	&		16.22	\\
A7614	&	2449386.3	&		16.29	\\
\enddata
\label{LVVULMAGS}
\end{deluxetable} 

\clearpage

\begin{deluxetable}{lrcrc}
\tabletypesize{\scriptsize}
\tablecolumns{3}
\tablewidth{0pc}
\tablecaption{DSS Magnitudes}
\tablehead{\colhead{Nova} & \colhead{$m_{pre}$} & \colhead{source\tablenotemark{a}} & \colhead{$m_{post}$} & \colhead{source\tablenotemark{a}}}
\startdata
LS And	&	20.52 	&	1	&	20.39	&	2\\
OS And 	&	18.44	&	1	&	17.52	&	2\\
V1229 Aql 	&	18.10	&	1	&	18.21	&	2\\
V705 Cas		&	16.64	&	2	&	16.9\tablenotemark{b}	& 3\\
V723 Cas 	&	18.76	&	2	&	15.75\tablenotemark{c} & 4 \\
IV Cep 	&	16.31	&	1	&	16.49	&	2 \\
V1330 Cyg 	&	18.74	&	1	&	17.57	&	2 \\
V1500 Cyg	&	21.5	&	5	&	18.79\tablenotemark{d}	& 	6\\
V1668 Cyg 	&	20.81	&	1	&	20.61	&	2 \\
V1974 Cyg	&	$>$21	&	2	&	16.88\tablenotemark{e}	&	3 \\
V827 Her 	&	18.16	&	1	&	17.82	&	2 \\
V838 Her	&	19.27	&	2	&	19.1\tablenotemark{f}	& 7\\
V400 Per 	&	19.56	&	1	&	19.55	&	2 \\
HZ Pup 	&	16.84	&	1	&	17.00	&	2 \\
V4633 Sgr &	$>$21	&	1	&	18\tablenotemark{g}		&	8	\\
V992 Sco 	&	18.23	&	2	&	18.38\tablenotemark{h} &  9 \\
FH Ser 	&	16.65	&	1	&	16.48	&	2 \\
RW UMi	&	$>$21	&	1	&	18.33	&	2\\
NQ Vul 	&	17.70	&	1	&	17.32	&	2 \\
PW Vul 	&	17.19	&	1	&	16.84	&	2 \\
QU Vul 	&	19.58	&	1	&	18.4\tablenotemark{i} &  10\\
\enddata
\tablenotetext{a}{1. USNO-B1.0 B1mag.  2.  USNO-B1.0 B2mag.  3. AAVSO. 4.  Goranskij et al. (2007).  5. Wade (1987). 6. Semeniuk et al (1995). 7. Szkody \& Ingram (1994). 8. Lipkin $\&$ Leibowitz (2008). 9.  Woudt \& Warner (2003).  10.  Shafter et al. (1995)} 
\tablenotetext{b}{USNO-B1.0 gives both the B2mag and the R2mag for V705 Cas. This gives us a B-R color of 0.72. We use this to estimate the B-V color to be $B-V\approx 0.4$mag.  This color correction is then applied to the AAVSO V-band magnitudes to get the tabulated B-band $m_{post}$ value.}
\tablenotetext{c}{B magnitude obtained by-eye from lightcurves in Goranskij et al. (2007). V723 Cas appears to have leveled off.} 
\tablenotetext{d}{Semeniuk et al. (1995) give the post eruption magnitude of V1500 Cyg to be V=18.0 mag. Applying a color term of $B-V=0.79$ (Szkody 1994), we take B=18.79.}
\tablenotetext{e}{The AAVSO database shows V1974 Cyg in quiescence in 2007. The yearly average of measurements taken in this year is V=16.63 mag. Shugarov et al. (2002) give a B-V curve, with $B-V\approx0.25$ once this color has leveled off. We therefore adopt B=16.88 mag.}
\tablenotetext{f}{Szkody \& Ingram (1994) give the combined magnitude of V838 Her and its companion to be V=18.3. In addition, they give $E(B-V)=0.6$ mag. With a typical $B-V=0.2$ mag for quiescent novae with low extinction (Szkody 1994), we expect V838 Her to have $B-V\approx 0.8$ mag. We take the average B-band magnitude to be 19.1 mag.}
\tablenotetext{g}{ A light curve was presented by Lipkin $\&$ Leibowitz (2008) that indicated that V4633 Sgr may still be on decline, however, our observations at McDonald Observatory on 14 July 2009 show that this decline has stopped.}
\tablenotetext{h}{Our recent measures of V992 Sco with the SMARTS 1.3-m telescope on Cerro Tololo have average B=18.77 and B-V=0.93. Warner and Woudt (2003) gives an average V=17.1 on seven nights, and hence B=18.0. These measures average to B=18.38.}
\tablenotetext{i}{Warner (1995) gives $A_V=1.0$ mag so we would expect $E(B-V)=0.3$ mag or so.  With a typical $B-V=0.2$ mag for quiescent novae with low extinction (Szkody 1994), we expect QU Vul to have $B-V\approx 0.5$ mag.  Shafter et al. (1995) measured the V-band magnitude on four nights with an average $V=17.9$ mag outside of eclipse.  So we take the average B-band magnitude in post-eruption quiescence to be 18.4 mag.}
\label{DSSSUM}
\end{deluxetable}

\clearpage

\begin{deluxetable}{cccr}
\tabletypesize{\scriptsize}
\tablewidth{0pc}
\tablecaption{Nova Behavior as Tied to the Eruption}
\tablehead{\colhead{Eruption Event} & \colhead{Pre-Eruption Anticipation?} & \colhead{Variability Change?} & \colhead{$\Delta m$}}
\startdata
LS And	&	\ldots	&	\ldots	&	+0.13	\\
OS And	&	\ldots	&	\ldots	&	+0.92	\\
CI Aql -- 1917	&	\ldots	&	No	&	+0.10	\\
CI Aql -- 1941	&	\ldots	&	No	&	-0.11	\\
CI Aql -- 2000	&	\ldots	&	No	&	+0.09	\\
V368 Aql	&	No	&	No	&	-0.37	\\
V1229 Aql	&	\ldots	&	\ldots	&	-0.11	\\
QZ Aur	&	No	&	No	&	+0.03\\
V705 Cas	&	\ldots	&	\ldots	&	-0.26	\\
V723 Cas	&	\ldots	&	\ldots	&	+3.01	\\
IV Cep	&	\ldots	&	\ldots	&	-0.18	\\
T CrB -- 1946	&	Yes -- Dip	&	No	&	-1.09\\
V1330 Cyg	&	\ldots	&	\ldots	&	+1.17	\\
V1500 Cyg	&	Yes -- Rise	&	\ldots	& +2.71	\\
V1668 Cyg	&	\ldots	&	\ldots	&	+0.20	\\
V1974 Cyg	&	\ldots	&	\ldots	&	$>$ 4.12	\\
HR Del	&	No	&	No	& +0.23 \\
DQ Her	&	No	&	No	&	-0.33	\\
V446 Her	&	No	&	No	&	-0.24	\\
V533 Her	&	Yes -- Rise	&	No	&	+0.47	\\
V827 Her	&	\ldots	&	\ldots	&	+0.34	\\
V838 Her	&	\ldots	&	\ldots	&	+0.17	\\
CP Lac	&	No	&	No	&	+0.37	\\
BT Mon	&	No	&	No	&	-0.09	\\
RS Oph -- 1933	&	No	&	No	&	-0.17	\\
RS Oph -- 1945	&	No	&	No	&	0.00\\
RS Oph -- 1958	&	No	&	No	&	+0.28\\
RS Oph -- 1967	&	No	&	No	&	-0.39\\
RS Oph -- 1985	&	No	&	No	&	+0.05\\
RS Oph -- 2006	&	No	&	No	&	+0.12 \\
GK Per	&	No	&	No	&	-0.07	\\
V400 Per	&	\ldots	&	\ldots	&	+0.01	\\
HZ Pup	&	\ldots	&	\ldots	&	-0.16	\\
T Pyx -- 1890	&	\ldots	&	\ldots	& -0.58	\\
T Pyx -- 1902	&	No	&	No	& -0.36	\\
T Pyx -- 1920	&	\ldots	&	No	& -0.14	\\
T Pyx -- 1944	&	No	&	No	& +0.16\\
T Pyx -- 1967	&	\ldots	&	No	& -0.84\\
V4633 Sgr	&	\ldots	&	\ldots	&	$>$ 3.0 \\
V3890 Sgr -- 1962	&	No	&	No	&	0.12	\\
U Sco -- 1979	&	No	&	\ldots	&	+0.14 \\
U Sco -- 1987	&	\ldots	&	No	&	-0.25 \\
U Sco -- 1999	&	\ldots	&	No	&	+0.07 \\
V992 Sco	&	\ldots	&	\ldots	&	-0.15\\
FH Ser	&	\ldots	&	\ldots	&	+0.17	\\
RW UMi	&	\ldots	&	\ldots	&	$>$ 2.67\\
LV Vul	&	No	&	No	&	+0.14	\\
NQ Vul	&	\ldots	&	\ldots	&	+0.38	\\
PW Vul	&	\ldots	&	\ldots	&	+0.35	\\
QU Vul	&	\ldots	&	\ldots	&	+1.18	\\
\enddata
\label{ROBSUM}
\end{deluxetable}

\clearpage

\clearpage

\clearpage

\clearpage

\clearpage

\clearpage

\clearpage

\clearpage

\clearpage

\clearpage

\clearpage

\begin{figure}[h!]
  \begin{center}
\includegraphics[scale=0.5]{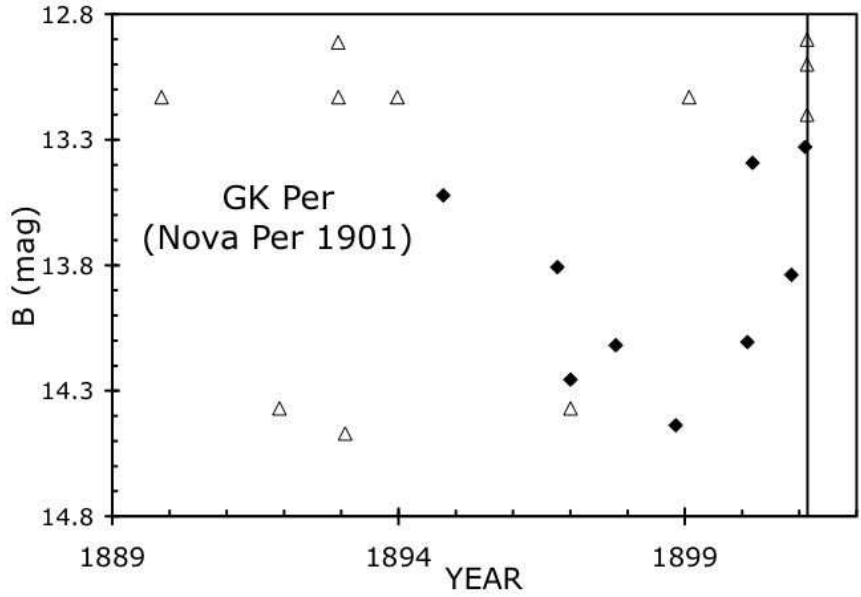}
  \end{center}
  \caption{GK Per light curve. The diamonds mark B magnitude measurements made of the nova, the triangles denote limiting magnitudes, and the eruption date is indicated by the vertical line.  We see that GK Per varies up-and-down between 13.4 mag and fainter than 14.5 mag, with this variation being typical of GK Per post-eruption and CVs in general at any time.}
  \label{GKPERLC}
\end{figure}

\clearpage

\begin{figure}[h!]
  \begin{center}
 \includegraphics[scale=0.5]{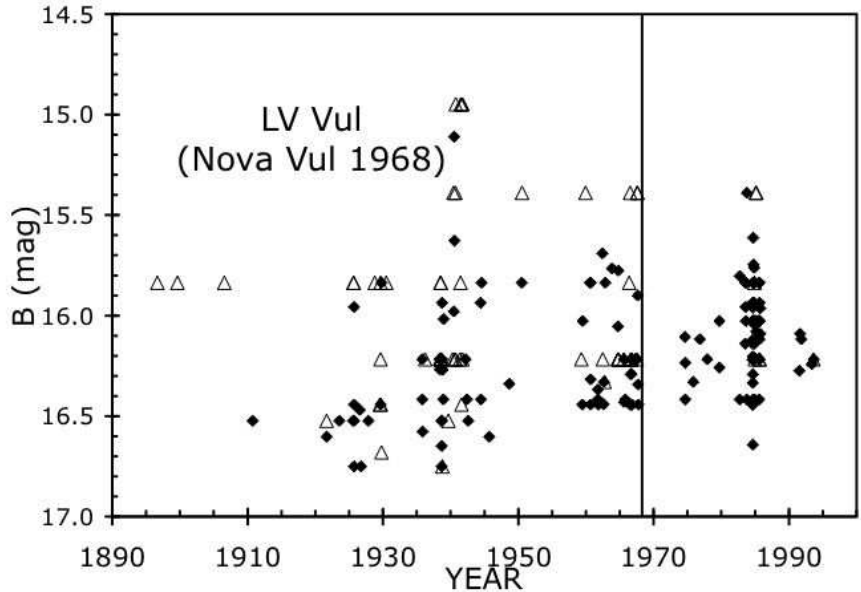}
  \end{center}
  \caption{LV Vul light curve. The diamonds denote measured B magnitudes and triangles denote limiting B magnitudes. The vertical line shows the date of the eruption of LV Vul. We observe no evidence of a pre-eruption rise of LV Vul, as the light curve is relatively flat (within the normal variability of the nova). Before the eruption, LV Vul averaged a brightness of 16.24, an RMS of 0.31 and a range of 1.64. After the eruption, we find LV Vul to average a brightness of 16.10, an RMS of 0.21 and a range of 1.25 mag. The nova also shows no significant change in its variability after the eruption compared to the variability shown before the eruption. Finally, we see no significant change in the nova's brightness after the eruption.}
  \label{LVVULLC}
\end{figure}

\clearpage

\begin{figure}[h!]
  \begin{center}
  \includegraphics[scale=0.5]{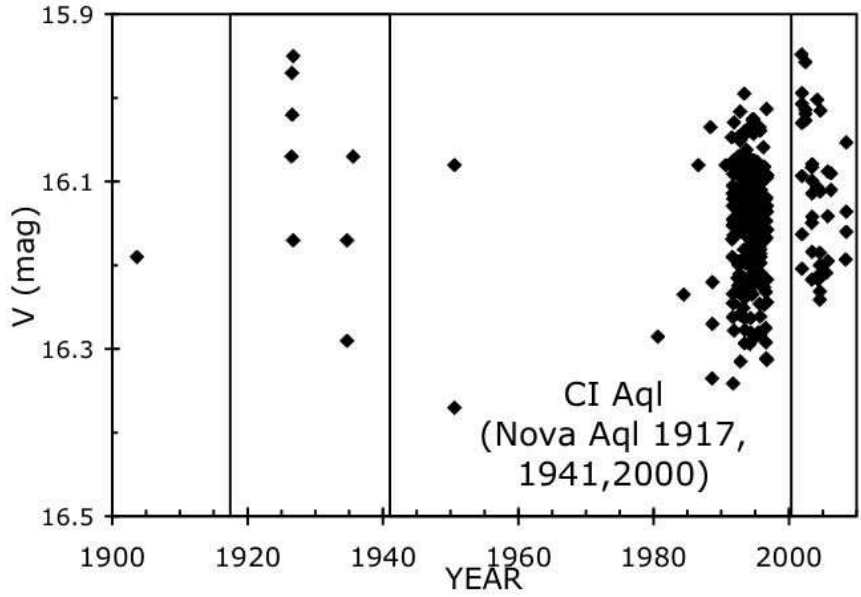}
  \end{center}
  \caption{Light curve of CI Aql in quiescence. The plot is of V-band magnitudes (with B-band magnitudes converted using B - V = 1.03 mag as from Schaefer (2009b)) taken from the Harvard plates, Schmidt Sky Surveys, RoboScope, McDonald Observatory, and Cerro Tololo. Eclipses are excluded from the light curve. The vertical lines mark observed outbursts of the recurrent nova. What we are left with is a flat light curve during quiescence.}
  \label{CIAQLLC}
\end{figure}

\clearpage

\begin{figure}[h!]
  \begin{center}
\includegraphics[scale=0.5]{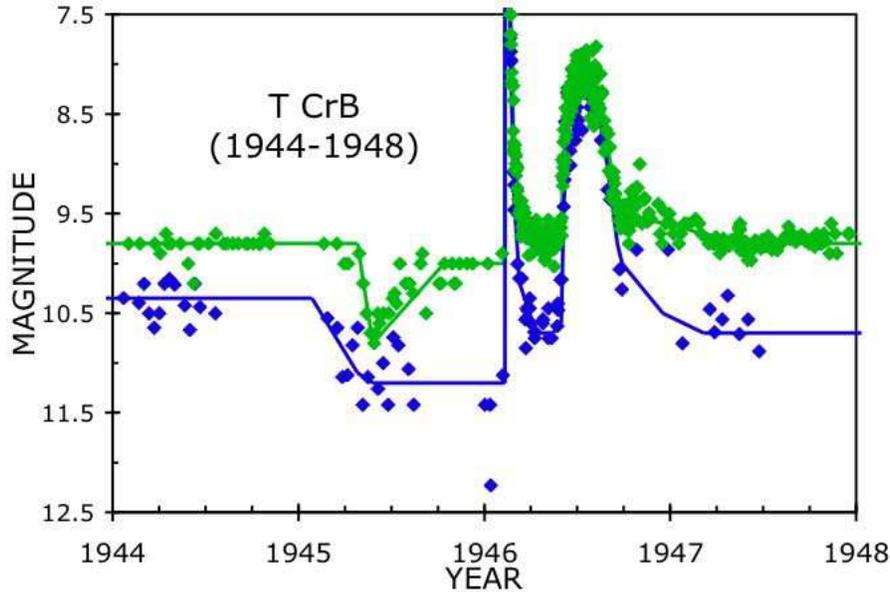}
  \end{center}
  \caption{T CrB between 1944 and 1948. The grey diamonds represent measurements in the V band and the black diamonds represent measurements in the B band. The lines for each set of data model the behavior of the nova in each band for details, see Schaefer (2009b). Here we observe two different phenomena. First, the light curve shows a significant and long-duration pre-eruption \textit{dip}, with the colors in the dip varying throughout the change in the system's brightness. Such an event suggests that the accretion rate onto the star went down. As with the pre-eruption rise observed in V533 Her, if we believe that this change in brightness was causally connected to the eruption, need an explanation of how. It is not clear how a lower accretion rate would somehow cause an eruption to occur. The second interesting feature of T CrB is that after eruption it appears to return to quiescence for nearly 50 days in both the V and B bands before spontaneously rising to eighth magnitude. Such a rise has not been seen in any other nova event. The cause of this event is also unclear.}
  \label{TCRBLC}
\end{figure}

\clearpage

\begin{figure}[h!]
  \begin{center}
\includegraphics[scale=0.5]{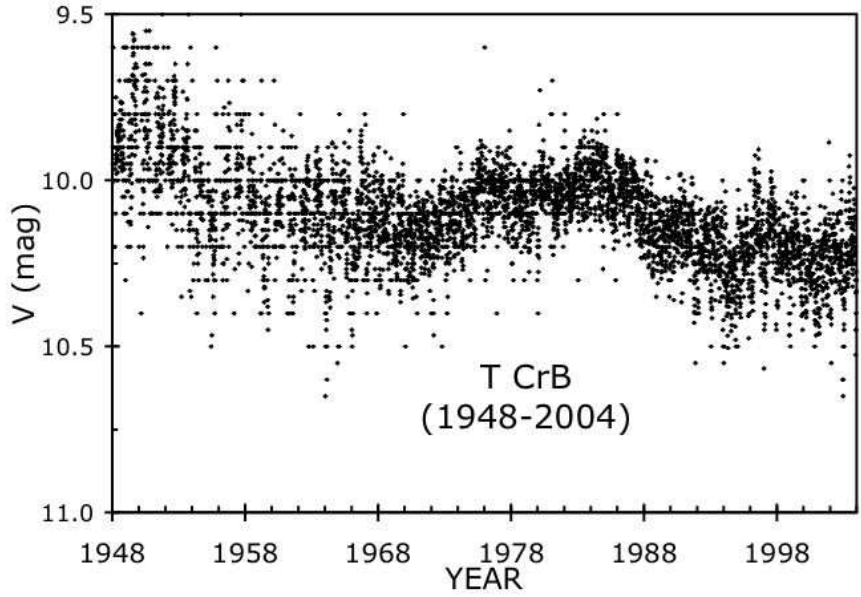}
  \end{center}
  \caption{T CrB in the 60 years of quiescence following the 1946 eruption. This light curve is composed of nearly 80,000 observations made by AAVSO observers of T CrB binned into 0.01 year intervals. The result is a complicated series of variations on a decadal time scale with an amplitude of roughly 0.4 mag as well as faster variations with an amplitude of $\approx$ 0.25 mag.}
  \label{TCRBLC2}
\end{figure}

\clearpage

\begin{figure}[h!]
  \begin{center}
 \includegraphics[scale=0.5]{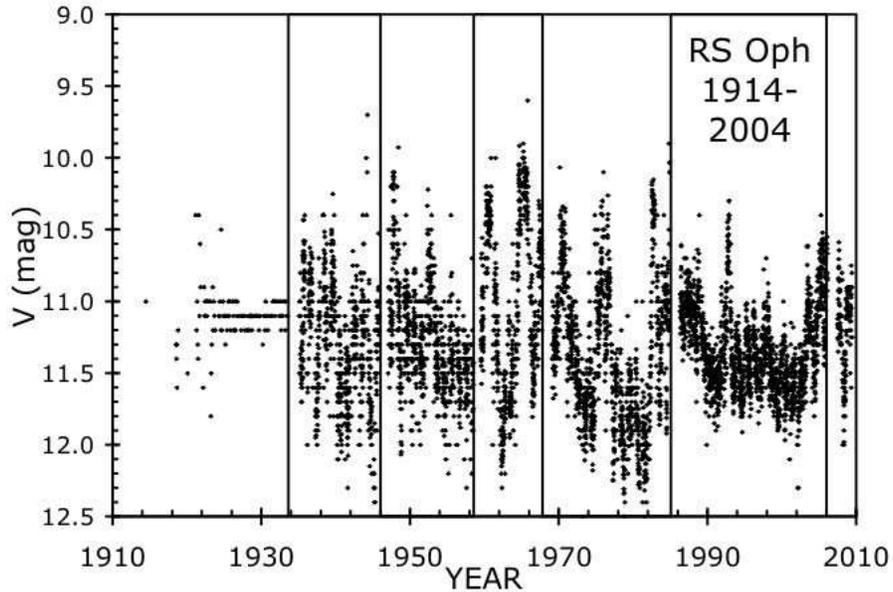}
  \end{center}
  \caption{RS Oph from 47,000 AAVSO V-band magnitudes measured between 1934 and 2004. The measurements are binned in time into 0.01 year intervals. The vertical lines denote observed outbursts of the nova. Here we have \textit{removed} eruptions and known eclipses so as to only observe the nova in quiescence. The light curve is very messy, showing variations on all time scales, and decadal episodes of relative calm and flares. While some eruptions may give in the indication of a pre-eruption rise or dip, the system shows no behavior that is abnormal. That is, since the system shows variability on all time scales, we believe that any rises or dips before an eruption are just a consequence of normal behavior for the novae, and not causally linked to an eruption.}
  \label{RSOPHLC}
\end{figure}

\clearpage

\begin{figure}[h!]
\begin{center}
\includegraphics[scale=0.25]{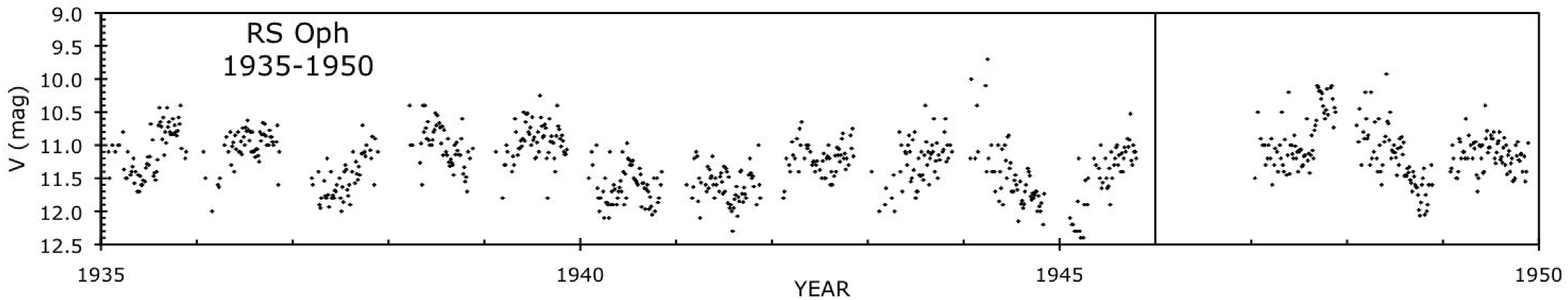}
\includegraphics[scale=0.25]{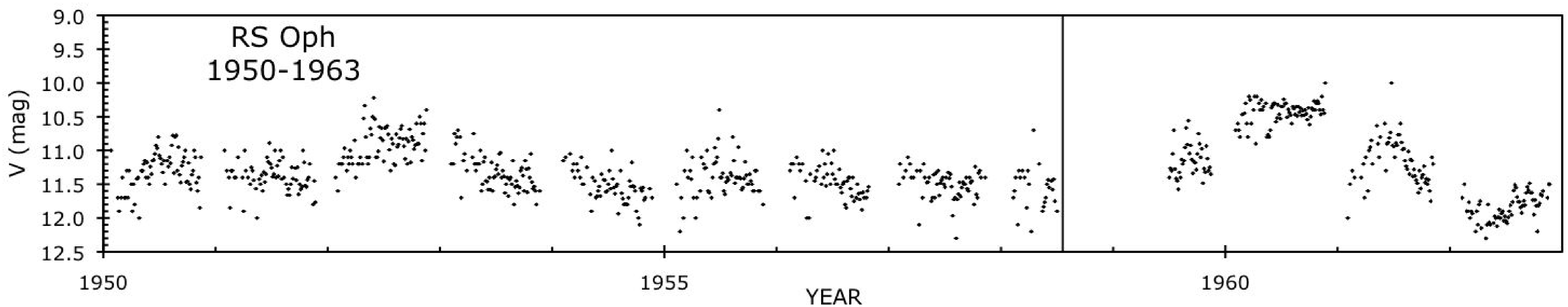}
\includegraphics[scale=0.25]{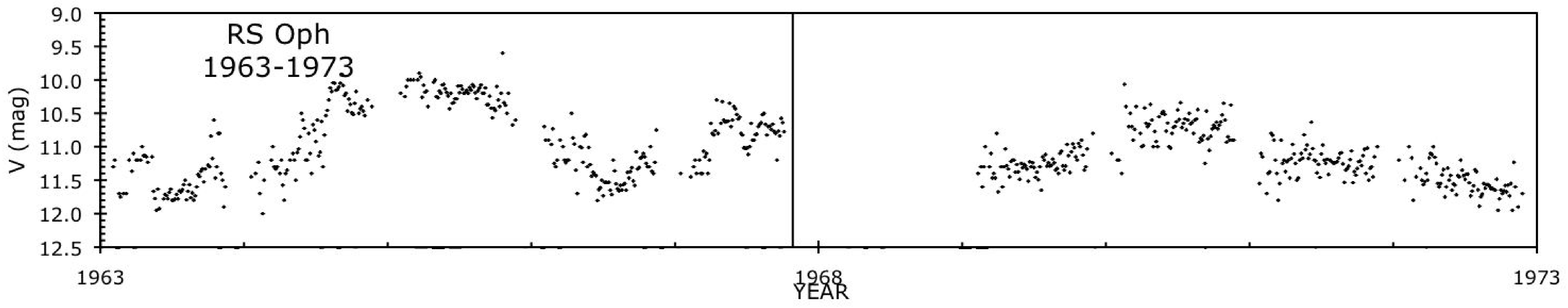}
\includegraphics[scale=0.25]{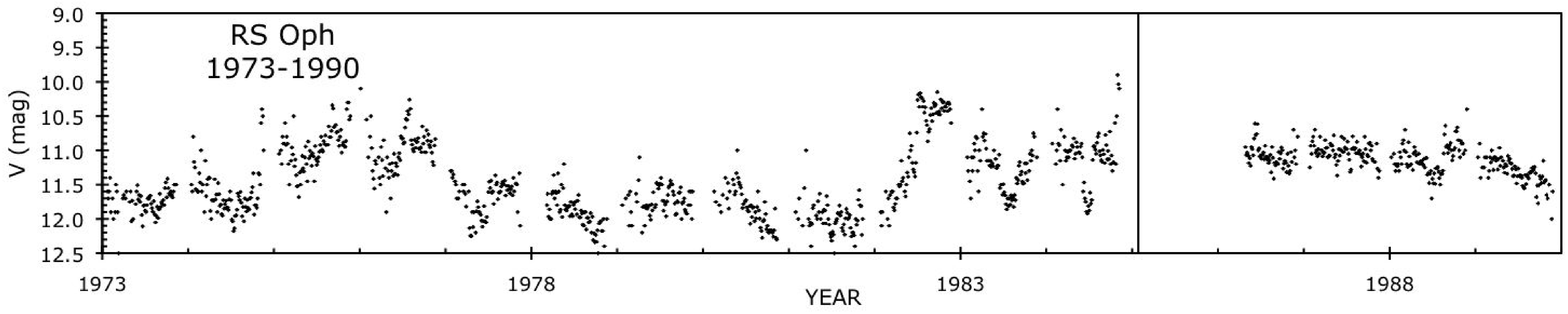}
\includegraphics[scale=0.25]{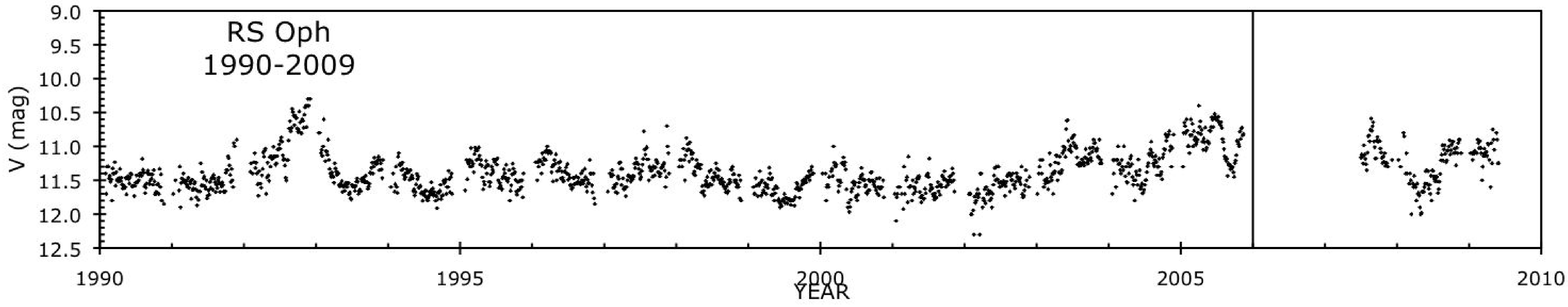}
\end{center}
\caption{Break down of the previous RS Oph light curve, zoomed in on the years leading up to and following the 1945, 1958, 1967, 1985 and 2006 eruptions of RS Oph. The large amount of variability is present throughout. There is no evidence of a pre-eruption anticipation event in RS Oph during any of these times.}
\label{RSOPHBREAKDOWN}
\end{figure}

\clearpage

\begin{figure}[h!]
  \begin{center}
 \includegraphics[scale=0.5]{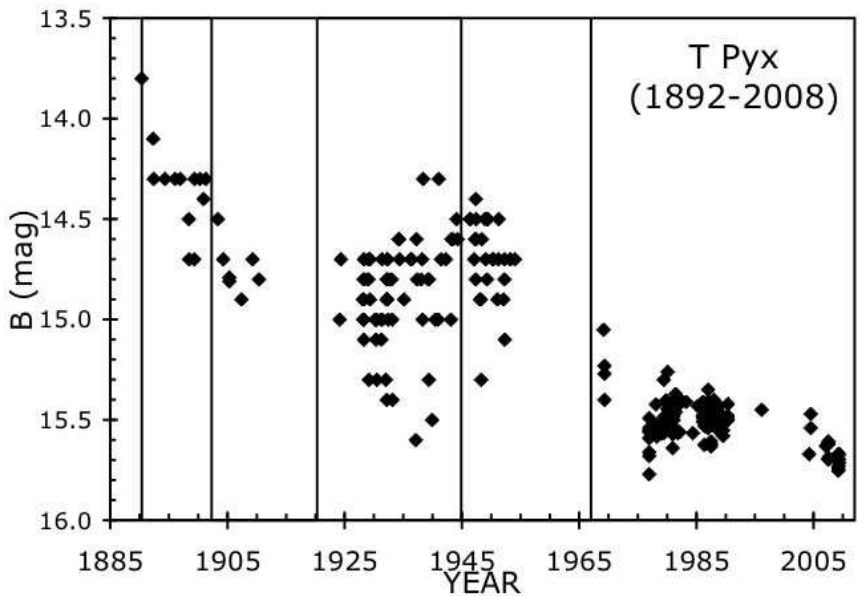}
  \end{center}
  \caption{T Pyx over the last 116 years.  Each diamond represents an individual measurement of the system, and the vertical lines represent outbursts. The light curve has substantial scatter, which is characteristic of usual variability. However, this variability is superimposed upon a more long term variation. T Pyx shows a systematic dimming since 1892, going from 13.8 to 15.7 mag. This is another example of how novae can and do have large-amplitude variability on decade to century time scales. This means that the accretion rate of matter onto the accretion disk must be changing on comparably long-term scales. We also note that T Pyx has shorter intervals between eruptions when it is brighter. This is easily explained by the requirement that the trigger for a nova occurs when some constant amount of matter has accumulated. Therefore, when T Pyx is bright, the matter accretes faster, so the next eruption happens sooner. Finally, the accretion rate diminished greatly after the eruption in 1967, so it will take a much longer time before the next eruption is triggered.}
 \label{TPYXLC}
\end{figure}

\clearpage

\begin{figure}[h!]
\begin{center}
\includegraphics[scale=0.5]{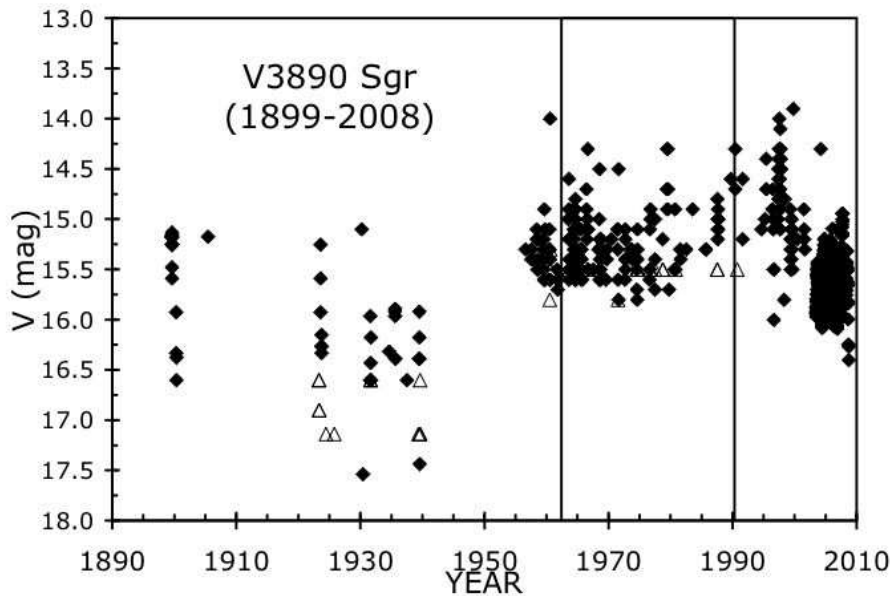}
\end{center}
\caption{V3890 Sgr in quiescence. In order from left to right, the long-term light curve is constructed from magnitudes corrected to the V band from Harvard plates, the Maria Mitchell plates, the AAVSO, and from SMARTS CCD data taken by Schaefer. Diamonds represent a measurement of the system, and the triangles represent limiting magnitude measurements. The vertical lines denote the dates of a known eruption of the nova. One problem in seeking secular changes is that the first three data sets have detection thresholds cutting off the distribution. The correction for these truncation effects will only increase the amplitude of variations for the first three data sets, and the large range of variability is already inconsistent with the small range observed recently in the SMARTS data. This inconsistency is likely caused by amplitude variations being smaller in red than in blue. No evidence of a pre-eruption rise or dip is visible. No change in brightness trend is visible following eruptions either. If we focus on variability of the nova just after it returns to quiescence, it does not appear that the variability of the nova is affected by the eruptions. The recent CTIO data does show a smaller variability than what has previously been seen, however, this change is too far after the 1990 eruption to be causally linked. }
\label{V3890SGRLC}
\end{figure}

\clearpage

\begin{figure}[h!]
  \begin{center}
\includegraphics[scale=0.5]{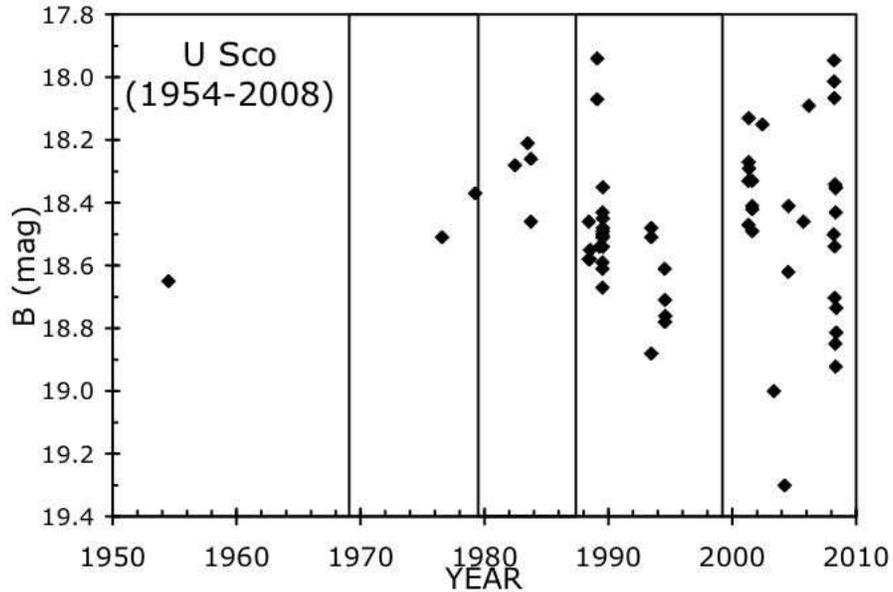}
  \end{center}
  \caption{The quiescent light curve for U Sco composed of data from UK Schmidt plates, CTIO measurements, Harvard Plates, Sonneberg Plates, AAVSO measurements, and the literature. Schaefer (2009b) gives a detailed account of each data point and its source. Diamonds mark measurements in the B band, and vertical lines denote each eruption of U Sco. We see U Sco varying on the order of $\sim$ 1 magnitude. This variation shows no apparent connection to the outbursts, nor does it appear to change as the result of an outburst. For the one event for which we have pre-eruption data close to the eruption, 1979, we do not see any evidence of a pre-eruption rise. We also observe no change in the brightness of U Sco on a long-term scale.}
  \label{USCOLC}
\end{figure}

\end{document}